\documentclass[useAMS,usenatbib]{mn2e}

\usepackage{amsmath}
\usepackage{amssymb}
\usepackage{graphicx}
\usepackage{graphics}
\usepackage{float}
\usepackage{epstopdf}
\usepackage{multicol}
\usepackage{breakurl}
\usepackage{textcomp}
\usepackage{enumitem}
\usepackage{subfigure}
\usepackage[bf,labelsep=period,font=small]{caption}
\usepackage{subfloat}
\usepackage{subfigure}
\usepackage{gensymb}
\usepackage{supertabular}
\usepackage{longtable,ltxtable}
\usepackage{afterpage}
\usepackage[flushleft]{threeparttable}
\title[mCP Stars Within 100pc I]{A Volume-Limited Survey of mCP Stars Within 100pc \\I: Fundamental Parameters and Chemical Abundances}
\author[J. Sikora et al.]
	{J.~Sikora,$^{1,2}$ G.~A.~Wade,$^{2}$ J.~Power,$^{1,2,4}$ C.~Neiner$^3$\\
$^{1}$Department of Physics, Engineering Physics \& Astronomy, Queen's University, Kingston, ON Canada, K7L 3N6\\
$^{2}$Department of Physics and Space Science, Royal Military College of Canada, PO Box 17000 Kingston, Ontario K7K 7B4, Canada\\
$^3$LESIA, Observatoire de Paris, PSL University, CNRS, Sorbonne Université, Univ. Paris Diderot, Sorbonne Paris Cité, \\\hspace{1cm}5 place Jules Janssen, 92195 Meudon, France\\
$^4$Large Binocular Telescope Observatory, 933 North Cherry Avenue, Tucson, AZ 85721, USA}

\begin{document}

\date{Accepted 2018 Nov. 13}

\pagerange{\pageref{firstpage}--\pageref{lastpage}} \pubyear{2018}

\maketitle

\label{firstpage}

\begin{abstract}
We present the first results of a volume-limited survey of main sequence (MS) magnetic chemically 
peculiar (mCP) stars. The sample consists of all identified intermediate-mass MS stars (mCP and non-mCP) 
within a heliocentric distance of $100\,{\rm pc}$ as determined using Hipparcos parallaxes. The two 
populations are compared in order to determine the unique properties that allow a small fraction of MS 
stars with masses $\gtrsim1.4\,M_\odot$ to host strong, large scale magnetic fields. A total of 52 
confirmed mCP stars are identified using published magnetic, spectroscopic, and photometric 
observations along with archived and newly obtained spectropolarimetric (Stokes $V$) observations. We 
derive the fundamental parameters (effective temperatures, luminosities, masses, and evolutionary 
states) of the mCP and non-mCP populations using homogeneous analyses. A detailed analysis of the mCP 
stars is performed using the {\sc llmodels} code, which allows observed spectral energy distributions to 
be modeled while incorporating chemical peculiarities and magnetic fields. The surface gravities and 
mean chemical abundances are derived by modelling averaged spectra using the {\sc gssp} and {\sc zeeman}
spectral synthesis codes. Masses and stellar ages are derived using modern, densely calculated 
evolutionary model grids. We confirm a number of previously reported evolutionary properties associated 
with mCP stars including a conspicuously high incidence of middle-aged MS stars with respect to the 
non-mCP subsample; the incidence of mCP stars is found to sharply increase with mass from 
$0.3$~per~cent at $1.5\,M_\odot$ to $\approx11$~per~cent at $3.8\,M_\odot$. Finally, we identify clear 
trends in the mean photospheric chemical abundances with stellar age.
\end{abstract}

\begin{keywords}
Stars: early-type, Stars: chemically peculiar, Stars: magnetic
\end{keywords}

\section{Introduction}\label{sect:intro}

Various studies of magnetic chemically peculiar (mCP) stars have been carried out over the past several 
decades revealing a wide range of characteristic properties. As their name suggests, these objects are 
defined by (1) the presence of enhanced or deficient abundances of specific elements (relative to solar 
abundances) that are often inhomogeneously distributed within their atmospheres and (2) by the presence 
of strong, large-scale surface magnetic fields \citep[typically $\gtrsim100\,{\rm G}$, 
e.g.][]{Landstreet1982,Auriere2007} that exhibit significant dipole components 
\citep[e.g.][]{Stibbs1950,Landstreet1992,Silvester2015}.

The origin of the magnetic fields hosted by mCP stars is still unclear 
\citep[e.g.][]{Moss2003a,Moss2004}; however, it is generally accepted that the observed fields are most 
likely not being actively generated by dynamo processes such as those believed to be taking place in 
the envelopes of cool main sequence (MS) stars 
\citep[e.g.][]{Charbonneau2001,Charbonneau2005,Donati2009}. Arguably, the most plausible explanation is 
given by the \emph{fossil field} model \citep{Cowling1945}, which states that the fields are 
slowly-decaying remnants (possibly originating from the interstellar medium or a pre-MS dynamo) from an 
earlier phase during the star's evolution. This theory is consistent with certain properties of the mCP 
stars' fields such as their typically simple structures and their long-term stability. The studies 
carried out by \citet{Braithwaite2006a} and \citet{Duez2010} also provide important theoretical 
evidence in support of the fossil field theory: the two studies demonstrate that stable field 
configurations within the radiative envelopes of upper MS stars -- which the fossil field model 
demands -- are a natural consequence of the magnetohydrodynamic relaxation of arbitrary 
initial stochastic fields.

Despite the successes of the fossil field theory, a number of questions regarding the origin of tepid (and 
hot) star magnetism remain unanswered. For instance, assuming that the fields do originate within the 
interstellar medium prior to the star's formation, it is unclear why only a small fraction of MS A-type 
stars are (strongly) magnetic \citep[e.g.][]{Shorlin2002,Auriere2007}. Moreover, it is noted that 
sub-surface convection zones, meridional circulation, and differential rotation are present to varying 
degrees in early F-, A-, and B-type MS stars. As discussed by \citet{Braithwaite2013}, the extent to 
which these processes may disrupt a fossil field remains an important, unanswered question.

There are also unanswered questions regarding the processes that lead to the horizontal 
\citep[e.g.][]{Kochukhov2015,Yakunin2015,Kochukhov2017a} and vertical 
\citep[e.g.][]{Babel1994,Ryabchikova2005,Rusomarov2015} chemical abundance stratification in mCP 
stars. In non-magnetic chemically peculiar (CP) stars, it is believed that sufficiently stable 
environments (established by slow rotational velocities, suppressed convection, etc.) enable radiation 
pressure to become dominant thereby allowing specific elements to accumulate within the star's 
atmosphere \citep[e.g.][]{Michaud1970,Michaud1976,Michaud1980}. The magnetic fields present in the 
atmosphere's of mCP stars are also expected to contribute to the environment's stability; however, it 
is only relatively recently that magnetic fields have been incorporated into atomic diffusion models 
\citep{Leblanc2004,Leblanc2009,Alecian2010,Alecian2015a}. \citet{Kochukhov2017} found that the most 
recent of these theories \citep{Stift2016} cannot reproduce the average surface abundances of Fe-peak 
elements observed in mCP stars \citep[e.g.][]{Sargent1967,Ryabchikova2005b,Ryabchikova2008}.

The fundamental, chemical, and magnetic properties of mCP stars that have been tested against 
various models are often inferred from inherently biased surveys: they are either biased towards the 
brightest objects or those exhibiting the strongest, and thus, most easily detectable magnetic 
fields. One way in which the influence of observer biases can be reduced is by carrying out a 
volume-limited survey. Such a survey was initiated by \citet{Power2007_MSc}, who used the Hipparcos 
catalogue and the Catalogue of Ap, HgMn and Am stars \citep{Renson1991,Renson2009} to identify a sample 
of intermediate-mass MS stars within a heliocentric distance of $100\,{\rm pc}$. Their work involved 
the acquisition of a large number of spectropolarimetric Stokes $V$ measurements using the (now retired) 
MuSiCoS instrument. These observations allowed for the magnetic parameters of approximately $50\,\%$ of 
the mCP stars in the sample to be derived. More recently, we have completed this survey primarily using 
Stokes $V$ measurements obtained with ESPaDOnS along with a small number obtained with NARVAL.

Here we present the first results from the volume-limited spectropolarimetric survey of mCP stars 
located within $100\,{\rm pc}$ that was initiated in 2007. The goal of this survey is to constrain 
various fundamental, chemical, and magnetic properties of this stellar population while attempting to 
minimize observational biases. Moreover, since the volume includes a large, effectively complete sample 
of non-mCP stars with accurate photometric measurements, we can directly compare the magnetic and 
non-magnetic populations. This provides a unique opportunity to identify the fundamental properties 
that distinguish the mCP and non-mCP stars. The results presented here will serve as a starting point 
for the magnetic analysis that will be presented in a subsequent publication, hereinafter referred to 
as ``Paper II".

In Section \ref{sect:sample}, we discuss the sample selection including the identification of both 
confirmed and candidate mCP stars. In Section \ref{sect:obs}, we briefly summarize the 
spectropolarimetric observations that were used to (1) confirm whether or not certain candidate mCP 
stars were magnetic and (2) constrain surface gravities and mean chemical abundances. In Section 
\ref{sect:complete}, we assess the completeness of the mCP sample. In Sections \ref{sect:fund_param} 
and \ref{sect:results}, we present the details of the analysis along with our results. Finally, in 
Section \ref{sect:discussion}, we summarize and discuss the main results from this work.

\section{Sample selection}\label{sect:sample}

\subsection{Hipparcos Sample}\label{sect:hipp_sample}

The sample of MS early-F-, A-, and B-type stars was selected from the Hipparcos 
catalogue \citep{ESA1997}, which consists of $118\,218$ sources (i.e. single stars, visual 
multi-star systems, and gravitationally bound multi-star systems) within the solar neighbourhood. It is 
complete down to magnitudes of $V=7.3-9.0\,{\rm mag}$ depending on the Galactic latitude and spectral 
type \citep{Perryman1997a}. The Tycho catalogue associated with the Hipparcos mission is essentially 
complete down to magnitudes of $V\lesssim9\,{\rm mag}$ regardless of sky coordinates. 
\citet{Mignard2000} estimated the completeness of the Hipparcos sample as a function of Galactic 
latitude and $V$ magnitude \citep[Fig. 4 of][]{Mignard2000} by comparing the number of sources listed 
in the Hipparcos and Tycho catalogues. These results can be used to assess the level of completeness of 
our volume-limited sample of MS early-F, A-, and B-type stars taken from the Hipparcos Catalogue by 
identifying the stars in our sample having faintest $V$ magnitudes.

The coolest stars in our sample have effective temperatures of $T_{\rm eff}\approx7\,000\,{\rm K}$ with 
corresponding absolute $V$ magnitudes of $M_V\approx3.2\,{\rm mag}$. At a distance of $100\,{\rm pc}$, 
such stars will have apparent magnitudes of $V\approx8.2\,{\rm mag}$. Comparing with Fig. 4 of 
\citet{Mignard2000}, it is apparent that, at $V=8.2\,{\rm mag}$, the Hipparcos Catalogue exhibits a 
minimum completeness of $\approx50$~per~cent near the Galactic plane. With this in mind, we carried out 
the following analysis in order to better evaluate the completeness of our sub-sample of MS stars within 
the Hipparcos Catalogue that are known to have distances $d<100\,{\rm pc}$.

Motivated by \citet{Mignard2000}, we estimated the completeness of our $100\,{\rm pc}$ sample by 
comparing (1) the number of stars in the Hipparcos Catalogue that satisfy our selection criteria 
(i.e. MS stars with $T_{\rm eff}\gtrsim7\,{\rm kK}$ and $d<100\,{\rm pc}$) and (2) the number of 
stars in the Tycho catalogue that either likely or potentially satisfy these criteria. In order to 
potentially satisfy the criteria, a star in the Tycho catalogue must have $V<8.2\,{\rm mag}$, 
$T_{\rm eff}\gtrsim7\,{\rm kK}$, and a distance that is either known to be $<100\,{\rm pc}$ or that 
could be $<100\,{\rm pc}$ and exhibit a luminosity greater than that of a zero-age MS star (the minimum 
$T_{\rm eff}$ and minimum luminosity values are based on evolutionary models generated by 
\citet{Ekstrom2012} and \citet{Mowlavi2012} and are discussed in more detail in Sect. \ref{sect:HRD}). We 
roughly estimated $T_{\rm eff}$ and the bolometric corrections (BCs) for all of the stars in the Hipparcos 
and Tycho catalogues \citep{ESA1997} with known $B$ and $V$ magnitudes by applying the $T_{\rm eff}$ and 
BC calibrations of \citet{Gray2005_SP} and \citet{Balona1994}, respectively. The luminosities for those 
stars with available parallax (distance) measurements were then derived using the estimated $T_{\rm eff}$ 
and BC values. For the Tycho catalogue stars without available parallax measurements, we assigned 
$L(T_{\rm eff})$ values based on the ZAMS associated with the evolutionary tracks generated by 
\citet{Ekstrom2012} and \citet{Mowlavi2012}. Adopting this value of $L(T_{\rm eff})$ then yields the 
minimum distance the star could have such that it would remain in our sample after applying the cuts 
discussed in Sect. \ref{sect:HRD}. We find that the estimated completeness of the Hipparcos Catalogue as 
it pertains to our volume-limited survey of MS stars increases approximately monotonically from 
87~per~cent at $T_{\rm eff}\sim6.5\,{\rm kK}$ to 100~per~cent at $T_{\rm eff}\sim19\,{\rm kK}$. A 
similar trend is found by deriving the masses associated with the estimated $T_{\rm eff}$ and $L$ 
values (see Sect. \ref{sect:HRD} for a description of the method by which the masses were derived): 
the completeness increases from $\sim87$~per~cent for masses of $M\sim1.4\,M_\odot$ to 
$\sim100$~per~cent for $M\gtrsim3\,M_\odot$. We conclude that our results are not likely to be 
significantly impacted by objects that are missing from the Hipparcos Catalogue (we discuss this in 
Sect. \ref{sect:mass_dist} in the context of the incidence rates of mCP stars as a function of mass).

Out of the $118\,218$ sources in the Hipparcos Catalogue, $23\,046$ have associated parallax angles -- 
as reported by \citet{VanLeeuwen2007} -- of ${\rm\pi}>10\,{\rm mas}$ corresponding to distances of 
$d<100\,{\rm pc}$. Referring to the Double and Multiple Systems Annex, we find that $5\,714$ of these 
sources are either composed of or are members of visual or gravitationally bound multi-star systems; we 
categorize these systems based on whether or not each of its components have equivalent parallax angles, 
which are taken from \citet{ESA1997} when no values are reported by \citet{VanLeeuwen2007}. The subset 
of $23\,046$ sources are then found to contain $2\,584$ ($611$) astrometrically resolved (unresolved) 
gravitationally bound multi-star systems. After taking into account the individual components of the 
resolved systems, a total of $25\,720$ sources within $100\,{\rm pc}$ were extracted from the Hipparcos 
catalogue.

The volume-limited survey presented here is not subject to the Malmquist bias associated with 
magnitude-limited surveys; however, since the sample selection is based on trigonometric parallax 
measurements, the sample is affected by the Lutz-Kelker (Trumpler-Weaver) bias 
\citep{Trumpler1953_oth,Lutz1973}. The effect of this bias is to systematically shift the inferred 
absolute magnitudes for a sample of stars from their true absolute magnitudes. The degree of this shift 
($\Delta M$) depends on the sample's average relative parallax uncertainty, 
$\langle\sigma_{\rm \pi}/{\rm \pi}\rangle$ \citep{Francis2013,Francis2014}. The $25\,720$ stars within 
$100\,{\rm pc}$ that were extracted from the Hipparcos Catalogue exhibit 
$\langle\sigma_{\rm\pi}/{\rm\pi}\rangle=0.078$; using the quadratic approximation for $\Delta M$ 
recommended by \citet{Francis2013}, we obtain $\Delta M=0.007\,{\rm mag}$. Therefore, the bias is 
considered to be negligible and no correction was applied.

\subsection{mCP Subsample}\label{sect:mCP_sample}

The most definitive method by which all of the mCP stars within $100\,{\rm pc}$ can be identified is by 
obtaining costly spectropolarimetric measurements of each early-F-, A-, and B-type MS star. This is 
unfeasible considering that the sample consists of several thousand such objects. We primarily relied 
on the Catalogue of Ap, HgMn and Am stars \citep{Renson2009} in order to identify both candidate 
and confirmed mCP stars; this allowed us to prioritize specific stars for which new magnetic measurements 
would be obtained. The catalogue, first compiled by \citet{Renson1991} and updated several times to 
include new objects, contains $3\,652$ stars classified as ``definite", ``probable", or ``doubtful" Ap 
stars, where ``Ap" is used broadly to include the magnetic Fp, Ap, Bp, He-weak, and He-strong stars. 
These classifications are based on whether or not (1) surface magnetic fields have been detected 
\cite[e.g.][]{Borra1980,Wade2000a,Bagnulo2006}, (2) characteristic flux abnormalities have been detected 
based, for instance, on $\Delta a$ or $\Delta(V_1-G)$ photometric colour indices 
\cite[e.g.][]{Maitzen1998,Bayer2000,Paunzen2005}, or (3) chemical peculiarities such as enhanced 
Si, Cr, or Sr have been reported \cite[e.g.][]{Cowley1969,Cucchiaro1978,Jaschek1980}.

We cross-referenced the list of $25\,720$ Hipparcos sources with the $3\,652$ definite Ap (definite 
mCP), probable Ap (probable mCP), and doubtful Ap (doubtful mCP) stars compiled by \citet{Renson2009} 
yielding a total of 141 potential mCP stars within $100\,{\rm pc}$: 47 of these stars are classified as 
definite, 27 as probable, and 67 as doubtful mCP stars. We also searched the Catalogue of Stellar 
Spectral Classifications compiled by \citet{Skiff2014} for stars which had been assigned a `p' 
classification (e.g. F3p, A0p, etc.) at some point in time. Cross-referencing this catalogue with the 
$25\,720$ Hipparcos sources yielded 65 stars not included in the most recently published Catalogue of 
Ap, HgMn and Am stars. Our tentative list of candidate mCP stars therefore consists of 206 stars.

\begin{table}
	\caption{List of the 52 confirmed mCP stars and 3 candidate mCP stars within the sample. Columns 1 
	to 3 list each stars' HD~number, identifier (if available), and mCP membership satus: confirmed mCP 
	stars are indiciated by a `Y', candidate mCP stars are indicated by a `U'.}
	\label{tbl:mCP_list}
	\begin{center}
	\begin{tabular}{@{\extracolsep{\fill}}l c c @{\hskip 0.59cm} c c r@{\extracolsep{\fill}}}
		\noalign{\vskip-0.4cm}
		\hline
		\hline
		\noalign{\vskip0.5mm}
		HD~& ID & mCP? & HD~& ID & mCP? \\
		\noalign{\vskip0.5mm}
		\hline	
		\noalign{\vskip0.5mm}
		3980   & $\xi$ Phe         & Y & 112413  & $\alpha^2$ CVn & Y \\
		11503  & $\gamma^2$ Ari    & Y & 117025  &                & Y \\
		12447  & $\alpha$ Psc A    & Y & 118022  & o Vir          & Y \\
		15089  & $\iota$ Cas       & Y & 119213  & CQ UMa         & Y \\
		15144  &                   & Y & 120198  & 84 UMa         & Y \\
		15717  &                   & U & 124224  & CU Vir         & Y \\
		18296  & 21 Per            & Y & 128898  & $\alpha$ Cir   & Y \\
		24712  & DO Eri            & Y & 130559  & $\mu$ Lib      & Y \\
		27309  & 56 Tau            & Y & 137909  & $\beta$ CrB    & Y \\
		29305  & $\alpha$ Dor      & Y & 137949  & 33 Lib         & Y \\
		32576  &                   & U & 140160  & $\chi$ Ser     & Y \\
		38823  & V1054 Ori         & Y & 140728  & BP Boo         & Y \\
		40312  & $\theta$ Aur      & Y & 148112  & $\omega$ Her   & Y \\
		49976  & V592 Mon          & Y & 148898  & $\omega$ Oph   & Y \\
		54118  & V386 Car          & Y & 151199  &                & Y \\
		56022  & L01 Pup           & Y & 152107  & 52 Her         & Y \\
		62140  & 49 Cam            & Y & 170000  & $\phi$ Dra     & Y \\
		64486  &                   & Y & 176232  & 10 Aql         & Y \\
		65339  & 53 Cam            & Y & 187474  & V3961 Sgr      & Y \\
		72968  & 3 Hya             & Y & 188041  & V1291 Aql      & Y \\
		74067  & NY Vel            & Y & 201601  & $\gamma$ Equ   & Y \\
		83368  &                   & Y & 203006  & $\theta^1$ Mic & Y \\
		96616  & 46 Cen            & Y & 217522  & BP Gru         & Y \\
		103192 & $\beta$ Hya       & Y & 217831  &                & U \\
		108662 & 17 Com            & Y & 220825  & $\kappa$ Psc   & Y \\
		108945 & 21 Com            & Y & 221760  & $\iota$ Phe    & Y \\
		109026 & $\gamma$ Mus      & Y & 223640  & i03 Aqr        & Y \\
		112185 & $\varepsilon$ UMa & Y &         &                &   \\
		\noalign{\vskip0.5mm}
		\hline \\
	\end{tabular}
	\end{center}
\end{table}

Using both published magnetic measurements and archived spectropolarimetric data, we conclude that 50 of 
the 206 stars are magnetic and that 8 are either non-mCP or are weakly-magnetic since no Zeeman 
signatures were detected (both possibilities suggest that they are not mCP stars). Additionally, no 
magnetic detections were yielded from measurements of 4 stars obtained by the BinaMIcS collaboration 
\citep{Binamics2013} and of 6 stars obtained by the BRITE-pol collaboration \citep{Britepol2017} 
(private communications). No magnetic measurements were available for 138 of the 206 mCP stars and 
therefore, they could not be confirmed or rejected as mCP members on the basis of magnetic 
measurements.

The characteristic chemical peculiarities associated with mCP stars can be reliably detected using both 
UV and optical spectroscopic measurements having relatively low dispersion/resolution 
\citep[e.g.][]{Cowley1973,Cucchiaro1976,Abt1979}. Moreover, mCP stars can be effectively identified 
using photometric measurements to detect abnormal flux depressions in the UV near $4\,100$, $5\,200$, 
and $6\,300\,{\rm \AA}$ \citep{Kodaira1969,Adelman1975a,Adelman1979}. The $\Delta a$ photometric system 
introduced by \citet{Maitzen1976} was specifically designed to be sensitive to the $5\,200\,{\rm \AA}$ 
flux depression. A comparable sensitivity to the UV flux depressions can also be obtained from the 
$\Delta(V_1-G)$ and $Z$ indices derived from filters associated with the Geneva photometric system 
\citep{Golay1972,Hauck1982,Cramer1999}.

For the remaining 138 candidate mCP stars, we compiled photometric- and spectroscopic-based evidence 
of mCP membership from the literature. In a number of cases, stars classified as Fp, Ap, or Bp were done 
so based solely on abnormally weak individual metal lines (e.g. weak $\lambda4481$ Mg~{\sc ii}) or were 
likely misclassified Fm, Am, or $\lambda$ Bootis stars \citep[e.g.][]{Abt1995}. We used the criteria 
for detecting CP stars proposed by \citet{Paunzen2005}, which are based on $\Delta a$, $\Delta(V_1-G)$, 
and $Z$ measurements. The criteria were then applied to the available $\Delta a$, $\Delta(V_1-G)$, and 
$Z$ measurements published by \citet{Rufener1988} and \citet{Paunzen2005}. In total, 128 stars were 
found to (1) exhibit photometric measurements consistent with normal or non-mCP stars or (2) were 
reported to have atmospheric chemical compositions that are not typically associated with mCP stars 
(e.g. non-peculiar, enhanced Hg and Mn, etc.) and were therefore removed from the list of candidate mCP 
stars.

Following our derivation of the fundamental parameters (presented in Section \ref{sect:fund_param}), we 
identified three candidate mCP stars in our sample (HD~107452, HD~122811, and HD~177880B) which are likely 
located beyond the $100\,{\rm pc}$ distance limit despite the parallax values reported in the Hipparcos 
Catalogue. These stars were therefore removed from the sample; this decision is justified in more detail 
in Section \ref{sect:HRD}.

In summary, prior to obtaining new magnetic measurements, we identified 50 definite mCP stars and 7 
candidate mCP stars. Only one of the candidate mCP stars, HD~15717, was not found in the catalogue 
compiled by \citet{Renson2009} and was instead identified using the spectral types compiled by 
\citet{Skiff2014}.

\section{Spectropolarimetric observations}\label{sect:obs}

Aside from the 7 candidate mCP stars that required magnetic measurements in order to confirm or 
reject their mCP membership, we also identified a significant fraction of the known mCP stars within 
the sample for which their magnetic field strengths and geometries could not be sufficiently 
constrained using previously published and archived magnetic measurements. We therefore proceeded to 
acquire circularly polarized (Stokes $V$) spectropolarimetric observations in an attempt to overcome 
these two deficiencies. 

The observations were primarily used to detect and measure the stars' longitudinal field strengths; however, 
the analysis of the fundamental and chemical parameters of the sample that we present here does depend 
on whether or not we were able to confirm or reject certain candidate mCP stars as being true mCP members. 
Furthermore, this analysis uses a number of the unpolarized (Stokes $I$) spectra that were acquired with 
the Stokes $V$ measurements. Therefore, we briefly summarize the observations here and identify which 
stars we confirmed as mCP members. The observations are discussed in more detail in Paper II in which we 
present the magnetic analysis that was performed on the confirmed mCP stars in the sample.

We obtained 118 new Stokes $V$ observations of 40 mCP and candidate mCP sample stars using the twin 
spectropolarimeters ESPaDOnS and NARVAL installed at the Canada-France-Hawaii Telescope (CFHT) and 
the T\'elescope Bernard Lyot (TBL), respectively. These instruments have a resolving power of 
$R\sim65\,000$ and yield optical spectra within a wavelength range of 
$3\,600\lesssim\lambda\lesssim10\,000\,{\rm \AA}$. Additionally, we analyzed 151 archival Stokes $V$ 
observations acquired using the now retired MuSiCoS spectropolarimeter 
($R\sim35\,000$, $3\,900\lesssim\lambda\lesssim8\,700\,{\rm \AA}$) that was installed at TBL prior to 
NARVAL's installation. The MuSiCoS observations were obtained from Feb. 12, 1998 to June 8, 2006.

Based on the new MuSiCoS, ESPaDOnS, and NARVAL Stokes $V$ measurements, we confirmed that 2 of the 
candidate mCP stars are magnetic and that 2 are likely not mCP stars since no detections were obtained. 
In Table \ref{tbl:mCP_list} we list the 52 confirmed mCP stars located within $100\,{\rm pc}$ along with 
the 3 candidate mCP stars which we did not observe.

\section{Distribution and completeness}\label{sect:complete}

As discussed in Section \ref{sect:hipp_sample}, the Hipparcos Catalogue of early-F-, A-, and B-type MS 
stars can reliably be considered to be complete within $100\,{\rm pc}$; however, the completeness of 
the subsample of mCP stars needs to be assessed. Magnetic measurements, which provide the 
most conclusive evidence of mCP membership, have not been obtained for the vast majority of the larger 
sample of Hipparcos stars. On the other hand, spectroscopic and photometric measurements, some of which 
can be used to distinguish between mCP stars and non-mCP stars, are available for many of the stars in the 
sample. We estimated the completeness of the mCP subsample by determining the fraction of the total sample 
for which such mCP-sensitive measurements have been reported in the literature. Note that the following 
estimates correspond to the sample of early-F-, A-, and B-type MS stars having masses $>1.4\,M_\odot$ 
(derived in Section \ref{sect:HRD}), which consists of $3\,254$ stars.

\begin{figure}
	\centering
	\includegraphics[width=1.0\columnwidth]{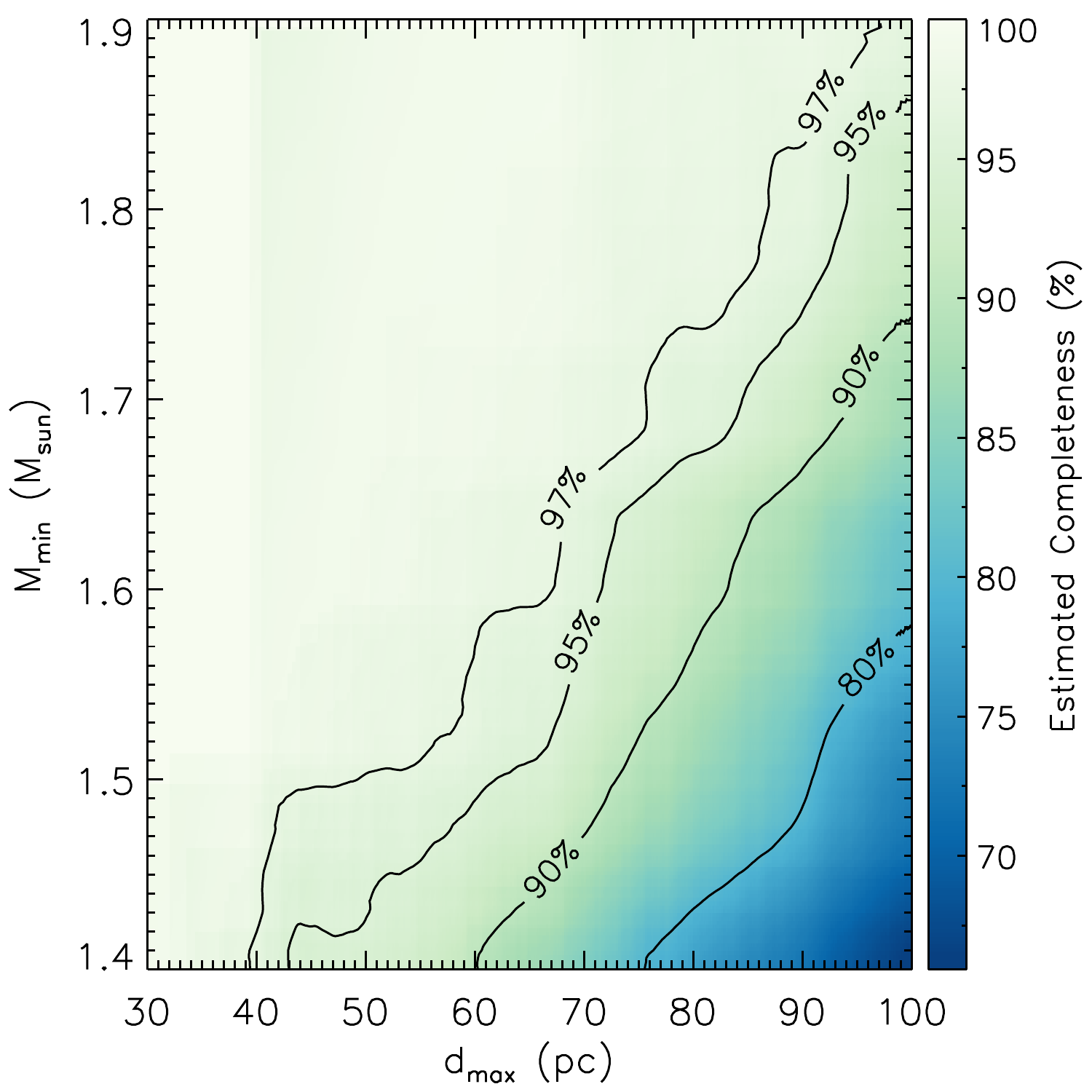}
	\caption{Estimated completeness of the mCP subsample within $100\,{\rm pc}$. The values are 
	based on the fraction of the sample's early-F-, A-, and B-type MS stars for which suitable 
	photometric or spectroscopic measurements have been obtained (i.e. those capable of distinguishing 
	between mCP and non-mCP stars). $d_{\rm max}$ and $M_{\rm min}$ correspond to the adopted distance 
	limit and the minimum stellar mass of the sample. The fraction decreases for larger volumes (greater 
	distance limits) and for lower masses as a result of an increasing sample size.}
	\label{fig:comp_dm}
\end{figure}

\citet{Paunzen2005} found that mCP stars can be identified with an efficiency of $\approx91$~per~cent 
based on $\Delta a$ measurements and $\approx83$~per~cent based on $\Delta(V_1-G)$ measurements. Using 
the published catalogues, we find that 9~per~cent of the sample's $3\,254$ stars have available 
$\Delta a$ measurements \citep{Maitzen1998,Vogt1998,Paunzen2005} while 46~per~cent have available Geneva 
measurements \citep{Rufener1988}. Eighty-four~per~cent of the sample stars have spectral classifications 
listed in the Catalogue of Stellar Spectral Classifications \citep{Skiff2014}; however, many of the 
spectroscopic measurements used to derive these classifications may be insensitive to chemical 
peculiarities due to insufficient resolution. For those classifications derived from optical spectra, we 
only considered those measurements obtained with dispersions $>90\,{\rm \AA\,mm}^{-1}$ 
\citep{Young1973,Abt1979} or resolutions $<3\,{\rm \AA}$. Those classifications derived from the UV 
spectra obtained using the S2/68 instrument onboard the TD1 satellite were also considered reliable 
despite the low resolution of $\approx37\,{\rm \AA}$ \citep{Wilson1972}. This is based on the fact that, 
similar to the $\Delta a$ and $\Delta(V_1-G)$ photometric measurements, the S2/68 measurements are 
sensitive to UV flux depressions characteristic of many mCP stars 
\citep{Cucchiaro1976,Cucchiaro1977,Cucchiaro1978,Cucchiaro1978a,Cucchiaro1980}.

The strength of the photometric peculiarity indices induced by the presence of strong surface magnetic 
fields is known to decrease with decreasing $T_{\rm eff}$ \citep{Maitzen1976,Maitzen1980,
Maitzen1983,Masana1998}. Moreover, as $T_{\rm eff}$ decreases, a greater number of mCP stars is 
found to exhibit non-peculiar $\Delta a$ and $\Delta(V_1-G)$ values \citep[e.g. Fig. 4 
of][]{Maitzen1976}. We find that our full sample contains 9 stars classified as non-mCP based solely on 
$\Delta a$ measurements and 299 based solely on $\Delta(V_1-G)$ measurements. We estimated the 
fraction of these 308 stars that may actually be undetected mCP stars by referring to Table 5 of 
\citet{Paunzen2005}: the authors find that $1.4$~per~cent and $2.2$~per~cent of their sample's mCP 
stars exhibit non-peculiar $\Delta a$ and $\Delta(V_1-G)$ measurements, respectively. Therefore, we 
estimate that our sample contains no additional undetected mCP stars based on $\Delta a$ 
based classifications (i.e. $9\times0.014<1$) and 6 additional undetected mCP stars based on 
$\Delta(V_1-G)$ based classifications (i.e. $299\times0.022<7$). Although we cannot provide 
precise constraints on the most-probable temperatures of these stars, it is likely that they are 
cooler stars with $T_{\rm eff}\lesssim7\,{\rm kK}$ based on the fact that a greater number of mCP stars 
are found to exhibit peculiarity indices that are largely indistinguishable from non-mCP stars at lower 
temperatures \citep[e.g.][]{Maitzen1976}.

Considering the total sample, we find that $1\,924$ of the $3\,254$ stars (67~per~cent) have been 
observed using spectroscopic and photometric methods capable of distinguishing between mCP and non-mCP 
stars. However, this number changes significantly based on the adopted lower mass limit ($M_{\rm min}$) 
and the adopted distance limit ($d_{\rm max}$) associated with the sample; in Fig. \ref{fig:comp_dm}, we 
show the estimated completeness as a function of $d_{\rm max}$ and $M_{\rm min}$. Evidently, the 
$100\,{\rm pc}$ sample of confirmed and candidate mCP stars has a completeness of $\approx90$~per~cent 
for masses $\gtrsim1.7\,M_\odot$. This completeness is more representative of the sample compared to 
the 67~per~cent value derived using $M_{\rm min}=1.4\,M_\odot$ given that 91~per~cent of the sample of 
confirmed and candidate mCP stars have masses $\gtrsim1.7\,M_\odot$.

\begin{figure}
	\centering
	\includegraphics[width=1.0\columnwidth]{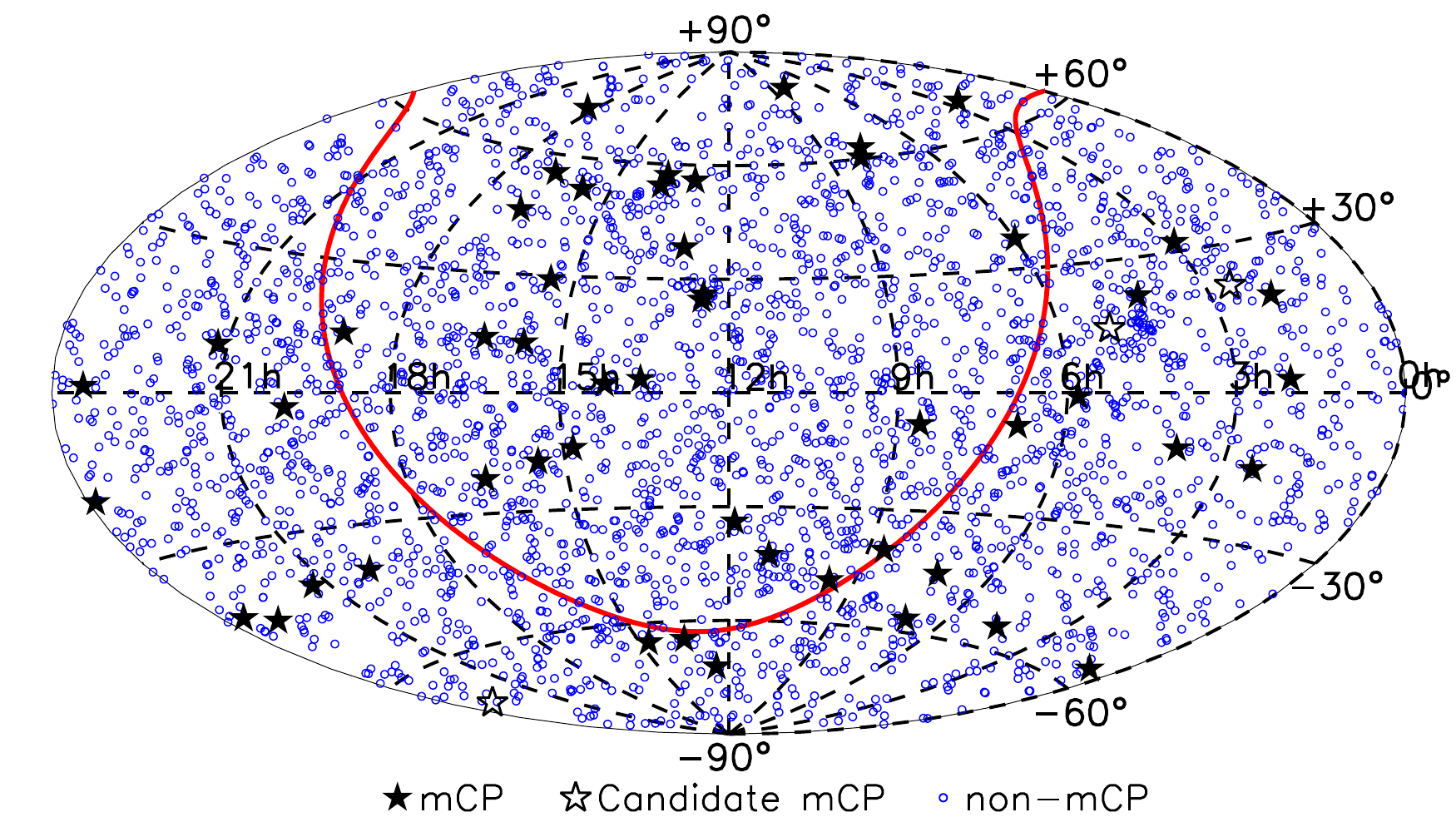}
	\caption{Equatorial coordinates of the mCP stars (black, filled stars), candidate mCP stars (black, 
	open stars) and non-mCP stars (blue circles). The solid red curve corresponds to the Galactic 
	plane.}
	\label{fig:sky_dist}
\end{figure}

We also performed two basic tests to look for obvious indications that the sample of mCP stars is 
incomplete. Both of the tests are based on the assumption that the spatial distribution of these 
particular objects within the sample's $100\,{\rm pc}$ distance limit is approximately random and 
uniform. Under this assumption, we would expect nearly the same number of mCP stars to be located in 
both the northern and sourthern hemispheres. In Fig. \ref{fig:sky_dist}, we plot the sky coordinates of 
the confirmed mCP stars, candidate mCP stars, and the non-mCP stars using an equal-area 
Hammer-Aitoff projection. Considering the 52 confirmed mCP stars in the sample, we find a ratio of the 
number of northern stars ($N_{\rm N}$) to southern stars ($N_{\rm S}$) of 1.08; including the 3 candidate 
mCP stars yields $N_{\rm N}/N_{\rm S}=1.08$ as well. The significance of these ratios can be assessed 
using a Monte Carlo (MC) simulation. We generated $10^5$ distributions containing 52 and 55 randomly 
assigned latitude values according to $\theta=\pi/2-\cos^{-1}(2x-1)$, where $x$ is a uniformly 
distributed random number between 0 and 1 \citep[e.g.][]{Kochukhov2013}. The simulated 
$N_{\rm N}/N_{\rm S}$ distributions suggest that the values derived from the subsample of 52 confirmed 
mCP stars and the combined subsample of 52 confirmed and 3 candidate mCP stars are consistent with 
random distributions at 68~per~cent and 79~per~cent confidence levels, respectively.

\begin{figure}
	\centering
	\includegraphics[width=0.8\columnwidth]{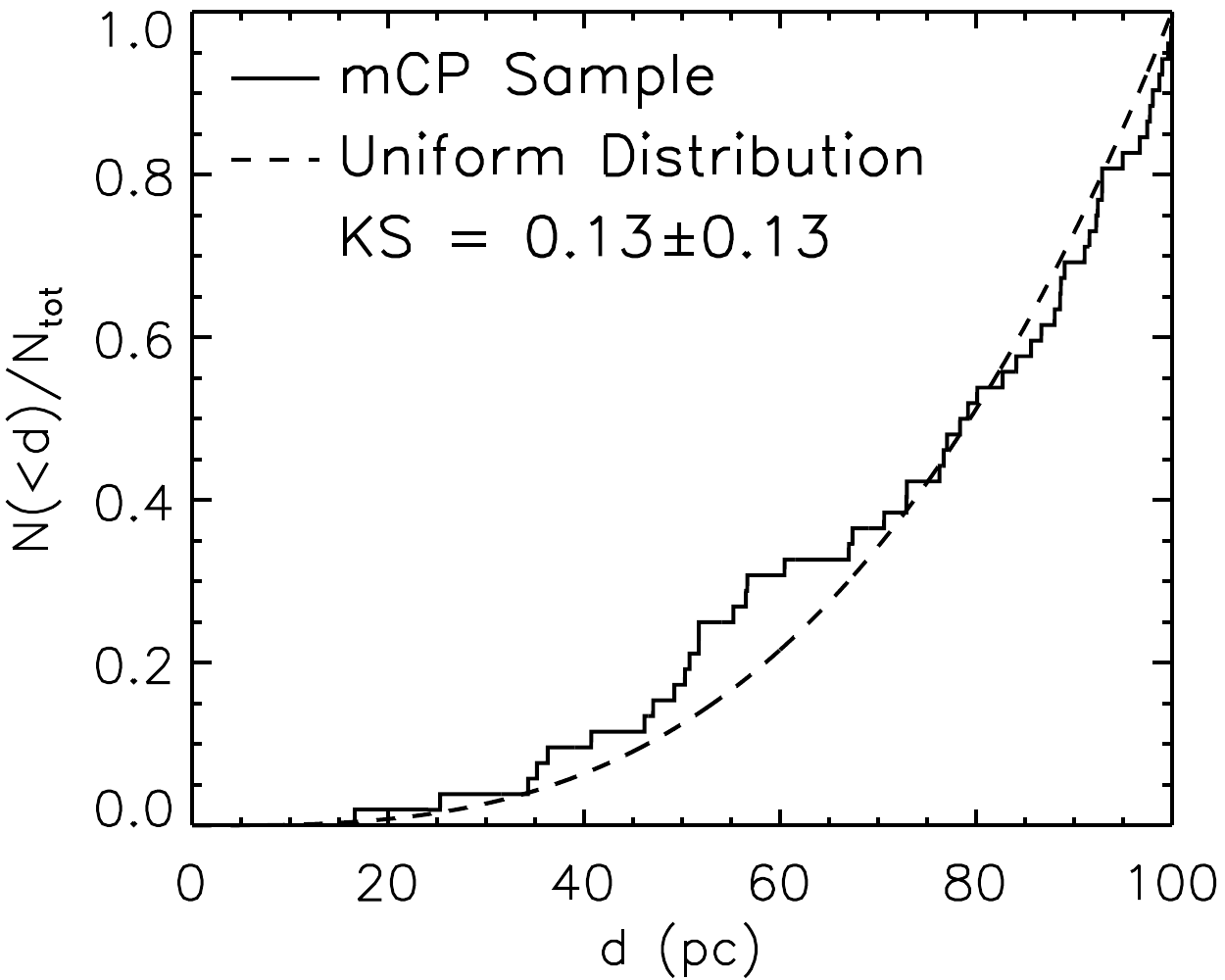}
	\caption{Cumulative distribution function of the mCP star distances (solid black) normalized 
	to the total number of mCP stars. The dashed black curve corresponds to the CDF associated with a 
	spatially uniform distribution. Comparing the two distributions yields a Kolmogorov-Smirnov test 
	statistic of $0.13\pm0.13$.}
	\label{fig:d_cdf}
\end{figure}

The second test simply involves comparing the distribution of the mCP subsample's distances to that of 
a spatially uniform distribution. This is shown in Fig. \ref{fig:d_cdf} in which the cumulative 
distribution function (CDF) of the distances to the mCP stars is plotted along side that expected from a 
sample with a uniform spatial distribution. Although the sample of confirmed and candidate mCP stars 
exhibits a slight excess between 50 and $70\,{\rm pc}$, the Kolmogorov-Smirnov (KS) test statistic 
implies that the distance distribution is consistent with that of a uniform distribution: we calculated 
a KS statistic of $0.13\pm0.13$ where the uncertainty corresponds to $3\sigma$ (the uncertainty was 
derived using the bootstrapping method of case resampling in which the KS statistic was calculated for 
$1\,000$ distance distributions generated by randomly sampling the empirical distance distribution).

\section{Derivation of fundamental parameters}\label{sect:fund_param}

Several techniques were employed in order to derive fundamental parameters (effective temperature, 
surface gravity, mass, radius, and age) for all of the stars in the sample. For the non-mCP stars 
and 3 candidate mCP stars in the survey, we used various photometric calibrations depending on the 
available archival measurements. A more detailed analysis of both photometric and spectroscopic 
measurements was carried out for the 52 confirmed mCP stars.

Archival broad-band photometric measurements were obtained from the General Catalogue of Photometric 
Data \citep{Mermilliod1997}. The catalogue consists of measurements of over $200\,000$ stars that were 
obtained using a wide range of photometric systems. In addition to the Johnson $V$, Hipparcos $H_p$, and 
Tycho $V_T$ and $B_T$ observations \citep{ESA1997}, we also used observations obtained with 
Geneva $UB_1BB_2V_1VG$ filters \citep{Rufener1988}, Johnson $UVBRI$ filters 
\citep{Hoffleit1991,Ducati2002}, Str\"omgren $uvby$ filters \citep{Hauck1997}, and 2MASS $JHK_s$ 
filters \citep{Cohen2003}. In some cases, no uncertainties for these observations were reported and we 
adopted values of 1~per~cent. Since the sample is limited to stars with $d<100\,{\rm pc}$, we assumed 
that -- for both the mCP and non-mCP stars -- photometric reddening caused by interstellar extinction is 
negligible \citep[${\rm E}\lbrack b-y\rbrack\lesssim0.02$,][]{Vergely1998}.

\subsection{non-mCP subsample}\label{sect:FP_nonmag}

We used three photometric temperature calibrations defined within the Johnson, Geneva, and Str\"omgren 
systems in order to derive $T_{\rm eff}$ for the non-mCP stars. The calibrations published by 
\citet{Gray2005_SP} were applied to the available Johnson $B$ and $V$ filter measurements that are 
appropriate for both ``cool" MS stars ($T_{\rm eff}\lesssim10^4\,{\rm K}$ or $B-V>0.0$, Eqn. 14.17) and 
``hot" MS stars ($T_{\rm eff}\gtrsim10^4\,{\rm K}$ or $B-V<0.0$, Eqn. 14.16). For those stars for which 
measurements obtained using the Geneva system are available, we applied the calibrations of 
\citet{Kunzli1997}, which are suitable for MS stars with $T_{\rm eff}\gtrsim5\,000\,{\rm K}$. Lastly, for 
those cases in which measurements obtained using the Str\"omgren system are available, we applied the 
calibrations of \citet{Balona1994}, which are suitable for MS stars with 
$5\,500\,{\rm K}<T_{\rm eff}<35\,000\,{\rm K}$.

\begin{figure}
	\centering
	\includegraphics[width=1.0\columnwidth]{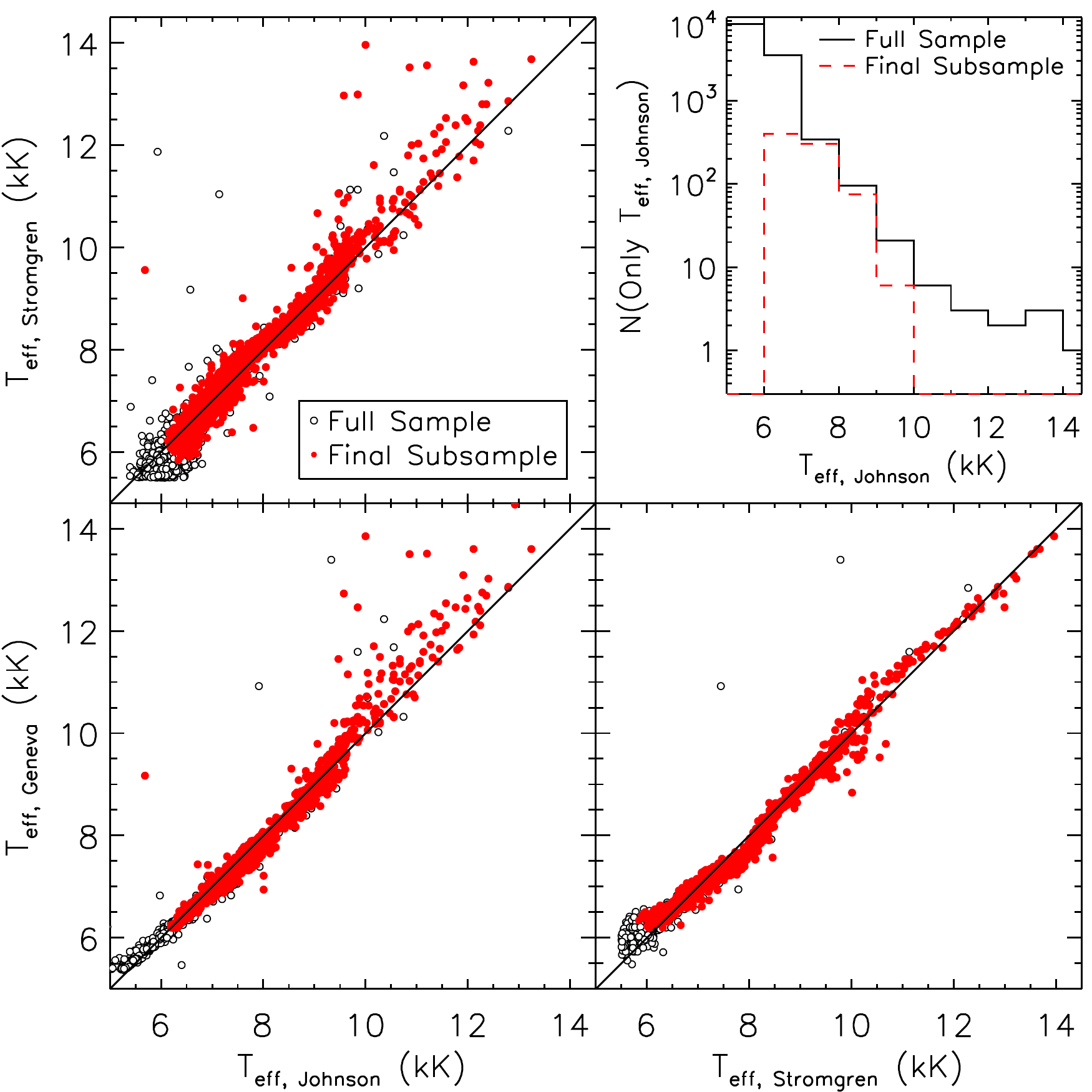}
	\caption{\emph{Top left and bottom row:} Comparison between the $T_{\rm eff}$ values of the non-mCP 
	stars derived using three calibrations: \citet{Kunzli1997} (Geneva filters), \citet{Balona1994} 
	(Str\"omgren filters), and \citet{Gray2005_SP} (Johnson filters). The open black circles 
	correspond to the full sample of non-mCP stars extracted from the Hipparcos Catalogue; the filled 
	red circles correspond to the subsample remaining after applying several cuts to the full sample 
	(described in Sect. \ref{sect:HRD}). \emph{Top right:} Distribution of non-mCP stars in the sample 
	for which only Johnson photometric measurements are available.}
	\label{fig:non_mCP_Teff}
\end{figure}

In Fig. \ref{fig:non_mCP_Teff} we compare $T_{\rm eff}$ derived using the three calibrations (for 
those stars for which more than one calibration could be applied). In general, we find reasonable 
agreement between the calibrations for $T_{\rm eff}\lesssim10^4\,{\rm K}$. At higher temperatures, both 
the calibrations involving Geneva filters ($T_{\rm eff}^{\rm Geneva}$) and Str\"omgren filters 
($T_{\rm eff}^{\rm Str\ddot{o}mgren}$) tend to predict higher $T_{\rm eff}$ values compared to those 
involving Johnson $B$ and $V$ filters ($T_{\rm eff}^{\rm Johnson}$). In Fig. \ref{fig:non_mCP_Teff} (top 
right) we show the distribution of $T_{\rm eff}^{\rm Johnson}$ derived for those stars that only have 
Johnson $B$ and $V$ measurements available (i.e. no $T_{\rm eff}^{\rm Geneva}$ or 
$T_{\rm eff}^{\rm Str\ddot{o}mgren}$ values could be derived). It is evident that the majority of the 
non-mCP stars in the sample ($\approx62$~per~cent) fall into this category; however, only 24 of those 
stars exhibit $T_{\rm eff}>10^4\,{\rm K}$ (i.e. the maximum temperature at which 
$T_{\rm eff}^{\rm Johnson}$ may be considered reliable). We note that, after applying several cuts to 
the total sample as described in Sect. \ref{sect:HRD}, all of the hotter stars without available Geneva 
or Str\"omgren photometry were removed.

When more than one calibration was applied, the final effective temperatures of each non-mCP star were 
taken to be the average of these values. If $T_{\rm eff}^{\rm Johnson}$ was found to differ significantly 
from $T_{\rm eff}^{\rm Geneva}$ and/or $T_{\rm eff}^{\rm Str\ddot{o}mgren}$, we removed 
$T_{\rm eff}^{\rm Johnson}$ prior to averaging. A uniform uncertainty of 5~per~cent was adopted for 
each of the final $T_{\rm eff}$ values based on the dispersion in the 
$T_{\rm eff}^{\rm Johnson}-T_{\rm eff}^{\rm Geneva}$ and 
$T_{\rm eff}^{\rm Johnson}-T_{\rm eff}^{\rm str\ddot{o}mgren}$ planes shown in Fig. 
\ref{fig:non_mCP_Teff}.

\begin{figure*}
	\centering
	\includegraphics[width=2.1\columnwidth]{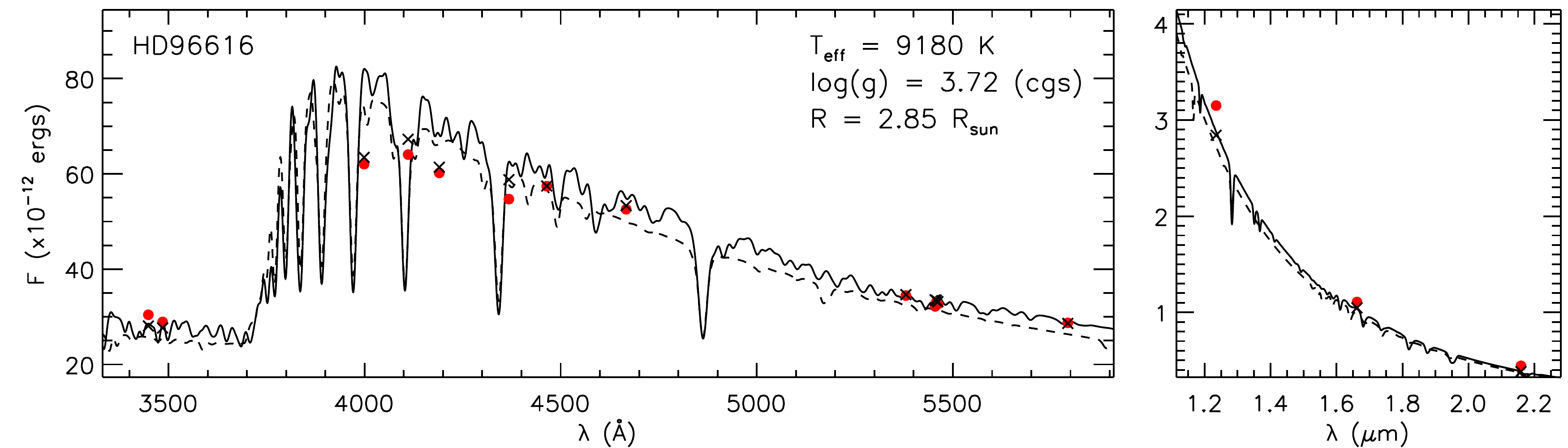}
	\caption{An example of the final fit to the observed photometry (filled circles) of HD~96616. The 
	left panel shows the optical photometry (Johnson, Geneva, and Str\"omgren) while the right panel 
	shows the 2MASS photometry. The solid black line is the detailed synthetic SED while the black `X' 
	symbols correspond to the synthetic flux of each photometric filter used in the comparisons with the 
	observations. The dashed black line shows the solar metallicity synthetic SED calculated without 
	taking into account chemical peculiarities or magnetic fields.  Both of the synthetic SEDs have been 
	convolved with a Gaussian function consistent with $v\sin{i}=1\,000\,{\rm km\,s^{-1}}$ for visual 
	clarity.}
	\label{fig:sed_ex1}
\end{figure*}

The luminosity ($L$) of each of the non-mCP stars was derived by first applying the bolometric 
correction (BC)-$T_{\rm eff}$ relation published by \citet{Balona1994}, which is applicable to MS stars 
with $T_{\rm eff}\gtrsim5\,500\,{\rm K}$. The absolute magnitude ($M_V$) values derived from the 
Hipparcos $V$ and ${\rm\pi}$ measurements \citep{ESA1997,VanLeeuwen2007} were then used in conjunction 
with the BC values to calculate each star's absolute bolometric magnitude ($M_{\rm bol}$), and thus, their 
luminosities (we used a solar absolute bolometric magnitude of 4.74 for this calculation). The stellar 
radii ($R$) were derived using the Stefan-Boltzmann relation with each star's $T_{\rm eff}$ and $L$. As 
with $T_{\rm eff}$, we adopted a uniform 5~per~cent uncertainty in $L$; the uncertainty in $R$ was derived 
by propogating the uniform $T_{\rm eff}$ and $L$ uncertainties.

\subsection{mCP subsample}\label{sect:FP_mag}

Various photometric temperature calibrations that are applicable to mCP stars are available in the 
literature. \citep[e.g.][]{Hauck1993,Stepien1994}. However, unlike for the non-mCP stars, we have 
high-resolution spectra available for nearly all of the mCP stars in this sample. We therefore opted 
to derive each mCP star's fundamental parameters by employing a more detailed method of iteratively 
fitting (1) the photometry to synthetic spectral energy distributions (SEDs) and (2) the observed 
spectra to model spectra \citep[a similar method is described by ][]{Silvester2015}. In this approach, 
we generated model atmospheres and their associated SEDs that took into account chemical peculiarities 
and strong surface magnetic fields. Chemical peculiarities and, to a lesser extent, surface magnetic 
fields can significantly modify a star's SED by redistributing flux from UV to redder wavelengths 
\citep{Kodaira1969,Stepien1978,Kupka2003}; therefore, it is often important to take these effects into 
account when deriving fundamental parameters by fitting SEDs.

Our analysis consisted of three iterations which we outline now and describe more thoroughly in the 
following sub-sections. In the first iteration, we obtained an estimate of $T_{\rm eff}$ and radius ($R$) 
by fitting the photometric observations to a grid of synthetic SEDs. Individual metallic spectral lines 
(e.g. Cr, Fe, Mg) were then isolated from various spectroscopic observations. These measurements were 
fit to models in order to derive the projected rotational velocity ($v\sin{i}$). Next, we estimated 
$\log{g}$ by fitting the available H$\beta$ and H$\gamma$ observations. Finally, wide spectral regions 
containing a large number of He and metallic lines (i.e. no Balmer lines) were fit in order to derive 
estimates of the chemical abundances.

The first iteration yielded estimates of $T_{\rm eff}$, $\log{g}$, and the chemical abundances. For 
the second iteration, these parameters were used to generate a small grid of atmospheric models 
and synthetic SEDs, which took into account the effects of chemical peculiarities and strong magnetic 
fields (the magnetic field strengths are derived in Paper II). The grid consisted of models spanning a 
narrow range of $T_{\rm eff}$. The photometric measurements were then refit using this new grid, which 
yielded more refined values of both $T_{\rm eff}$ and $R$. With a new value of $T_{\rm eff}$ derived, 
the observed spectra were once again fit in order to refine $\log{g}$ and the chemical abundances.

The third and last iteration simply involved recalculating single atmospheric models and synthetic SEDs 
using the final values of $T_{\rm eff}$, $\log{g}$, the chemical abundances, and the magnetic field 
strength (if a sufficiently strong magnetic field was inferred). A final value of $R$ was then derived 
by fitting the photometry to the synthetic SED.

\subsubsection{SED fitting}\label{sect:mag_sed}

In order to fit the observed photometric measurements, we first generated a grid of model atmospheres 
with accompanying synthetic SEDs spanning a range of $T_{\rm eff}$ and $\log{g}$. We used the 
{\sc llmodels} code, which calculates plane-parallel model atmospheres under the local thermodynamic 
equilibrium (LTE) assumption \citep{Shulyak2004}. This code is well-suited to this study primarily 
because it allows for abundances of individual elements to be varied rather than simply scaling all 
abundances with [M/H]. Furthermore, it can also include the anomalous Zeeman effect induced by the 
presence of strong surface magnetic fields \citep{Kochukhov2005a}.

For the first iteration, the grid of model atmospheres and synthetic SEDs was calculated for 
$5\,500\,{\rm K}\lid T_{\rm eff}\lid10\,000\,{\rm K}$ in increments of $100\,{\rm K}$ and for 
$10\,000\,{\rm K}\lid T_{\rm eff}\lid20\,000\,{\rm K}$ in increments of $250\,{\rm K}$. We used a solar 
metallicity ([M/H]$=0.0$), which is based on the solar abundances reported by \citet{Asplund2009}. Both 
the microturbulence ($v_{\rm mic}$) and the surface gravity were fixed at $0\,{\rm km\,s^{-1}}$ and 
$4.0$ (cgs), respectively, while all other physical parameters were unmodified from their default 
values.

The output flux associated with the synthetic SEDs -- given in physical flux units of 
erg/s/cm$^2$/{\AA} -- was used to calculate the flux associated with various photometric filters. 
Transmission functions for the relevant filters -- Hipparcos, Johnson, Tycho \citep{Bessell2012}, Geneva 
\citep{Rufener1988}, Str\"omgren \citep{Bessell2011}, and 2MASS \citep{Cohen2003} -- were obtained 
from the literature. These functions were then normalized, interpolated to the wavelength abscissa of 
each synthetic SED using a spline method, and multiplied by the associated flux. The resulting curves 
were then integrated yielding the total stellar flux contributed to each filter.

We converted the observed magnitudes of each star to physical flux units using zero point values 
reported in the same publications that the filter transmission functions were obtained from. The 
best-fitting model SED associated with the observed flux values was then derived using a 
Levenberg-Marquardt algorithm (LMA) implemented in {\sc idl} with 2 free parameters: $T_{\rm eff}$ and 
the scaling factor $\alpha\equiv(R/d)^2$. The free parameters were allowed to vary continuously, which 
required an interpolation (we used a spline interpolation method) of the model grid in $T_{\rm eff}$.

A similar SED fitting procedure was applied during the second iteration using a narrower grid of 
synthetic SEDs. This grid consisted of between 4 and 6 models bracketing the current value of 
$T_{\rm eff}$ in increments of $500\,{\rm K}$ and included the derived chemical abundances (discussed in 
Section \ref{sect:abund_fit}) and magnetic fields (if the field strength derived in Paper II was found 
to exceed $5\,{\rm kG}$). For those elements whose abundances were not derived, we used the values 
associated with a solar metallicity. The final value of the scaling factor $\alpha=(R/d)^2$ was 
used to derive $R$; the final values of $T_{\rm eff}$ and $\alpha$ were used in conjunction with the 
Stefan-Boltzmann relation to derive each star's luminosity:
\begin{equation}
\frac{L}{L_\odot}=\alpha\frac{d^2}{R_\odot^2}\left(\frac{T_{\rm eff}}{T_{\rm eff,\odot}}\right)^4=\left(\frac{R}{R_\odot}\right)^2\left(\frac{T_{\rm eff}}{T_{\rm eff,\odot}}\right)^4.
\end{equation}

The uncertainties in $T_{\rm eff}$ and $\alpha$ were estimated through the method of residual 
bootstrapping. In this procedure, the residuals associated with the best-fitting synthetic SED are 
first scaled using each data point's estimated statistical leverage and subsequently centered by 
subtracting the mean residual \citep{Davison1997}. A distribution of $1\,000$ data sets was then 
generated by randomly sampling the scaled and centered residuals and adding them to the fitted flux 
values. Distributions of $T_{\rm eff}$ and $\alpha$ were then obtained by fitting synthetic SEDs to the 
$1\,000$ bootstrapped data sets thereby allowing 3 $\sigma$ uncertainties to be estimated from the 
distribution widths. Since the luminosity is expected to be correlated (to some extent) with both 
$T_{\rm eff}$ and $\alpha$, we estimated the uncertainty in $\log{L}$ using the bootstrapped 
distributions of $T_{\rm eff}$ and $\alpha$. Lastly, the reported uncertainties in the Hipparcos 
parallaxes \citep{VanLeeuwen2007} were factored into the uncertainties in $R$ and $\log{L}$ using a 
standard error propagation method.

An example of the final fit to the photometry that was obtained after two iterations of chemical 
abundance and $\log{g}$ derivations is shown in Fig. \ref{fig:sed_ex1}.

\subsubsection{Balmer line fitting}\label{sect:balmer_fit}

The shape of the broad Lorentzian wings of the Balmer lines is relatively sensitive to variations in 
the surface gravity, particularly for $T_{\rm eff}\gtrsim10\,000\,{\rm K}$ (we found that the 
sensitivity tends to decrease with decreasing $T_{\rm eff}$). We derived $\log{g}$ by fitting H$\beta$ 
and H$\gamma$ profiles. This analysis depends crucially on the normalization that is applied to these 
measurements; therefore, measurements obtained using certain instruments for which automatic 
normalization was carried out by the reduction pipeline (e.g. MuSiCoS) could not be used for this 
purpose. We used NARVAL and ESPaDOnS spectra -- both new and archival observations -- along with 
archived ELODIE, HARPS, SOPHIE, and UVES spectra.

\begin{figure*}
	\centering
	\subfigure{\includegraphics[width=2.1\columnwidth]{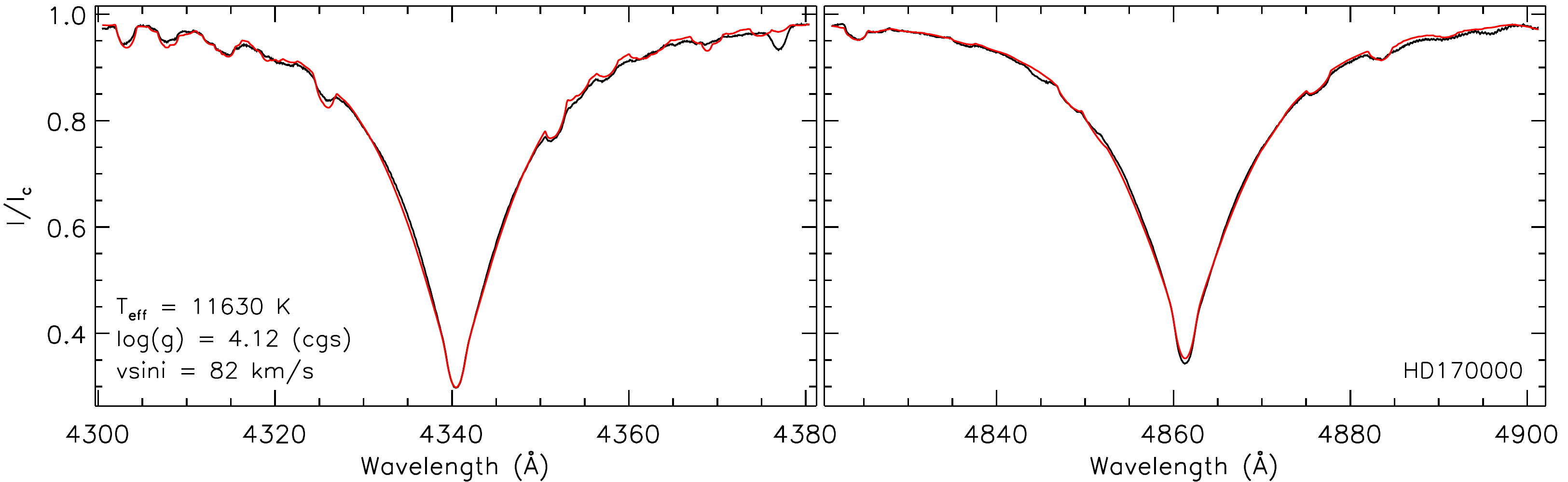}}
	\subfigure{\includegraphics[width=2.1\columnwidth]{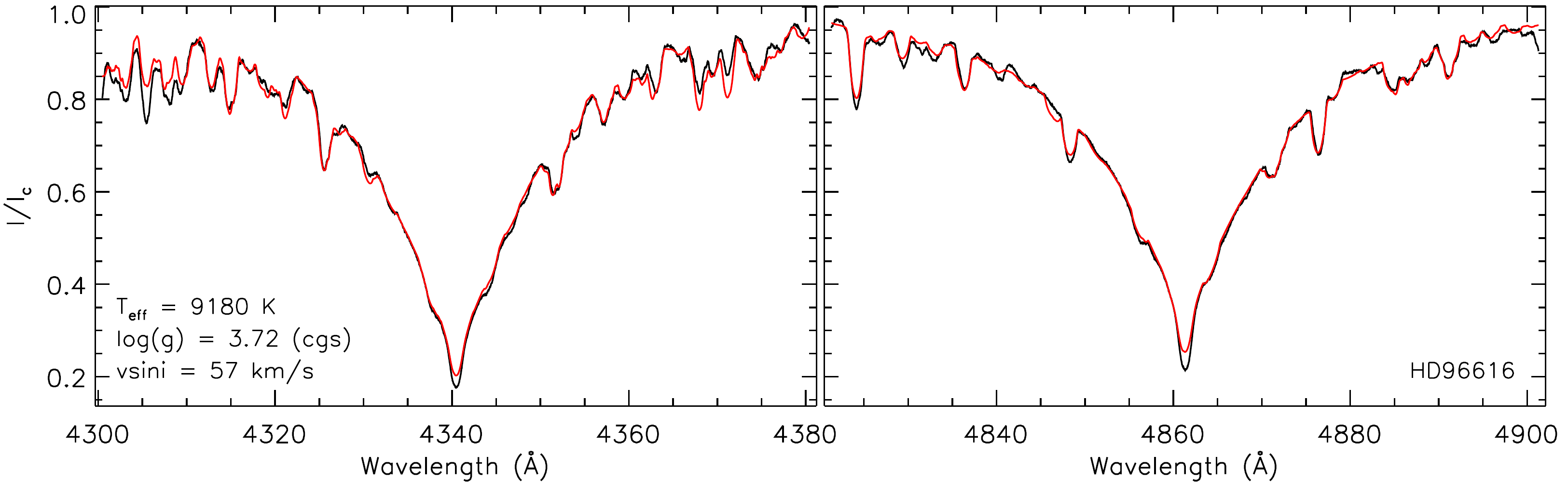}}
	\subfigure{\includegraphics[width=2.1\columnwidth]{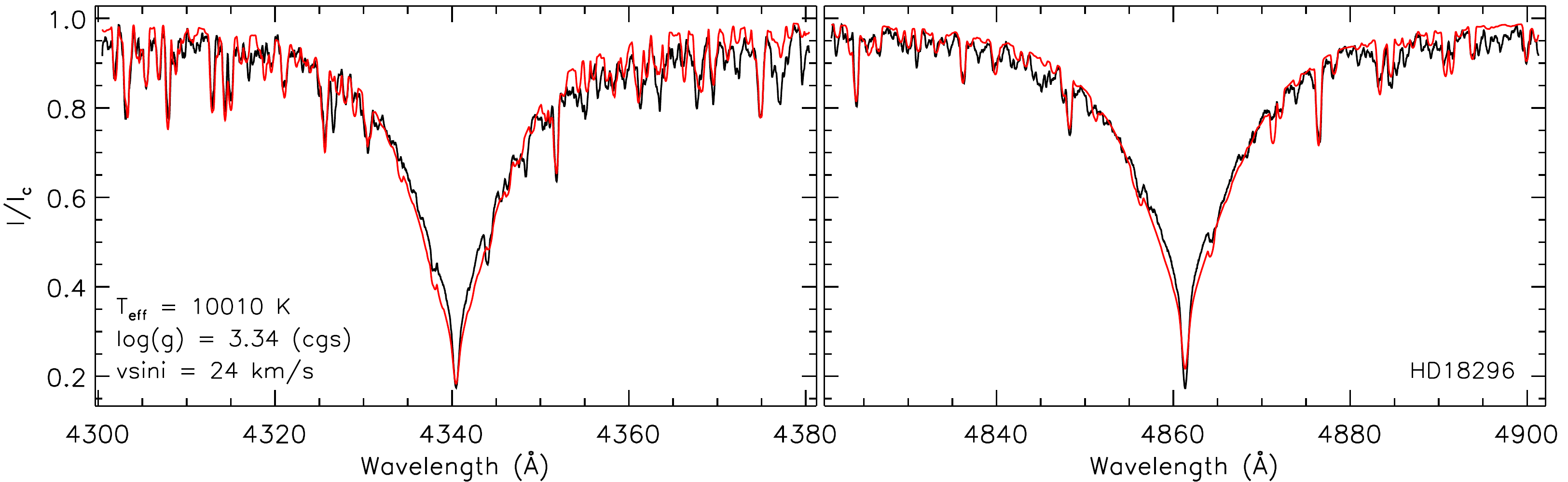}}
	\subfigure{\includegraphics[width=2.1\columnwidth]{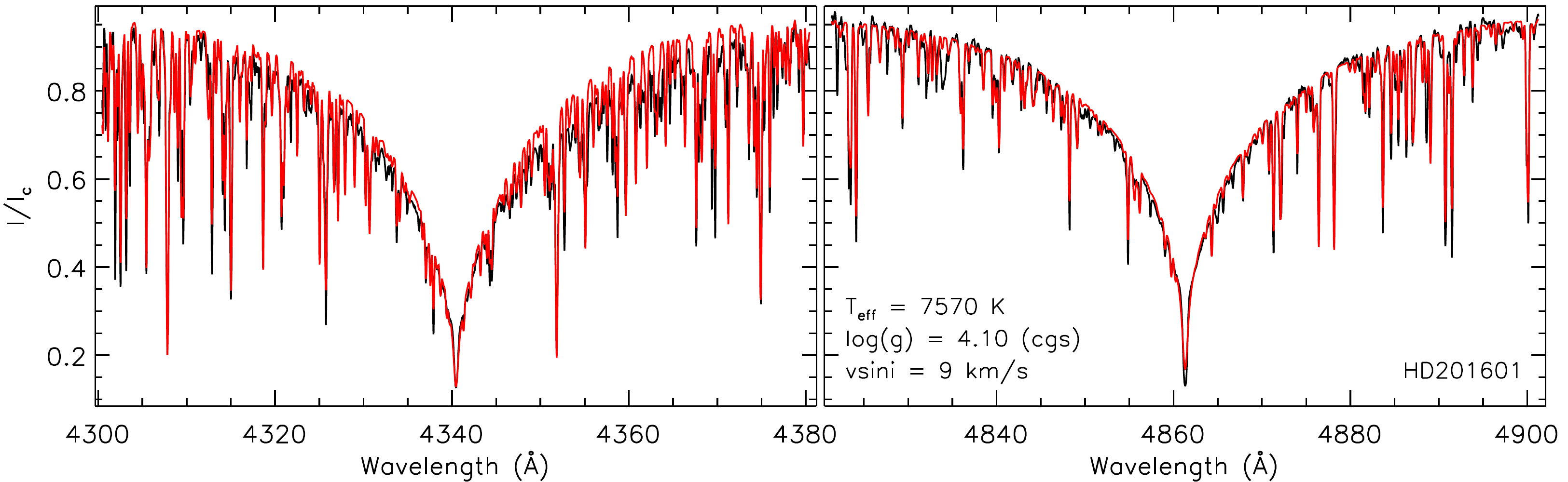}}
	\caption{Examples of the final fits to the observed H$\gamma$ (left) and H$\beta$ (right) profiles. 
	The black curves correspond to the averaged observed spectra while the red curves are the 
	best-fitting models calculated using {\sc gssp}.}
	\label{fig:Balmer_ex1}
\end{figure*}

The spectral orders extracted from cross-dispersed \'{e}chelle spectra -- such as ESPaDOnS and NARVAL -- 
are known to exhibit an intrinsic curvature. This can yield biased results when modelling the broad 
wings of the Balmer lines, particularly for those that are incident near the edges of the extracted 
orders (e.g. H$\beta$ in ESPaDOnS and NARVAL spectra). For the A- and B-type stars composing our 
survey, the extracted orders exhibit continua that are approximately linear near H$\gamma$ and 
H$\beta$; therefore, fits to H$\gamma$ and H$\beta$ profiles that have been normalized using a local 
linear fit (described below) should be relatively unaffected by the instrinsic curvature. In Fig. 
\ref{fig:Balmer_ex1}, a number of examples of the final fits to H$\gamma$ and H$\beta$ lines are 
shown where, in same cases, relatively small discrepancies between the model and the observed spectra 
are apparent within the wings. This may be evidence that some of the ESPaDOnS and NARVAL spectra are 
noticeably affected by distortions in the continuum or normalization discrepancies. It is noted that 
any systematic error in $\log{g}$ introduced by this aspect of the observed spectra is likely less than 
the uncertainty introduced by the degeneracy between $T_{\rm eff}$ and $\log{g}$ (the estimation of 
$\log{g}$ uncertainties is discussed at the end of this section).

For each measurement, we first derived radial velocities by fitting a Gaussian function to the Doppler 
core of H$\beta$ and H$\gamma$; the wavelengths of the spectra were then adjusted to remove these 
radial velocities. Next, each of the Balmer line profiles were individually normalized by applying a 
linear fit to narrow continuum regions located on either side of the line's core at 
$\approx\pm35\,{\rm \AA}$ relative to the central wavelength ($\lambda_0$). Lastly, the normalized 
measurements of each star's H$\beta$ and H$\gamma$ profiles were grouped into sets based on the 
instrument with which they were obtained. The measurements within each of these sets were then 
interpolated onto a common wavelength abscissa -- ranging from $-40$ to $+40\,{\rm \AA}$ relative to 
$\lambda_0$ -- and the associated flux values were averaged. Care was taken to ensure that, prior to 
averaging, the flux values near the continuum regions at $\lambda_0\pm35\,{\rm \AA}$ were approximately 
equal.

We fit the averaged H$\beta$ and H$\gamma$ profiles using the Grid Search in Stellar Parameters 
({\sc gssp}) code developed by \citet{Tkachenko2015}. This code uses atmospheric models provided by 
{\sc llmodels} and chemical line lists provided by the Vienna Atomic Line Database (VALD) 
\citep{Piskunov1995,Kupka2000} to generate model spectra associated with a set of input parameters 
(e.g. $T_{\rm eff}$, $\log{g}$, $v\sin{i}$, $v_{\rm mic}$, and individual atmospheric chemical 
abundances) under the LTE assumption. {\sc gssp} differs from similar spectral line fitting codes 
\citep[e.g. {\sc sme},][]{Valenti1996} in that it generates grids of models corresponding to the range 
and increments of the specified input parameters and compares $\chi^2$ values in order to determine the 
best-fitting model.

The method used by {\sc gssp} is computationally expensive since a large number of models are 
typically produced. We reduced the total computation time by first roughly estimating $\log{g}$ by eye, 
then allowing the chemical abundances to be fit individually to within $1.0$ dex. The abundances were 
then fixed and {\sc gssp} was used to determine the best-fitting value of $\log{g}$ to within $0.1$ 
(cgs). The specific elements that were included in each fit were selected by referring to a VALD line 
list generated using an Extract Stellar request with the current estimate of $T_{\rm eff}$ and 
$\log{g}$; all elements exhibiting lines within the wavelength region to be fit having normalized depths 
$\gid0.1$ were then identified and their abundances were allowed to vary. The second iteration was 
carried out in a similar fashion; however, narrower abundance and $\log{g}$ grids were adopted having 
increments of $0.1$ dex and $0.01$ (cgs), respectively. Instrumental broadening was included in the 
model calculations by specifying the resolving power of the instrument used to obtain the observation. 
The microturbulence was fixed at $0\,{\rm km\,s^{-1}}$ throughout the analysis while $v\sin{i}$ was 
fixed at the values derived during the first iteration of He and metallic line fitting (see Section 
\ref{sect:abund_fit}). Several examples of the final fits to the H$\beta$ and H$\gamma$ lines are shown 
in Fig. \ref{fig:Balmer_ex1} for a range of $v\sin{i}$ values.

We typically found small discrepancies between the values of $\log{g}$ derived from the modelling of 
each star's H$\gamma$ and H$\beta$ measurement. These discrepancies provide an estimate of the 
uncertainty in $\log{g}$ resulting from statistical noise and continuum normalization inaccuracies. By 
far, the largest contribution to the uncertainty in $\log{g}$ results from the slight degeneracy 
between $\log{g}$ and $T_{\rm eff}$: similar-quality fits to the Balmer lines can be derived using a 
range of $\log{g}$ and $T_{\rm eff}$ values. We derived uncertainties in $\log{g}$ by carrying out the 
fitting routine using both the minimal and maximal values of $T_{\rm eff}$ associated 
$\sigma_{T_{\rm eff}}$. The derived $\log{g}$ values are listed in Table \ref{tbl:spec_tbl}.

\subsubsection{He and metallic line fitting}\label{sect:abund_fit}

Spectroscopic measurements obtained by various instruments -- including those which could not be used 
for the Balmer line fitting analysis -- were used to derive each mCP star's chemical abundances and 
$v\sin{i}$ values. As in Section \ref{sect:balmer_fit}, we opted to perform this analysis on averaged 
spectra obtained by combining multiple measurements. This was carried out by first calculating and 
subsequently removing the radial velocity shift of each measurement. The radial velocities were derived 
by fitting a Gaussian function to the Doppler cores of several Balmer lines (H$\alpha$, H$\beta$, and 
H$\gamma$ depending on the wavelength span of the spectra) and averaging the results.

We selected nine spectral regions to be fit from $5\,000$ to $5\,900\,{\rm \AA}$, each 
$100\,{\rm \AA}$ in width. These regions were selected primarily because of (1) the absence of Balmer 
lines and (2) the reasonably (but not excessively) high density of metallic lines; lower wavelength 
regions (e.g. $\lambda\sim4\,000\,{\rm \AA}$) exhibit a higher density of both Balmer and metallic 
lines, which tend to increase the number of blended lines and introduce systematic errors assoicated 
with the normalization procedure. The selected regions were individually normalized by fitting a 
multi-order polynomial to the continuum. Each set of spectra covering these regions that were obtained 
using the same instruments were then compared in order to ensure consistent normalization. They were 
then interpolated onto a common wavelength abscissa and averaged. The normalized average spectra were 
subdivided into smaller regions having widths of $25$ or $50\,{\rm \AA}$ depending on the number of 
lines found in each of the $100\,{\rm \AA}$ spectral regions: observations from stars having sharp 
lines, such as HD~217522 ($v\sin{i}\approx5\,{\rm km\,s^{-1}}$) were subdivided into $25\,{\rm \AA}$ 
regions while those exhibiting exceptionally broad lines such as HD~124224 
($v\sin{i}\approx150\,{\rm km\,s^{-1}}$) were either subdivided into $50\,{\rm \AA}$ regions or left as 
$100\,{\rm \AA}$ regions.

We proceeded to fit these $25$ to $100\,{\rm \AA}$ width spectral regions using an optimized version of 
the {\sc zeeman} spectrum synthesis code, which solves the polarized radiative transfer equations under 
the LTE assumption \citep{Landstreet1988,Wade2001}. These fits were initially carried out without 
including the effects of magnetic fields (see discussion below). The optimizations were implemented by 
\citet{Folsom2012}, who also developed a Levenberg-Marquardt algorithm (LMA) to be used in conjunction with 
{\sc zeeman} in order to determine a minimal $\chi^2$ solution for a given set of free parameters 
($T_{\rm eff}$, $\log{g}$, $v\sin{i}$, etc.). The relevant data for each spectral line (i.e. wavelength, 
depth, excitation energy, etc.) were obtained from VALD3 \citep{Ryabchikova2015a} using an Extract 
Stellar request for a range of $T_{\rm eff}$ values from $6\,000$ to $20\,000\,{\rm K}$, a detection 
threshold of $0.005$, and $\log{g}=4.0$ (cgs).

We first used the {\sc zeeman} LMA to derive the $v\sin{i}$ value associated with individual 
(non-averaged) spectra of each star. This was done by isolating a large number of individual metallic 
lines, which appeared to be unblended, and applying the fitting routine; $T_{\rm eff}$ was fixed at the 
value derived from the first iteration's fit to the photometry, $\log{g}$ and $v_{\rm mic}$ were fixed at 
$4.0$ (cgs) and $0\,{\rm km\,s^{-1}}$, respectively, while the abundance associated with the identified 
line and the value of $v\sin{i}$ were allowed to vary. Any resulting fits to these lines that were judged 
by eye to be inadequate -- likely as a result of systematic errors associated with non-uniform surface 
abundance distributions \citep[e.g.][]{Kochukhov2004} -- were removed from the analysis. The value and 
uncertainty of $v\sin{i}$ were then derived by taking the median and mean absolute deviation of the 
$v\sin{i}$ values associated with the accepted fits, which ranged in number from 7 to 85 per star. 

Depending in part on the strength of each star's magnetic field and the value of $v\sin{i}$, Zeeman 
broadening may have a non-negligible contribution to the width and overall profile of a given spectral 
line. Therefore, we repeated the $v\sin{i}$ derivation with the inclusion of a dipole surface magnetic 
field. The strength of the field ($B_{\rm d}$), obliquity angle ($\beta$), and inclination angle of the 
axis of rotation ($i$) were estimated using the method described in Paper II. These three parameters 
could not be derived for five stars due to an insufficient number of obtained observations and/or 
because of exceptionally long rotational periods, which generally prevents $v\sin{i}$ (and thus, $i$) 
from being accurately constrained. In these cases, we adopted $B_{\rm d}$ values estimated using Eqn.~6 
of \citet{Auriere2007} and $i=\beta=45\degree$.

For those stars for which both $v\sin{i}$ values (i.e. the values derived with and without including 
Zeeman broadening) were found to be in agreement, we adopted the value exhibiting a lower uncertainty. 
Similarly, if more precise published values that are also in agreement with our derived values are 
available, we adopted the published $v\sin{i}$ values (e.g. HD~65339, HD~176232). In Table 
\ref{tbl:spec_tbl}, we list the derived $v\sin{i}$ values along with values found in the literature; 
the final adopted values correspond to those listed in bold.

It is evident from Table \ref{tbl:spec_tbl} that significant discrepancies exist between $v\sin{i}$ 
values derived in this study and those available in the literature for six mCP stars in the sample 
(HD~3980, HD~56022, HD~109026, HD~148898, HD~188041, and HD~223640). In the case of HD~109026, 
the published $v\sin{i}$ value is associated with the primary non-magnetic component, as discussed in 
Sect. \ref{sect:HD109026} and by \citet{Alecian2014}. For both HD~3980 and HD~188041, 
\citet{Hubrig2007} report lower values of $v\sin{i}$ compared to the values derived here. The 
$v\sin{i}$ values derived with and without Zeeman broadening are comparable for these two stars. We 
note that HD~188041's radius ($R=2.32\pm0.19\,R_\odot$) derived here and its rotational 
period ($223.8\,{\rm d}$) and inclination angle ($70\degree$) reported by \citet{Landstreet2000} 
suggest that $v\sin{i}\sim0.5\,{\rm km\,s}^{-1}$. The published $v\sin{i}$ values of HD~56022, 
HD~148898, and HD~223640 were obtained by \citet{Abt1995} based on Gaussian fits of the 
$\lambda4476$~Fe~{\sc i} and $\lambda4481$~Mg~{\sc ii}lines. We computed synthetic models of these 
lines -- both with and without the inclusion of Zeeman broadening -- for each of the stars using the 
published $v\sin{i}$ values; comparing the models and the observed spectra used in our study suggest 
that the published values are incorrect.

The chemical abundance analysis was carried out during both the first and second iterations of our 
analysis using similar procedures. The effective temperature and surface gravity were fixed at the 
values derived from the fitting of the photometry (Sect. \ref{sect:mag_sed}) and of the Balmer lines 
(Sect. \ref{sect:balmer_fit}), respectively, that were performed during the first (current) iteration. 
The value of $v\sin{i}$ was fixed at the value previously derived or adopted. We opted to fix the 
microturbulence at $0\,{\rm km\,s^{-1}}$ for two reasons: (1) we found that, for those spectral regions 
in which certain lines appearing in the observed spectrum but not in the model spectrum, $v_{\rm mic}$ 
would tend to be significantly overestimated ($\gtrsim10\,{\rm km\,s}^{-1}$) resulting in poor overall 
fits; and (2) $v_{\rm mic}$ is known to be supressed in mCP stars by the presence of strong surface 
magnetic fields \citep[e.g.][]{Ryabchikova1997,Kochukhov2006a}. Magnetic fields were included 
for 14 stars during the second iteration of the chemical abundance analysis; the rational for this 
decision to not include magnetic fields for all of the stars is discussed below.

\begin{figure}
	\centering
	\includegraphics[width=1.0\columnwidth]{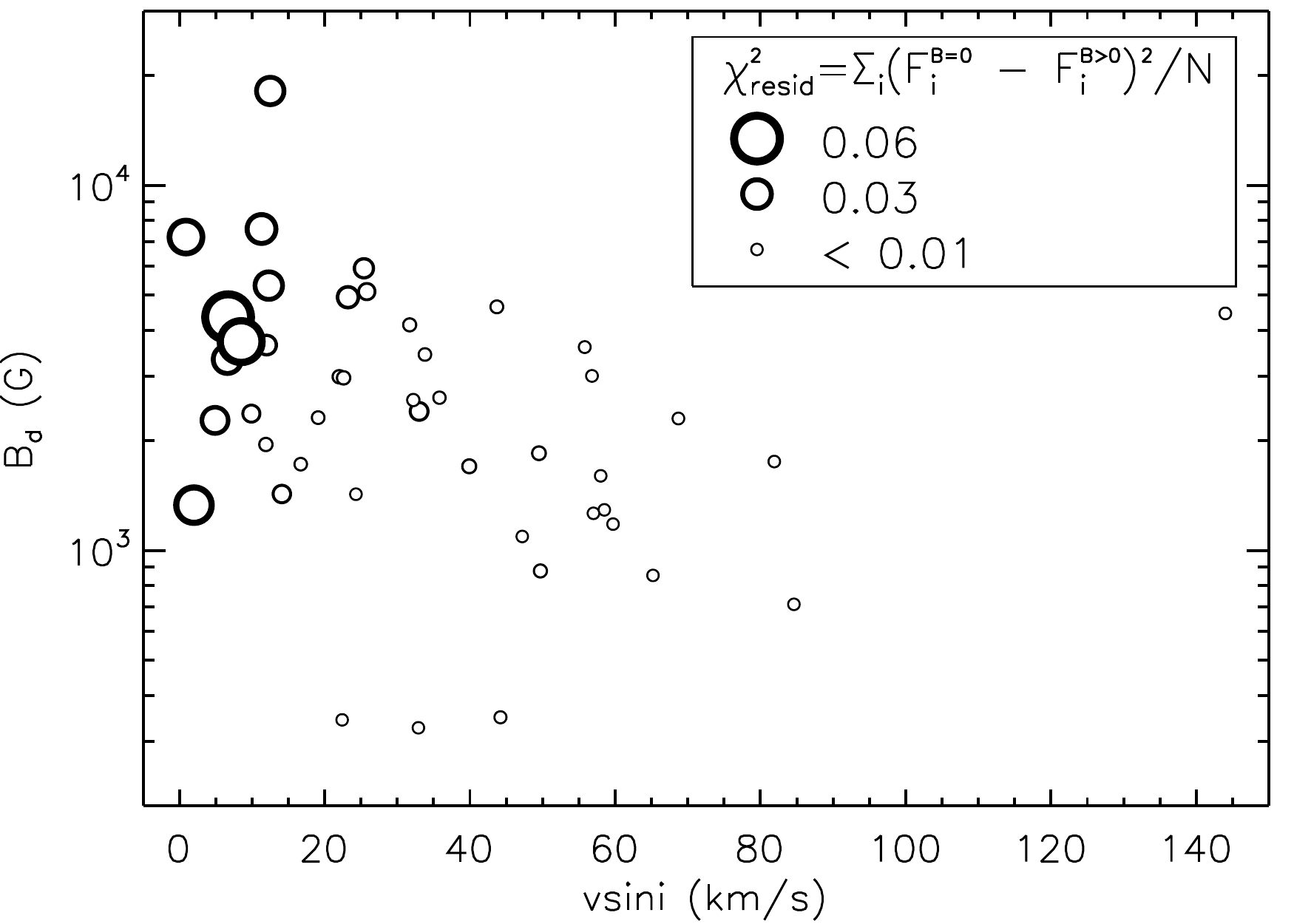}
	\caption{Comparisons between spectral models generated with {\sc zeeman} in which Zeeman broadening is 
	included ($F^{B>0}$) and models in which Zeeman broadening is not included ($F^{B=0}$). We find that, 
	for those fits in which $\chi^2_{\rm resid}\lesssim0.02$, Zeeman broadening is essentially negligible.}
	\label{fig:compare_zeeman}
\end{figure}

For each spectral region, we selected the elements whose abundances would be fit by referring to the VALD 
line list and identifying those lines having normalized depths $>0.03$. The abundances of a number of 
elements selected using this criterion were unable to be fit by the LMA within certain spectral regions, 
likely as a result of line blending. In these instances, that abundance was fixed at the element's 
solar value and the LMA was re-run. Typically the abundances of between four and six elements were fit 
for each spectral region. The final abundances and their uncertainties were derived by taking 
the median and mean absolute deviation of the values (not including those abundances that were fixed at 
solar values) derived from each of the 25 to $100\,{\rm \AA}$ width spectral regions; only those 
chemical abundances which were able to be fit in three or more spectral regions were included in this 
calculation.

The {\sc zeeman} LMA requires a prohibitively long time to carry out the polarized 
radiative transfer calculations for the full sample of mCP stars over spectral windows with widths 
$\gtrsim10\,{\rm \AA}$. We attempted to identify those stars that (1) host relatively weak surface 
magnetic fields and/or (2) exhibit relatively high $v\sin{i}$ values such that Zeeman broadening could 
be neglected without significantly impacting the fitting parameters. This was done by generating models 
using {\sc zeeman} with the fitting parameters from the first iteration of the analysis (chemical 
abundances, $v\sin{i}$, etc.) with the additional inclusion of a dipole magnetic field. The models were generated 
at the rotational phase corresponding to the maximum width of the lines (the phase at which the impact 
of Zeeman broadening is greatest). These models were then compared with those generated using the same 
fitting parameters but without the inclusion of Zeeman broadening. The differences between the synthetic 
model fluxes in which Zeeman broadening was ($F^{B>0}$) and was not ($F^{B=0}$) included was evaluated 
by first removing any points corresponding to the continuum ($F>0.99$) and subsequently calculating the 
normalized sum of the residuals squared: $\chi^2_{\rm resid}=\sum_i(F_i^{B=0}-F_i^{B>0})^2/N$ where the 
sum is carried out over the synthetic flux of each model calculated within a given spectral window and 
$N$ corresponds to the number of synthetic flux values within that spectral window.

\begin{figure*}
	\centering
	\subfigure{\includegraphics[width=2.0\columnwidth]{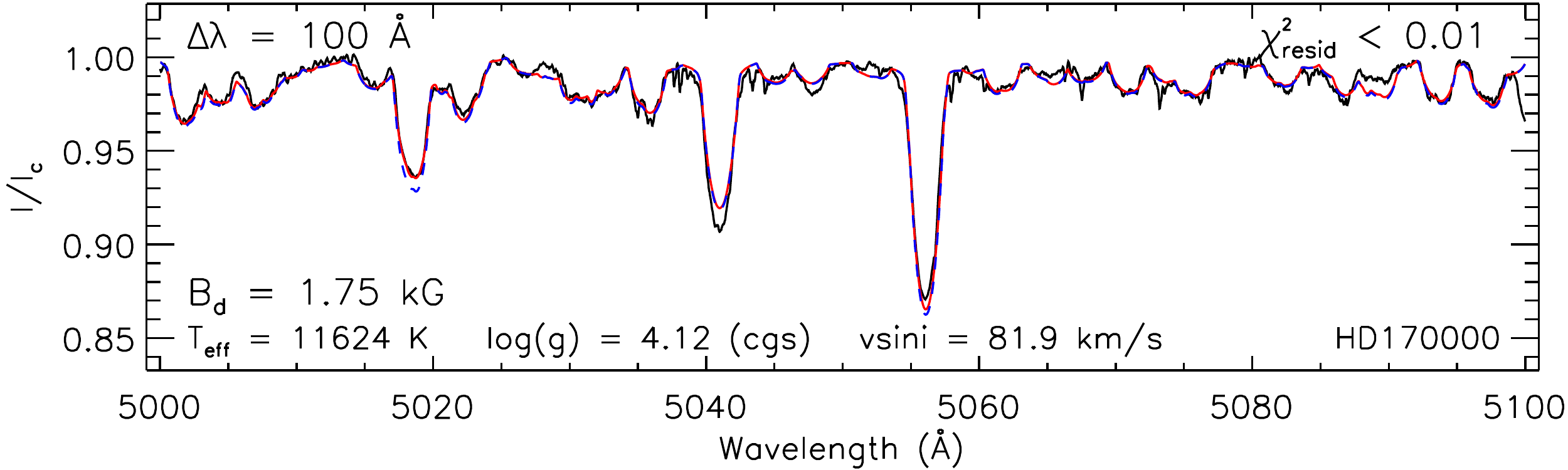}}
	\subfigure{\includegraphics[width=2.0\columnwidth]{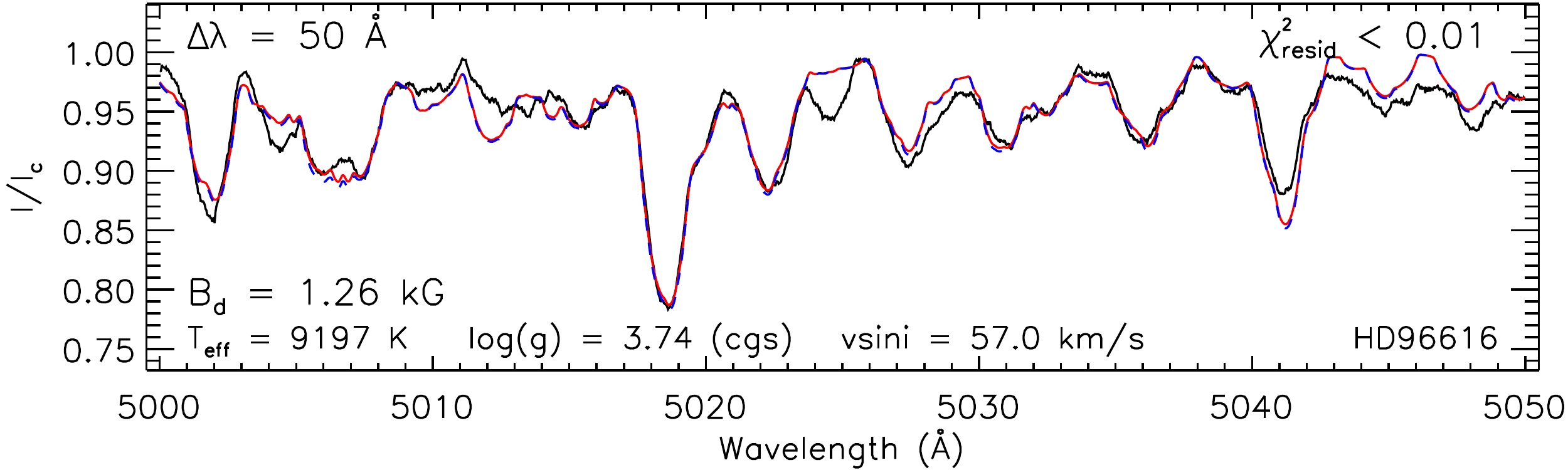}}
	\subfigure{\includegraphics[width=2.0\columnwidth]{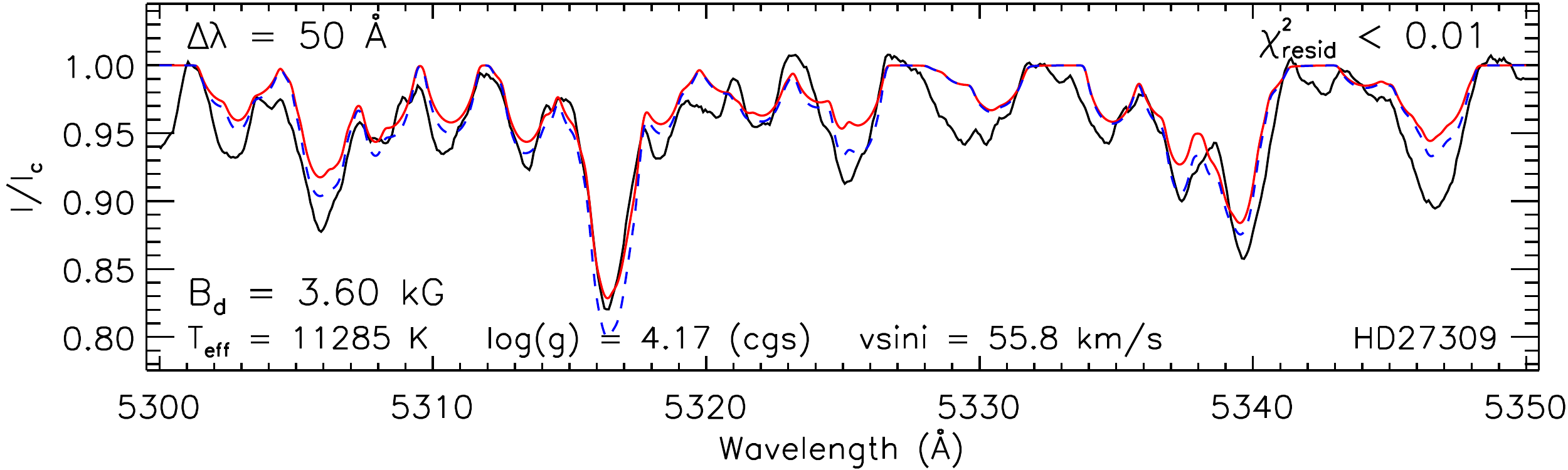}}
	\subfigure{\includegraphics[width=2.0\columnwidth]{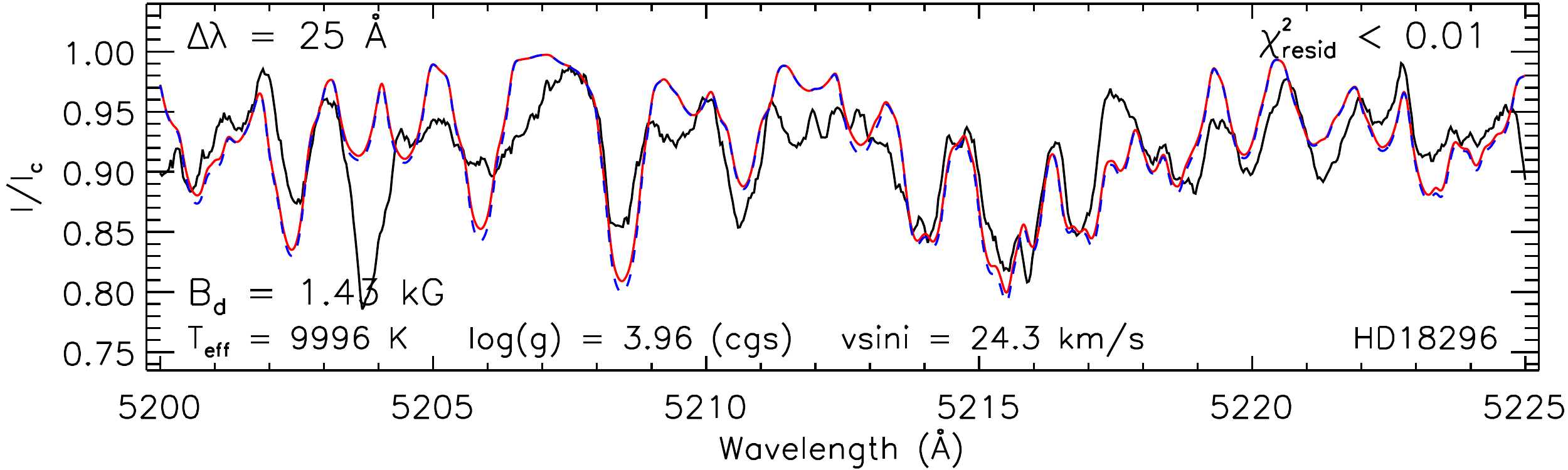}}
	\caption{Several examples of the fits generated as discussed in Sect. \ref{sect:abund_fit}. The 
	black curves correspond to the observed spectra, the red curves correspond to the best-fitting model 
	spectra derived using the {\sc zeeman} LMA without including Zeeman broadening, and the dashed-blue 
	curves correspond to the spectra generated using {\sc zeeman} with the same fitting parameters and 
	with Zeeman broadening included. A range of $v\sin{i}$ values are represented decreasing from top to 
	bottom. Note that the different widths of the spectral regions shown, which are labeled in the top 
	left of each figure.}
	\label{fig:metal_fit}
\end{figure*}

\begin{figure*}
	\centering
	\subfigure{\includegraphics[width=2.0\columnwidth]{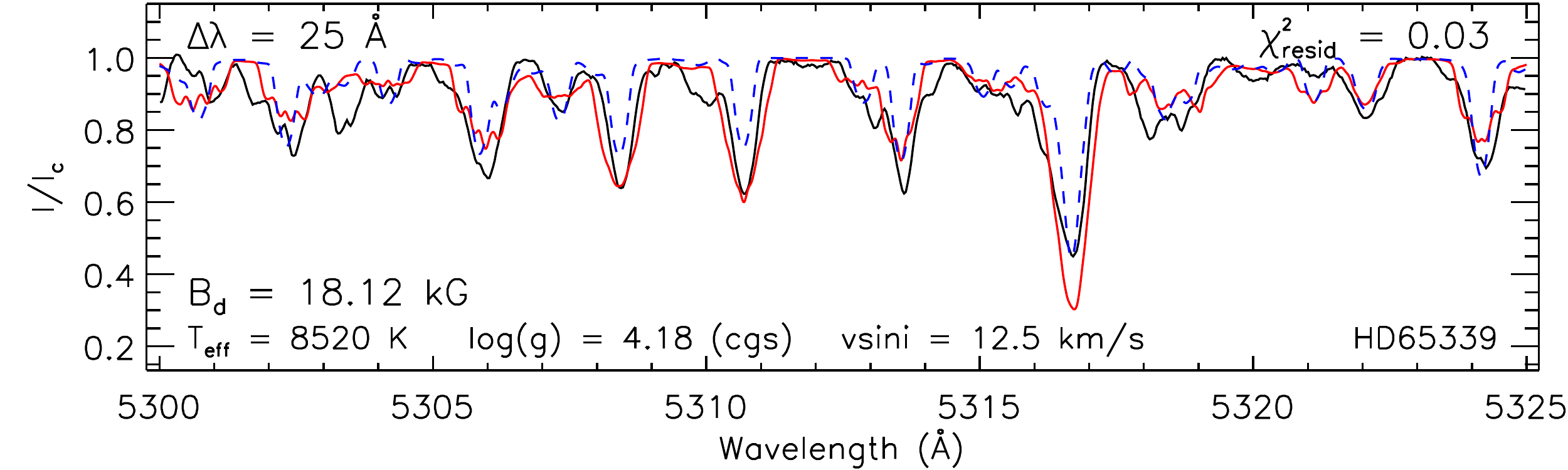}}
	\subfigure{\includegraphics[width=2.0\columnwidth]{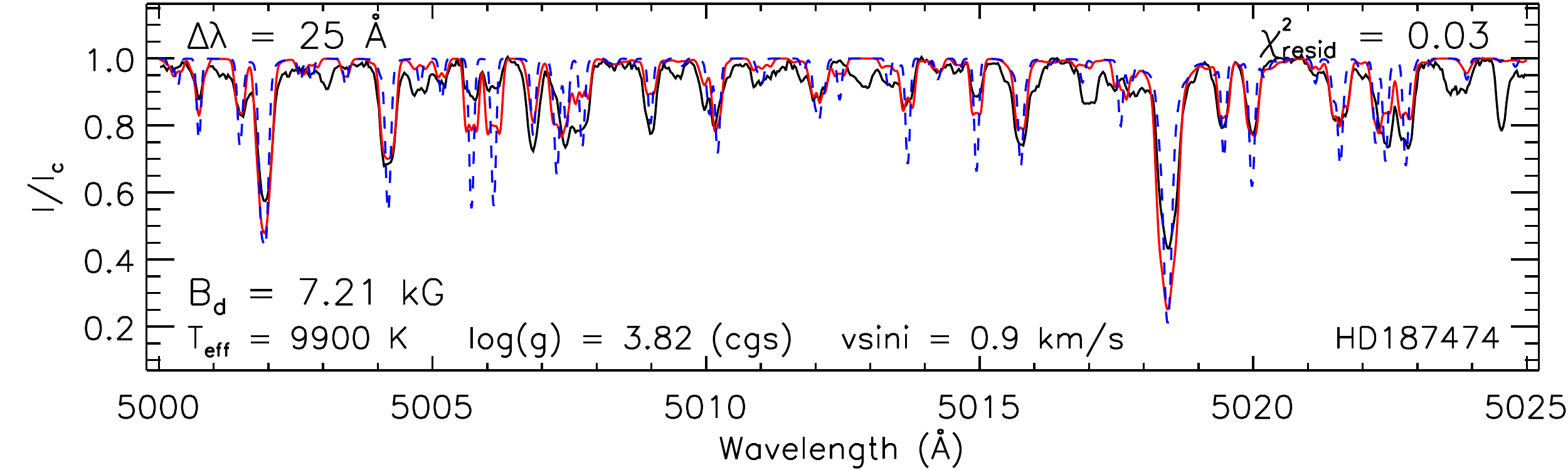}}
	\subfigure{\includegraphics[width=2.0\columnwidth]{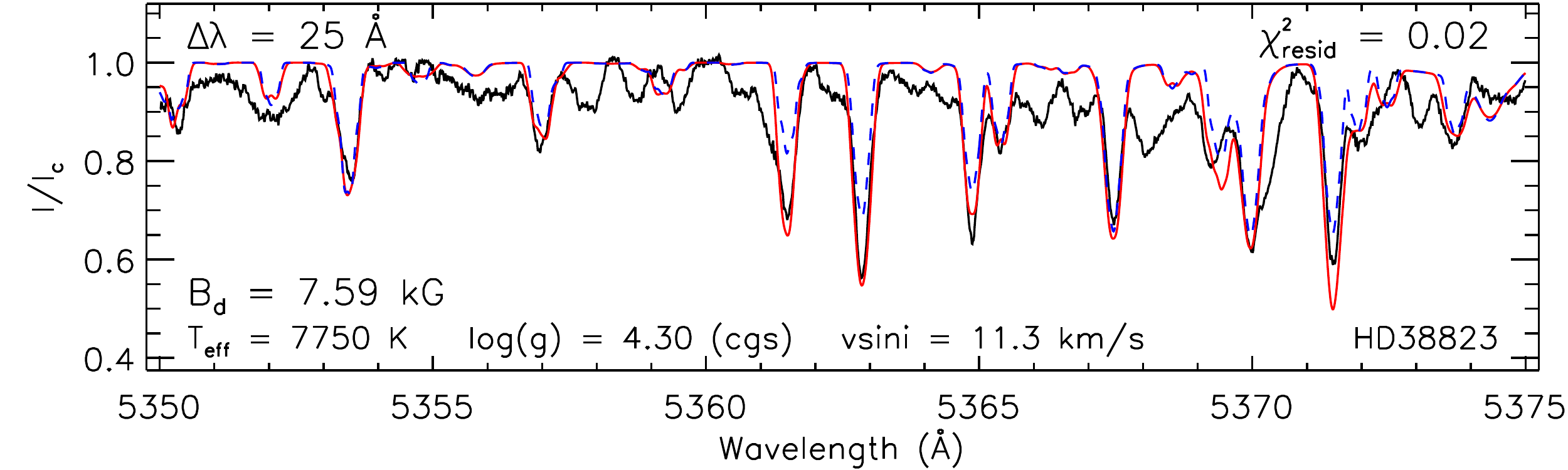}}
	\subfigure{\includegraphics[width=2.0\columnwidth]{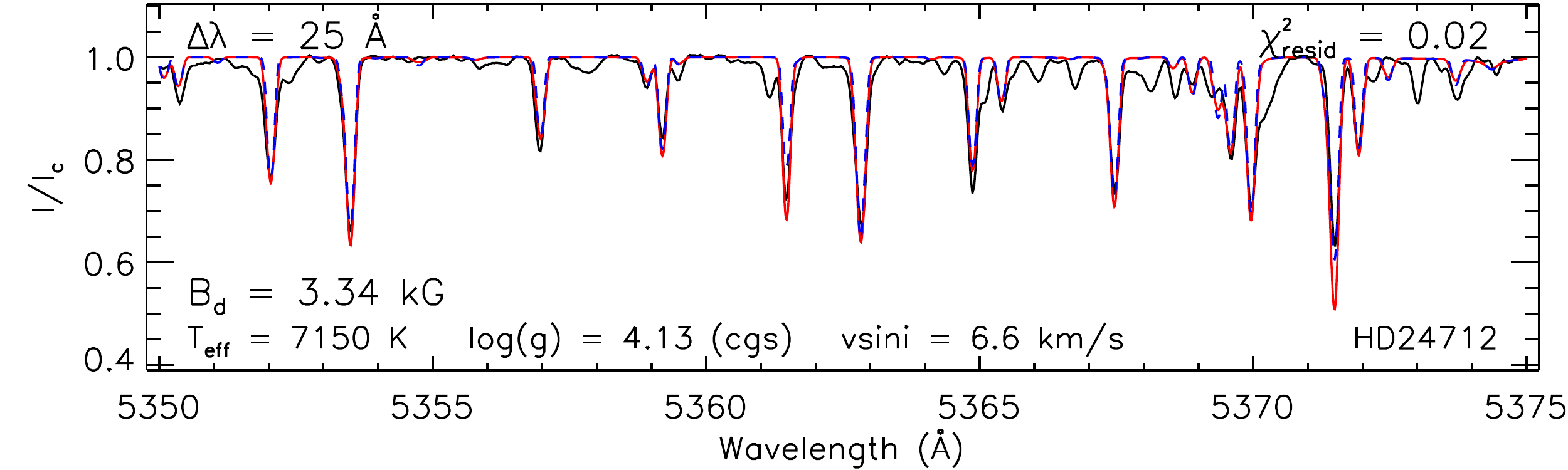}}
	\caption{Examples of fits generated where a dipole magnetic field was included in the model. The 
	black curves correspond to the observed spectra, the red curves correspond to the best-fitting 
	model spectra derived using the {\sc zeeman} LMA with Zeeman broadening included, and the dashed-blue 
	curves correspond to the spectra generated using {\sc zeeman} with the same fitting parameters and 
	no Zeeman broadening included. In each case, $\chi^2_{\rm resid}\ge0.02$ (as shown in Fig. 
	\ref{fig:compare_zeeman}) and Zeeman broadening is considered to be non-negligible.}
	\label{fig:metal_fit_zeeman}
\end{figure*}

In Fig. \ref{fig:compare_zeeman}, we show $\chi^2_{\rm resid}$ as a function of $v\sin{i}$ and 
$B_{\rm d}$. As expected, the models corresponding to the lowest $v\sin{i}$ and highest $B_{\rm d}$ values 
exhibit the largest $\chi^2_{\rm resid}$ values. In Fig. \ref{fig:metal_fit}, we show examples of fits 
obtained with and without the inclusion of magnetic fields. Based on these comparisons, we concluded that 
Zeeman broadening can be neglected (for the purposes of this investigation) for the 31 stars where 
$\chi^2_{\rm resid}\lesssim0.02$. In Fig. \ref{fig:metal_fit_zeeman}, we show examples of fits obtained 
where Zeeman broadening was found to be non-negligible.

\begin{figure*}
	\centering
	\includegraphics[width=1.9\columnwidth]{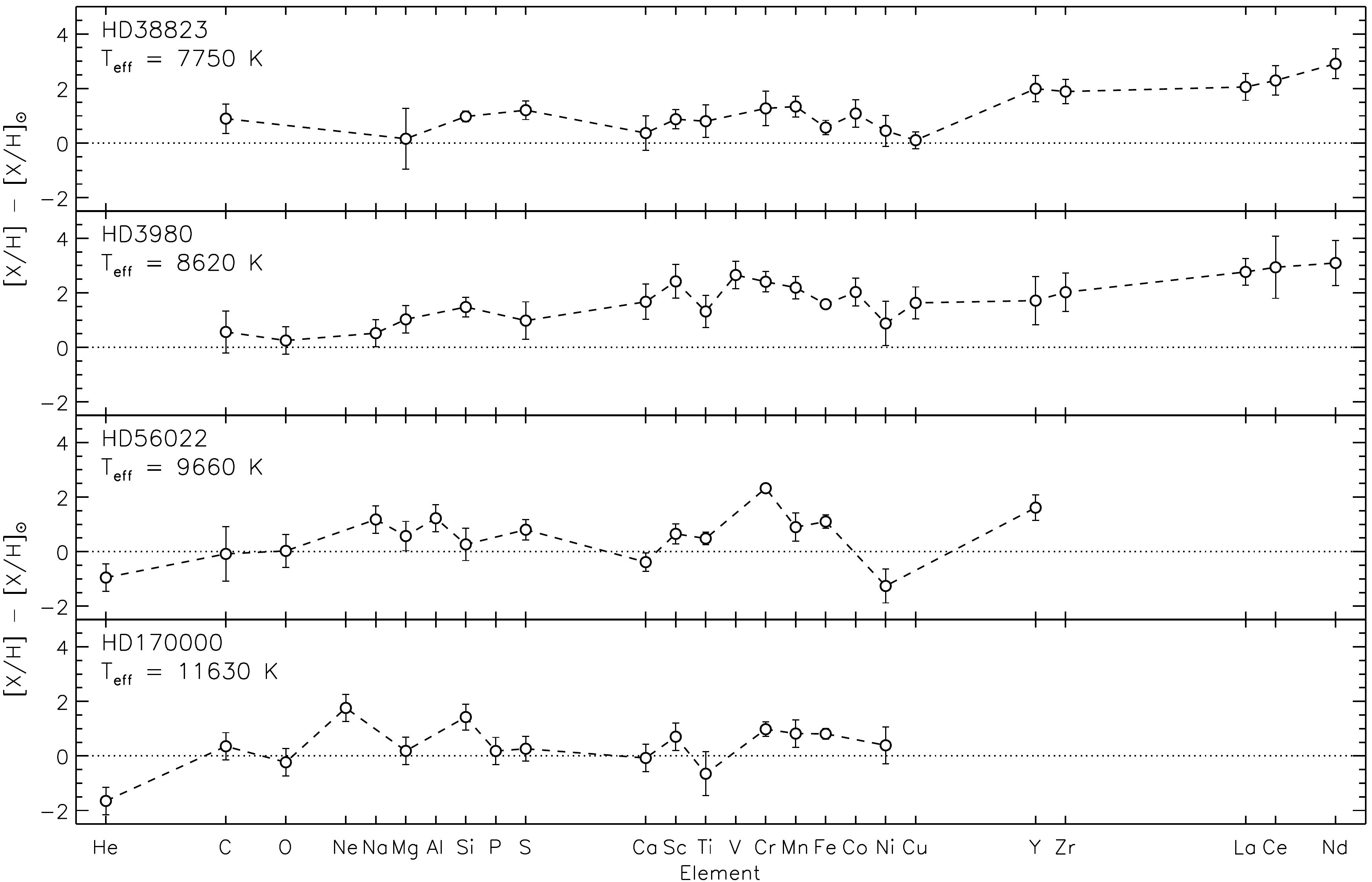}
	\caption{Examples of the tables of derived average surface chemical abundances. The examples shown 
	increase in $T_{\rm eff}$ from top to bottom.}
	\label{fig:abund_tbl}
\end{figure*}

We carried out the spectral modelling of the $5\,000\leq\lambda\leq5\,900\,{\rm \AA}$ regions for the 
majority of the confirmed mCP stars in the sample. No abundance analysis was carried out for the six 
stars without available new or archival spectra (HD~12447, HD~49976, HD~54118, HD~64486, HD~103192, and 
HD~117025 along with the three candidate mCP stars, HD~15717, HD~32576, and HD~217831). No abundance 
analysis of the SB2 system HD~109026 was carried out due to the significant contamination between the 
non-magnetic primary and magnetic secondary components in the HARPS spectra. In total, average surface 
abundances of various elements were derived for 45/52 of the mCP stars. Several examples of final 
spectral fits obtained after the first and second iterations are shown in Fig. \ref{fig:metal_fit}. The 
four examples span a range of $v\sin{i}$ values and approximately represent the range of the quality of 
the fits obtained during the analysis. The derived chemical abundances are listed in Table 
\ref{tbl:abund_tbl}. We also show four examples of the derived abundance tables in Fig. 
\ref{fig:abund_tbl}.

\subsection{HD 109026}\label{sect:HD109026}

HD~109026 is the only double-lined spectroscopic binary system identified amongst the 52 confirmed mCP 
stars in the sample. The primary component is a slowly pulsating B-type star, which exhibits 
pulsations with a period of $2.73\,{\rm d}$ \citep{Waelkens1998}. The cooler secondary component was 
first discovered by \citet{Alecian2014} based on spectra obtained using HARPSpol. The authors detected 
variable circularly polarized Zeeman signatures and variable metallic lines, which led them to classify 
the secondary component as an Ap star. No radial velocity shifts in either the primary or secondary 
component were reportedly detected over a period of seven nights indicative of an orbital period that 
is significantly longer than several days.

\begin{figure*}
	\centering
	\subfigure{\includegraphics[width=1.0\columnwidth]{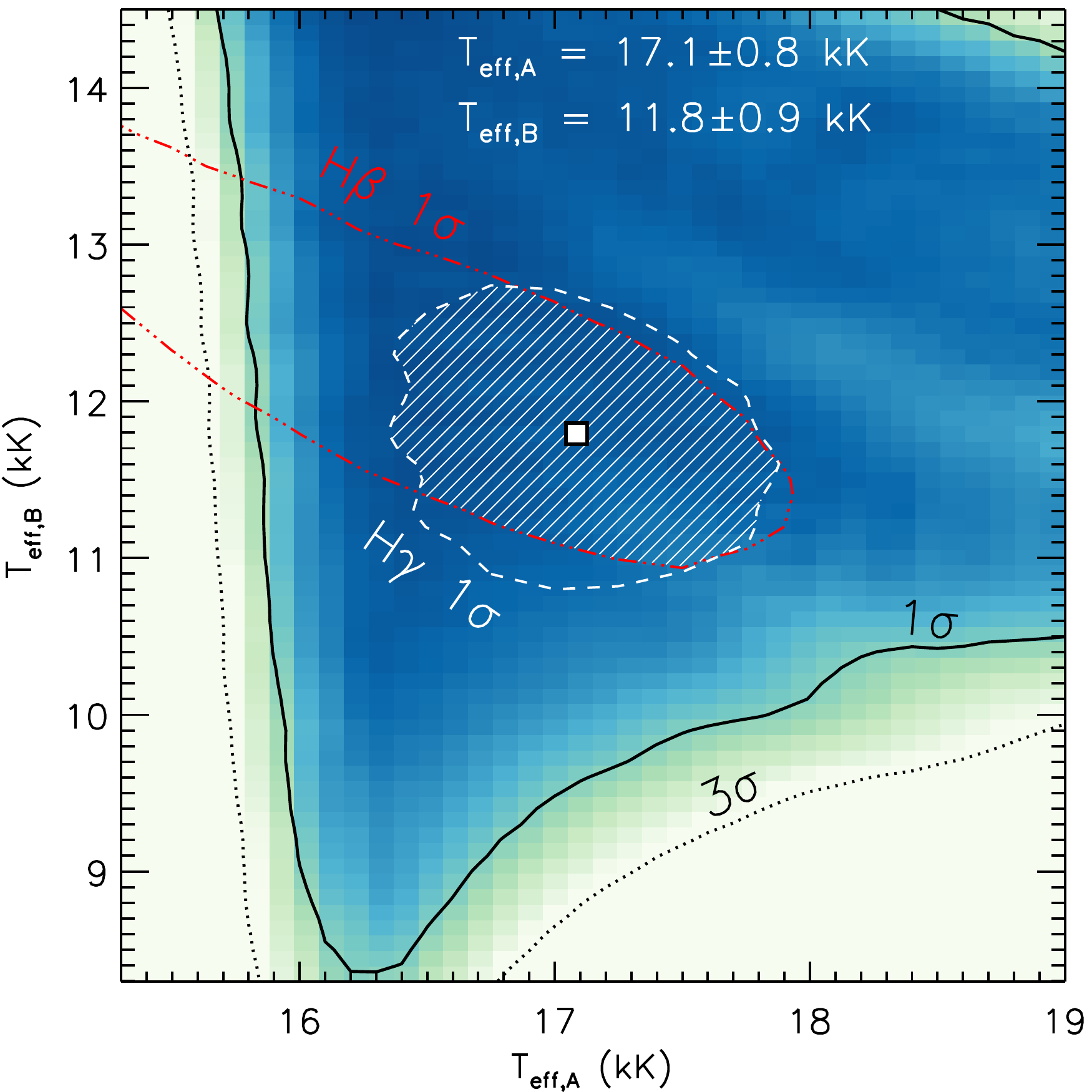}}
	\subfigure{\includegraphics[width=1.0\columnwidth]{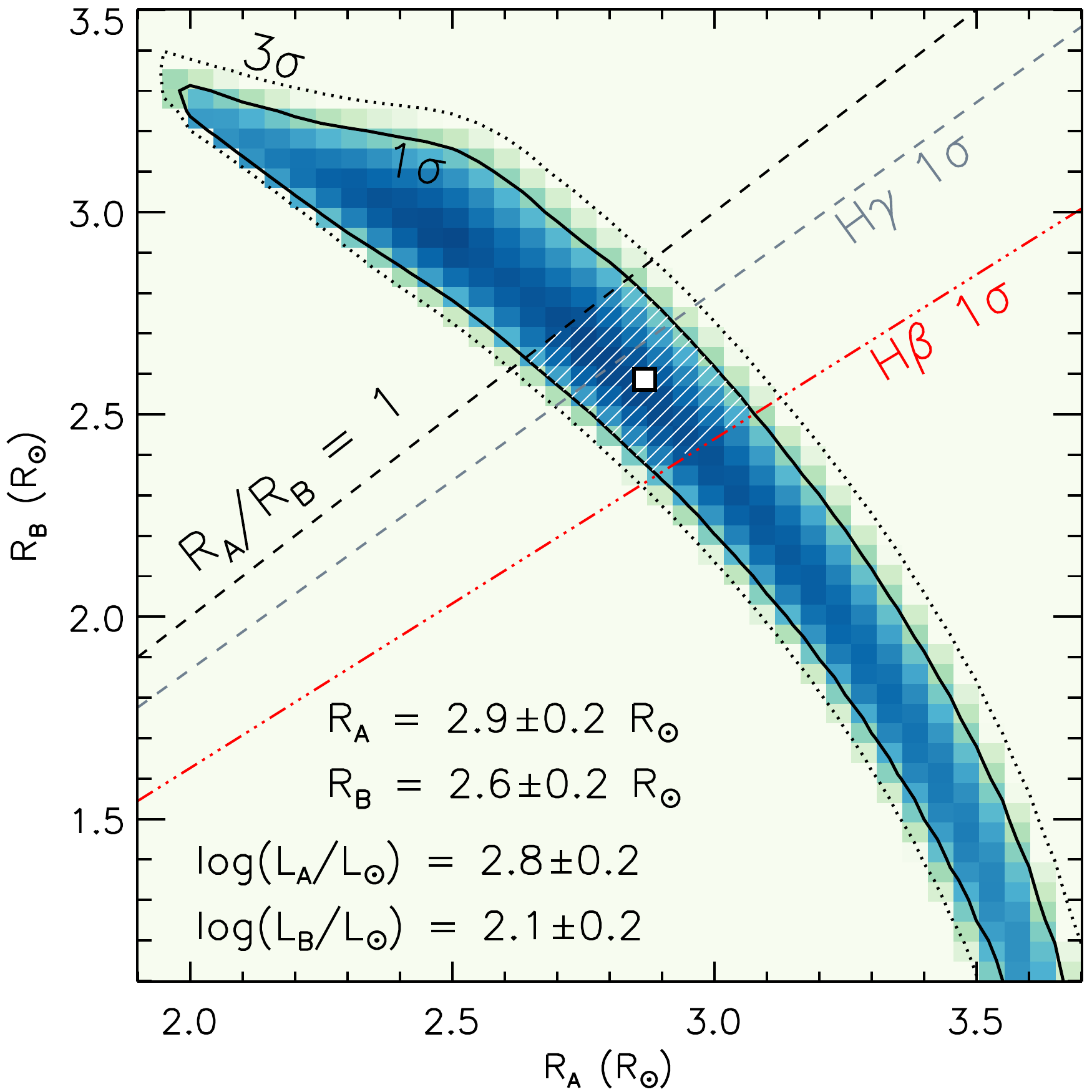}}
\caption{The $\chi^2$ distributions associated with the SED modelling (shown in Fig. 
\ref{fig:hd109026_sed}): the colours correspond to the $\chi^2$ value increasing from dark to light 
while the solid and dotted black contours correspond to the $1\sigma$ and $3\sigma$ limits, 
respectively. In both plots, the white hatched regions indicate the regions used to assign final 
values (marked by white squares) and uncertainties of the plotted parameters. \emph{Left:} Comparisons 
between the $T_{\rm A}-T_{\rm B}$ $\chi^2$ plane associated with the photometry and the 
Balmer line modelling. The red dot-dashed and white dashed contours correspond to the $1\sigma$ limits 
of the H$\beta$ and H$\gamma$ modelling shown in Fig. \ref{fig:hd109026_balmer}. \emph{Right:} The 
$R_{\rm A}-R_{\rm B}$ $\chi^2$ plane associated with the photometry. The Balmer 
line modelling yields constraints on the ratio of the radii ($R_{\rm A}/R_{\rm B}$); the $1\sigma$ 
upper limits for $R_{\rm A}/R_{\rm B}$ are indicated by the grey dashed and red dot-dashed lines 
while the dashed black line shows the lower limit of $R_{\rm A}/R_{\rm B}=1$ enforced during the 
fitting analysis.}\label{fig:hd109026_fp}
\end{figure*}

\begin{figure*}
	\centering
	\includegraphics[width=2.1\columnwidth]{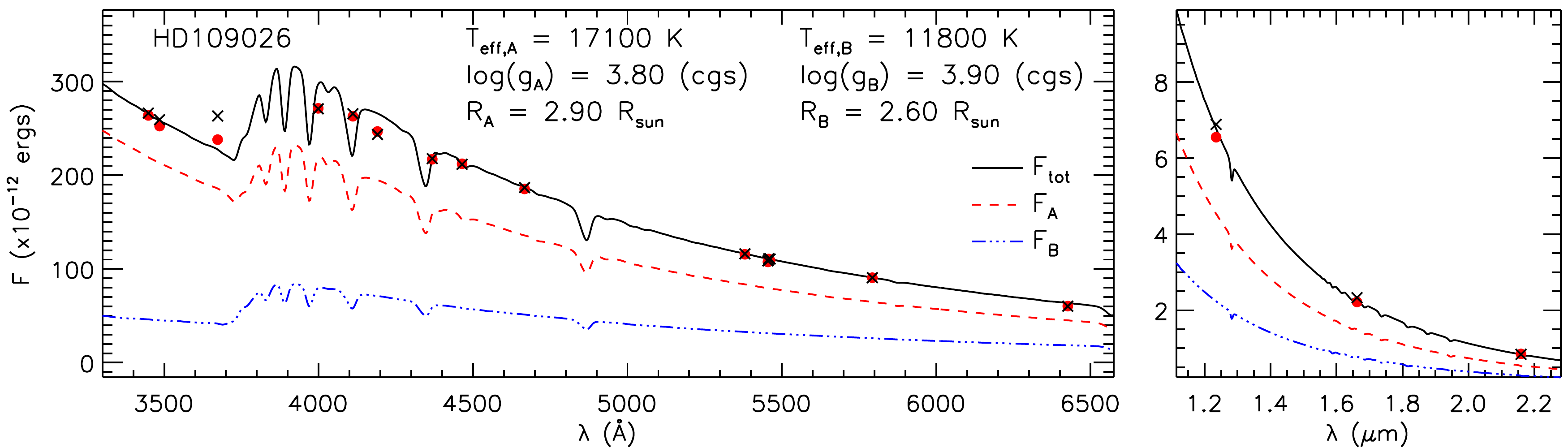}
	\caption{The final fit to the observed photometry (filled red circles) of the SB2 system HD~109026. 
	The left panel shows the optical photometry (Johnson, Geneva, and Str\"omgren) while the 
	right panel shows the 2MASS photometry. The solid black, dashed red, and dot-dashed blue lines 
	correspond to the total, primary component, and secondary (magnetic) component synthetic SEDs. 
	The black `X' symbols correspond to the total synthetic flux of each photometric filter.}
	\label{fig:hd109026_sed}
\end{figure*}

\begin{figure*}
	\centering
	\includegraphics[width=2.1\columnwidth]{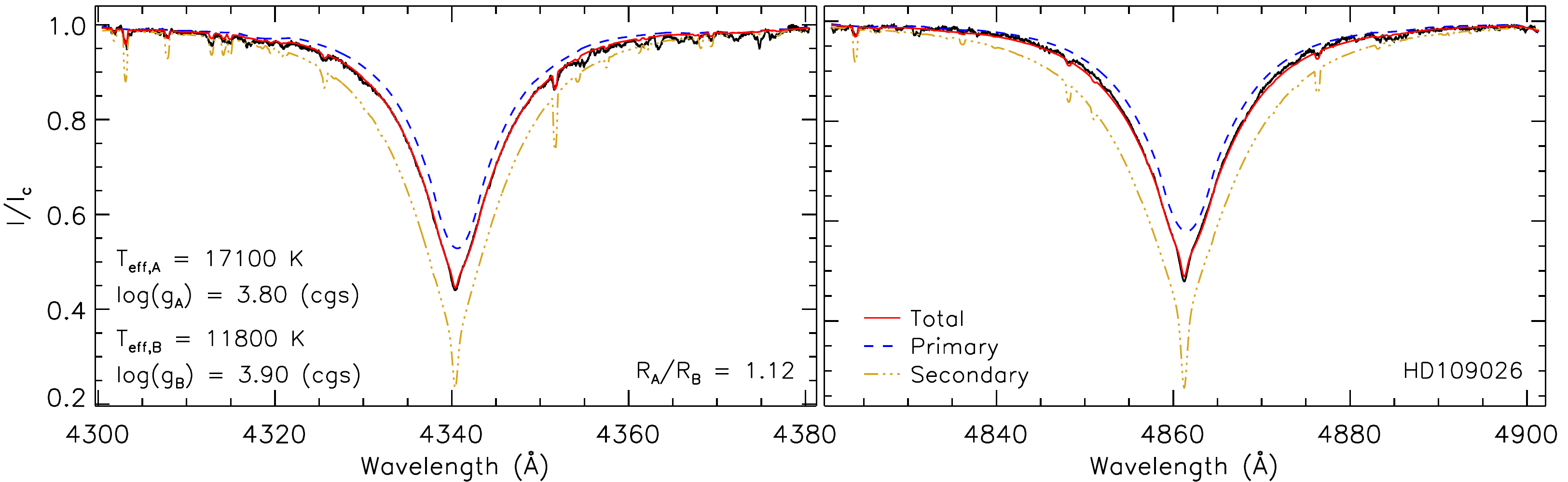}
	\caption{The final fits to the observed H$\gamma$ (left) and H$\beta$ (right) profiles (black 
	curves) of the SB2 system HD~109026. The dashed blue, dot-dashed yellow, and solid red curves 
	correspond to the primary, secondary (magnetic), and composite model spectra.}
	\label{fig:hd109026_balmer}
\end{figure*}

Various effective temperatures of the system ranging from $15\,{\rm kK}$ to $17.7\,{\rm kK}$ have been 
previously reported based on photometric calibrations \citep[e.g.][]{Molenda-Zakowicz2004,Kochukhov2006,
Zorec2009}. However, no $T_{\rm eff}$ of the cooler magnetic component could be found in the literature. 
\citet{Alecian2014} attribute their inability to constrain $T_{\rm eff}$ to the contamination of the 
primary component's moderately variable lines with the secondary component's strongly variable lines. We 
attempted to estimate $T_{\rm eff}$ of both components by combining the constraints yielded by (1) SED 
fitting and (2) spectral Balmer line modelling.

Johnson, Str\"omgren, Geneva, and 2MASS photometric measurements are available for HD~109026. We 
attempted to fit the observations using a grid search method in which $\chi^2$ is calculated for a 
grid of free parameters allowing the minimal $\chi^2$ solution to be identified. Unlike the SED fitting 
method discussed in Sect. \ref{sect:mag_sed}, we (1) included contributions from both a primary and 
secondary component and (2) we did not consider the effects of chemical peculiarities or surface 
magnetic fields. The metallicity and surface gravity of both components were fixed at [M/H]$=0$ and 
$\log{g}=4.0\,{\rm (cgs)}$, respectively. The effective temperatures and radii of both components were 
free parameters; the primary component's $T_{\rm eff}$ and $R$ were bounded by 
$T_{\rm eff, A}/{\rm kK}\in[14,20]$ and $R_{\rm A}/R_\odot\in[3,5]$ while the secondary component's 
$T_{\rm eff}$ and $R$ were bounded by $T_{\rm eff, B}/{\rm kK}\in[8,15]$ and 
$R_{\rm B}/R_\odot\in[3,5]$. Lastly, in order to reduce the number of grid points in the $\chi^2$ 
space (and thus, reduce the computation time), we restricted the primary component's luminosity 
(computed using the Stefan-Boltzmann relation) such that $L_{\rm A}/L_{\rm B}>1$. In Fig. \ref{fig:hd109026_fp} we show the two components' $\chi^2$ 
distributions in the $T_{\rm eff}$ and $R$ planes associated with the SED modelling. It is apparent 
that the constraints on the free parameters yielded by the SED modelling alone are poor: similar 
quality fits are obtained for a wide range of solutions ($16\lesssim T_{\rm eff,A}\lesssim20\,{\rm kK}$ 
and $9\lesssim T_{\rm eff,B}\lesssim15\,{\rm kK}$).

\renewcommand{\arraystretch}{1.2}
\begin{table*}
	\caption{Rotational broadening ($v\sin{i}$) and surface gravity $\log{g}$ derived from 
	spectroscopic measurements. Columns 2, 4, 6, and 8 list the $v\sin{i}$ and $\log{g}$ values 
	derived in this study; columns 3 and 7 list $v\sin{i}$ values obtained from the literature. The bold 
	$v\sin{i}$ values correspond to the final adopted values used in the final spectral models and in the 
	calculation of the magnetic parameters in Paper II. References for the published $v\sin{i}$ values 
	are listed in the table's footer.}
	\label{tbl:spec_tbl}
	\begin{center}
	\begin{tabular*}{2.0\columnwidth}{@{\extracolsep{\fill}}l c c c @{\hskip 0.75cm} | @{\hskip 0.75cm} c c c r@{\extracolsep{\fill}}}
		\noalign{\vskip-0.2cm}
		\hline
		\hline
		\noalign{\vskip0.5mm}
		HD  & $(v\sin{i})_{\rm obs}$ & $(v\sin{i})_{\rm pub}$ & $\log{(g)}_{\rm obs}$ & HD  & $(v\sin{i})_{\rm obs}$ & $(v\sin{i})_{\rm pub}$ & $\log{(g)}_{\rm obs}$ \\
		    & $({\rm km\,s^{-1}})$   & $({\rm km\,s^{-1}})$   & (cgs)                 &     & $({\rm km\,s^{-1}})$   & $({\rm km\,s^{-1}})$   & (cgs)                 \\
		(1) & (2)                    & (3)                    & (4)                   & (5) & (6)                    & (7)                    & (8)                   \\
		\noalign{\vskip0.5mm}
		\hline	
		\noalign{\vskip0.5mm}
3980   &     $\mathbf{25.4\pm1.7}$ &              $15.0\pm3.0^{\rm \,a}$ &    $4.04_{-0.42}^{+0.78}$ & 112413 &     $\mathbf{17.3\pm2.2}$ &                                   - &    $3.89_{-0.37}^{+0.77}$ \\
11502  &     $\mathbf{56.8\pm1.7}$ &                                   - &    $4.04_{-0.50}^{+0.96}$ & 117025 &                         - &                                   - &                         - \\
12446  &                         - &     $\mathbf{56.0\pm3.0^{\rm \,b}}$ &                         - & 118022 &     $\mathbf{12.0\pm0.7}$ &              $12.0\pm1.0^{\rm \,k}$ &    $4.11_{-0.42}^{+0.89}$ \\
15089  &     $\mathbf{49.5\pm2.8}$ &                                   - &                         - & 119213 &     $\mathbf{35.8\pm3.8}$ &                                   - &    $4.17_{-0.46}^{+0.83}$ \\
15144  &     $\mathbf{12.0\pm0.9}$ &              $13.0\pm1.0^{\rm \,c}$ &                         - & 120198 &     $\mathbf{58.0\pm1.5}$ &                                   - &    $3.83_{-0.17}^{+0.71}$ \\
18296  &     $\mathbf{24.3\pm2.4}$ &                                   - &    $3.33_{-0.33}^{+0.97}$ & 124224 &        $\mathbf{144\pm5}$ &                                   - &    $4.06_{-0.12}^{+0.36}$ \\
24712  &      $\mathbf{6.6\pm0.6}$ &                                   - &    $3.80_{-0.80}^{+1.20}$ & 128898 &     $\mathbf{14.1\pm0.5}$ &              $14.0\pm1.0^{\rm \,l}$ &                   $>4.22$ \\
27309  &     $\mathbf{55.8\pm2.4}$ &                                   - &                         - & 130559 &     $\mathbf{18.6\pm1.6}$ &                                   - &    $3.84_{-0.49}^{+1.09}$ \\
29305  &     $\mathbf{44.2\pm1.9}$ &                                   - &    $3.97_{-0.31}^{+0.47}$ & 137909 &      $\mathbf{9.5\pm0.4}$ &                                   - &    $3.25_{-0.25}^{+1.33}$ \\
38823  &     $\mathbf{11.3\pm1.9}$ &                                   - &    $4.30_{-1.29}^{+0.51}$ & 137949 &      $\mathbf{9.7\pm0.5}$ &                                   - &    $3.30_{-0.30}^{+1.38}$ \\
40312  &     $\mathbf{58.5\pm2.5}$ &             $57.0\pm10.0^{\rm \,d}$ &    $3.41_{-0.21}^{+0.49}$ & 140160 &     $\mathbf{59.7\pm2.1}$ &                                   - &    $4.00_{-0.16}^{+0.24}$ \\
49976  &                         - &     $\mathbf{31.0\pm3.0^{\rm \,e}}$ &                         - & 140728 &     $\mathbf{68.7\pm3.2}$ &                    $65.0^{\rm \,m}$ &    $3.69_{-0.21}^{+0.53}$ \\
54118  &                         - &     $\mathbf{34.0\pm2.0^{\rm \,f}}$ &                         - & 148112 &     $\mathbf{47.2\pm3.0}$ &                                   - &    $3.45_{-0.25}^{+0.67}$ \\
56022  &     $\mathbf{84.6\pm7.4}$ &              $28.0\pm8.0^{\rm \,b}$ &    $3.93_{-0.16}^{+0.64}$ & 148898 &     $\mathbf{32.2\pm1.4}$ &              $51.0\pm8.0^{\rm \,m}$ &    $3.85_{-0.69}^{+0.65}$ \\
62140  &     $\mathbf{25.8\pm1.0}$ &                                   - &    $3.70_{-0.70}^{+1.30}$ & 151199 &     $\mathbf{49.7\pm1.9}$ &                                   - &    $4.02_{-0.12}^{+0.22}$ \\
64486  &                         - &                                   - &                         - & 152107 &     $\mathbf{23.2\pm1.4}$ &                                   - &    $4.10_{-0.19}^{+0.61}$ \\
65339  &              $13.3\pm2.7$ &     $\mathbf{12.5\pm0.5^{\rm \,g}}$ &    $4.18_{-0.25}^{+0.67}$ & 170000 &     $\mathbf{81.9\pm1.6}$ &                                   - &    $4.11_{-0.13}^{+0.27}$ \\
72968  &     $\mathbf{16.1\pm0.8}$ &                                   - &    $3.83_{-0.33}^{+0.97}$ & 176232 &               $2.1\pm0.7$ &      $\mathbf{2.0\pm0.5^{\rm \,n}}$ &             $4.00\pm1.00$ \\
74067  &     $\mathbf{33.8\pm2.8}$ &              $33.0\pm1.0^{\rm \,h}$ &    $4.13_{-0.31}^{+0.69}$ & 187474 &      $\mathbf{0.9\pm0.6}$ &                                   - &    $3.82_{-0.58}^{+1.18}$ \\
83368  &              $33.8\pm1.0$ &     $\mathbf{33.0\pm0.5^{\rm \,g}}$ &    $3.56_{-0.56}^{+0.24}$ & 188041 &      $\mathbf{6.7\pm0.5}$ &                     $4.0^{\rm \,a}$ &    $4.36_{-0.76}^{+0.64}$ \\
96616  &     $\mathbf{57.0\pm2.5}$ &              $56.0\pm2.5^{\rm \,i}$ &    $3.72_{-0.16}^{+0.44}$ & 201601 &      $\mathbf{7.0\pm0.5}$ &                                   - &    $3.85_{-0.85}^{+0.51}$ \\
103192 &                         - &           $\mathbf{72.0^{\rm \,b}}$ &                         - & 203006 &     $\mathbf{43.7\pm1.7}$ &                                   - &    $3.73_{-0.23}^{+0.29}$ \\
108662 &     $\mathbf{22.0\pm1.7}$ &                                   - &    $3.72_{-0.16}^{+0.26}$ & 217522 &      $\mathbf{4.9\pm0.6}$ &                                   - &                   $<4.20$ \\
108945 &     $\mathbf{65.2\pm3.8}$ &                                   - &    $3.85_{-0.15}^{+0.11}$ & 220825 &     $\mathbf{39.9\pm2.2}$ &                                   - &    $3.94_{-0.18}^{+0.12}$ \\
109026 &     $\mathbf{12.0\pm1.5}$ &                $188\pm10^{\rm \,j}$ &             $3.90\pm0.30$ & 221760 &     $\mathbf{22.4\pm0.7}$ &                                   - &    $3.65_{-0.09}^{+0.41}$ \\
112185 &     $\mathbf{32.9\pm1.4}$ &                                   - &    $3.66_{-0.14}^{+0.96}$ & 223640 &     $\mathbf{31.7\pm1.4}$ &              $20.0\pm5.0^{\rm \,m}$ &    $3.92_{-0.22}^{+0.42}$ \\
\noalign{\vskip0.5mm}
\hline
\multicolumn{8}{l}{$^{\rm a\,}$\citet{Hubrig2007}, $^{\rm b\,}$\citet{Abt2001}, $^{\rm c\,}$\citet{Auriere2007}, $^{\rm d\,}$\citet{Rice1990}} \\
\multicolumn{8}{l}{$^{\rm e\,}$\citet{Pilachowski1974}, $^{\rm f\,}$\citet{Donati1997}, $^{\rm g\,}$\citet{Kochukhov2004}, $^{\rm h\,}$\citet{Royer2002}} \\
\multicolumn{8}{l}{$^{\rm i\,}$\citet{Hoffleit1995}, $^{\rm j\,}$\citet{Brown1997}, $^{\rm k\,}$\citet{Khalack2006}, $^{\rm l\,}$\citet{Reiners2003}} \\
\multicolumn{8}{l}{$^{\rm m\,}$\citet{Abt1995}, $^{\rm n\,}$\citet{Kochukhov2002}} \\
	\end{tabular*}
	\end{center}
\end{table*}
\renewcommand{\arraystretch}{1.0}

The available HARPS Stokes $I$ spectra of HD~109026 span the H$\gamma$ and H$\beta$ lines. We performed a 
similar spectral line modelling analysis to that presented in Sect. \ref{sect:balmer_fit} using the 
{\sc gssp composite} module described by \citet{Tkachenko2015}. This module functions in the same way 
as the {\sc gssp single} module used in Sect. \ref{sect:balmer_fit} (i.e. by generating a grid of 
solutions from which the minimal $\chi^2$ solution can be identified) but allows a spectrum consisting 
of two components to be fit. We defined a grid of $T_{\rm eff,A}/{\rm kK}\in[14,19]$, 
$T_{\rm eff,B}/{\rm kK}\in[10,15]$, $\log{g_{\rm A}}\in[3.0,5.0]$, and $\log{g_{\rm B}}\in[3.0,5.0]$ 
values with increments of $\Delta T_{\rm eff}=100\,{\rm K}$ and $\Delta\log{g}=0.1\,{\rm (cgs)}$. The 
$v\sin{i}$ values of each component were varied from $170-190\,{\rm km\,s^{-1}}$ in increments of 
$1\,{\rm km\,s^{-1}}$ (primary) and $5-25\,{\rm km\,s^{-1}}$ in increments of $0.5\,{\rm km\,s^{-1}}$ 
(secondary). We used fixed solar abundancies (i.e. no individual abundances were varied), and fixed 
microturbulence values of $2\,{\rm km\,s^{-1}}$ and $0\,{\rm km\,s^{-1}}$ for the primary and secondary 
components, respectively. The contribution of each component to the composite spectrum is primarily 
determined by the ratio of the luminosities 
($L_{\rm A}/L_{\rm B}=[T_{\rm A}/T_{\rm B}]^4[R_{\rm A}/R_{\rm B}]^2$); we allowed the ratio of the radii, 
$R_{\rm A}/R_{\rm B}$, to range from 1 to 3 in increments of 0.05.

The resulting $\chi^2$ grids associated with the fits to H$\gamma$ and H$\beta$ were then compared with 
the $\chi^2$ grids generated during the SED fitting analysis. In Fig. \ref{fig:hd109026_fp} (left), we 
show the resulting spectral modelling 1$\sigma$ contours associated with $T_{\rm eff,A}$ and 
$T_{\rm eff,B}$ overlayed onto the SED fitting $T_{\rm eff}$ $\chi^2$ distribution. In Fig. 
\ref{fig:hd109026_fp} (right), we show the spectral modelling 1$\sigma$ limits of $R_{\rm A}/R_{\rm B}$ 
overlayed onto the $R$ SED fitting $\chi^2$ distribution. By combining the constraints yielded by both the 
SED fitting and the Balmer line fitting, we obtain best fitting parameters -- corresponding to the centers 
of the overlapping 1$\sigma$ contours -- with large but reasonable 1$\sigma$ errors: 
$T_{\rm eff,A}=17.1\pm0.8\,{\rm kK}$, $T_{\rm eff,B}=11.8\pm0.9\,{\rm kK}$, 
$R_{\rm A}=2.9\pm0.2\,R_\odot$, and $R_{\rm B}=2.6\pm0.2\,R_\odot$. These values yield luminosities of 
$\log{L_{\rm A}/L_\odot}=2.8\pm0.2$ and $\log{L_{\rm B}/L_\odot}=2.1\pm0.2$. In Figures 
\ref{fig:hd109026_sed} and \ref{fig:hd109026_balmer}, we show the SED fits and Balmer line fits generated 
using the derived most probable parameters.

\subsection{Hertzsprung-Russell diagram}\label{sect:HRD}

With the values of $T_{\rm eff}$ and $L$ derived for all of the stars (mCP, candidate mCP, and 
non-mCP), their masses ($M$) and ages ($t_{\rm age}$) could be estimated by generating a theoretical 
Hertzsprung-Russell diagram (HRD) and comparing with evolutionary models. Various theoretical 
evolutionary tracks spanning a wide range of metallicities and rotation rates are available in the 
literature \citep[e.g.][]{Schaller1992,Schaerer1993,Georgy2013a}. Depending on the choice of models, 
the derived masses and, in particular, the derived ages of the stars can vary substantially.

Although our analysis includes a derivation of most of the mCP stars' surface chemical compositions, 
it is understood that these values are not representative of their bulk metallicites 
\citep[e.g.][]{Michaud1970,Michaud1980}. Studies of the Sun's metallicity as well as that of 
nearby B-type stars \citep{Przybilla2008} and star-forming H~{\sc ii} regions \citep[e.g.][]{Esteban2004} 
suggest that $Z\approx0.014$ within the solar neighbourhood \citep{Asplund2009}. The Geneva-Copenhagen 
survey \citep{Nordstrom2004,Casagrande2011}, which derived the metallicities of more than $16\,000$ F-, 
G-, and K-type stars in the Hipparcos Catalogue, report a similar value within the solar neighbourhood: 
the $9\,605$ stars within $100\,{\rm pc}$ with derived [M/H] values exhibit an approximately Gaussian 
distribution with mean and sigma values of $-0.04$ and $0.15$, respectively. No significant change in 
[M/H] is found as a function of distance within $100\,{\rm pc}$. Based on these results, we adopted 
metallicities of $\lbrack {\rm M/H}\rbrack=0.0$ with a 2 $\sigma$ uncertainty of $0.3$ ($Z=0.014\pm0.009$) 
for all of the stars (both mCP and non-mCP) in the sample.

Magnetic F-, A-, and B-type MS stars are typically known to have relatively low equatorial rotational 
velocities ($v_{\rm eq}$) \citep[e.g.][]{Abt1995a,Netopil2017}, however, the $v_{\rm eq}$ values 
exhibited by their non-mCP counterparts vary widely. We estimated the non-mCP sub-sample's typical ratio 
of $v_{\rm eq}$ to the critical rotational velocity (i.e. the break-up rotational velocity, 
$v_{\rm crit}$) by first compiling the $v\sin{i}$ and $T_{\rm eff}$ values of MS A-type stars reported 
by \citet{Ammler-vonEiff2012} and \citet{Zorec2012}. We then estimated $v_{\rm crit}$ for each of the 
$2\,835$ catalogue entries using the non-rotating, solar metallicity evolutionary tracks computed by 
\citet{Ekstrom2012}. A Monte Carlo (MC) simulation was then performed in which, for each of the $1\,000$ 
steps that were carried out, inclination angles were randomly assigned for each star under the assumption 
that the rotational axes are randomly oriented in space. Our results suggest that we can expect the 
subsample of non-mCP stars to have a median $v_{\rm eq}/v_{\rm crit}$ of approximately 0.3 and for 
68~per~cent of the sample to have $v_{\rm eq}/v_{\rm crit}\lesssim0.4$.

\begin{figure}
	\centering
	\includegraphics[width=1.0\columnwidth]{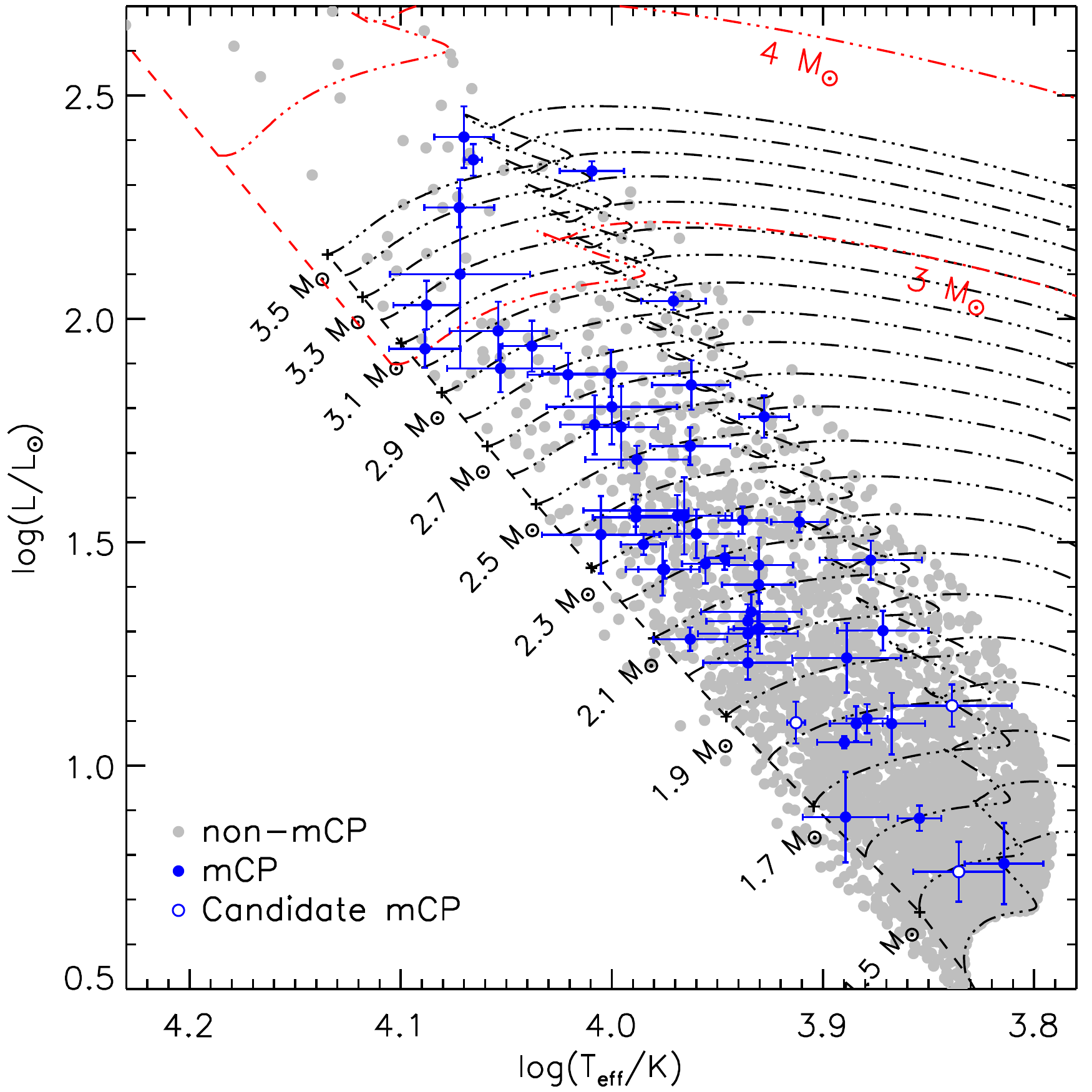}
	\caption{Theoretical HRD of the sample of non-mCP stars (grey points), the confirmed mCP stars (blue 
	filled circles), and the candidate mCP stars (blue open circles). The black and red dash-dotted 
	curves correspond to the solar metallicity low- and high-mass evolutionary tracks computed by 
	\citet{Mowlavi2012} and \citet{Ekstrom2012}, respectively.}
	\label{fig:hrd}
\end{figure}

We used several grids of evolutionary models computed with a range of metallicities to derive the 
stellar masses and ages along with their uncertainties. \citet{Mowlavi2012} computed 6 dense grids of 
evolutionary tracks corresponding to metallicities of $0.006$, $0.01$, $0.014$, $0.02$, $0.03$, and 
$0.04$; each grid contains 39 tracks with $0.5\leq M/M_\odot\leq3.5$. \citet{Ekstrom2012} computed 
four larger, lower density grids having metallicities of $0.002$ and $0.014$, both with rotation 
($v_{\rm eq}/v_{\rm crit}=0.4$) and without rotation; each grid contains 24 tracks corresponding to a 
wide range of masses ($0.8\leq M/M_\odot\leq120$). Comparing these grids, it is evident that 
$T_{\rm eff}$ of the terminal-age MS (TAMS) decreases significantly for those tracks associated with a 
high metallicity ($Z=0.03$) and/or fast rotation rate ($v_{\rm eq}/v_{\rm crit}=0.4$) relative to the 
non-rotating solar metallicity models. We linearly interpolated the grids computed by 
\citet{Mowlavi2012} in order to generate a $Z=0.023$ grid (corresponding to the adopted high metallicity 
limit inferred from the Geneva-Copenhagen survey) and similarly interpolated the grids computed by 
\citet{Ekstrom2012} in order to generate a solar metallicity, $v_{\rm eq}/v_{\rm crit}=0.3$ grid 
(corresponding to the estimated typical rotational velocity of the non-mCP stars). We found that the 
$Z=0.023$ models yielded a moderately cooler TAMS compared to the rotating models; therefore, since the 
width of the MS is more strongly affected by an increase in $Z$ compared to an 
increase in $v_{\rm eq}/v_{\rm crit}$, we opted to use only the available non-rotating models. We used 
the grids computed by \citet{Mowlavi2012} for the stars with $M<3.5\,M_\odot$ along with the 
non-rotating grids computed by \citet{Ekstrom2012} for the stars with $M\geq3.5\,M_\odot$. As a 
consequence, the derived uncertainties in $M$ and $t_{\rm age}$ associated with the lower-mass stars 
incorporate both low and high metallicities ($0.006\leq Z\leq0.023$) while those of the high-mass 
stars only incorporate low metallicites ($0.005\leq Z\leq0.014$).

At this point in the analysis, we applied three cuts to the sample of non-mCP stars. All non-mCP stars 
with derived masses $<1.4\,M_\odot$ were removed since this approximately corresponds to the least 
massive mCP star in the sample (HD~217522, which has a mass of $1.48\,M_\odot$). This cut was applied 
using the $Z=0.014$ grid of evolutionary tracks \citep{Ekstrom2012}. The next two cuts removed all 
stars positioned (1) below the zero-age MS \citep[ZAMS, identified using the $Z=0.023$ evolutionary 
models computed by][]{Mowlavi2012} -- these stars are discussed below -- and (2) above the coolest point 
along the MS \citep[identified using the $Z=0.014$ models computed by][]{Ekstrom2012}. These stars 
have evolved off of the MS and are associated with various phenomena (e.g. rapid structural changes) 
that are outside the scope of this study. The three cuts reduced the total number of non-mCP stars in 
the sample from $21\,665$ to $3\,141$. The HRD containing the confirmed mCP, candidate mCP, and non-mCP 
stars composing the sample is shown in Fig. \ref{fig:hrd} along with the adopted solar metallicity 
evolutionary tracks.

We note that $150$ non-mCP stars -- along with three candidate mCP stars (HD~107452, HD~122811, and 
HD~177880B) -- exhibit $\log{T_{\rm eff}/{\rm K}}\gtrsim3.9$ and $\log{L/L_\odot}\lesssim0.3$ and are 
therefore positioned well below the ZAMS associated with the $Z=0.006$ grid of evolutionary tracks. It 
is unlikely that errors in the derived $T_{\rm eff}$ and/or $\log{L}$ alone could reposition these 
stars below the MS this significantly. \citet{Wraight2012} note that HD~107452 is a binary system and 
suggest that the parallax reported within the Hipparcos Catalogue may have been affected by this. Based 
on the derived effective temperatures of these three candidate mCP stars, we find that the distances 
would need to be increased by factors of approximately 2.5 to 4.5 in order for them to be positioned 
near the ZAMS (this estimate does not consider the flux contributed from the binary companion, which 
would reduce the factor). An analysis of specific binary systems presented by \citet{Pourbaix2000} 
demonstrates that the parallax angles measured using Hipparcos can be overestimated by a factor 
$\lesssim3$ if the orbital motion of a star within the system is not accounted for, which suggests that 
the aformentioned stars in this sample may be similarly affected.

Each sample star's most probable $M$ and $t_{\rm age}$ were derived by linearly interpolating the 
adopted solar metallicity evolutionary tracks. Their associated uncertainties were estimated by 
carrying out the following MC analysis. A large number ($\gtrsim1\,000$) data points consisting of 
$T_{\rm eff}$, $L$, and $Z$ values were generated by randomly sampling from Gaussian distributions 
defined by each parameter's most probable values and uncertainties: $T_{\rm eff}$, 
$\sigma_{T_{\rm eff}}$, $L$, and $\sigma_{L}$ were derived in Section \ref{sect:mag_sed} while $Z$ and 
$\sigma_{Z}$ were inferred from the Geneva-Copenhagen survey (this MC HRD method corresponds to a 
simplified version of that carried out by Shultz et al., in prep). We then derived the corresponding 
masses and ages for each of those data points found to overlap with the evolutionary model grids. The 
uncertainties in $M$ and $t_{\rm age}$ were determined from the widths of the resulting distributions. 
No masses and ages could be derived for those non-mCP stars positioned below the solar metallicity 
evolutionary grid's MS using their most probable $T_{\rm eff}$ and $L$ values derived in Section 
\ref{sect:FP_nonmag}. In these cases, $M$ and $t_{\rm age}$ were assigned the values associated with 
the nearest valid $T_{\rm eff}$, $L$, and $Z$ obtained during the MC error analysis.

In Table \ref{tbl:fund_tbl}, we list (1) the $T_{\rm eff}$, $R$, and $\log{L}$ values derived from the 
spectroscopic and photometric observations, and (2) the $M$, $\log{g}$, $\log{t_{\rm age}}$, and 
$\tau$ values derived using the evolutionary models; the 52 confirmed mCP stars and the 3 candidate 
mCP stars are included in this table.

\section{Results}\label{sect:results}

\subsection{Confirmation of $T_{\rm eff}$}\label{sect:conf_Teff}

\begin{figure}
	\centering
	\includegraphics[width=1.0\columnwidth]{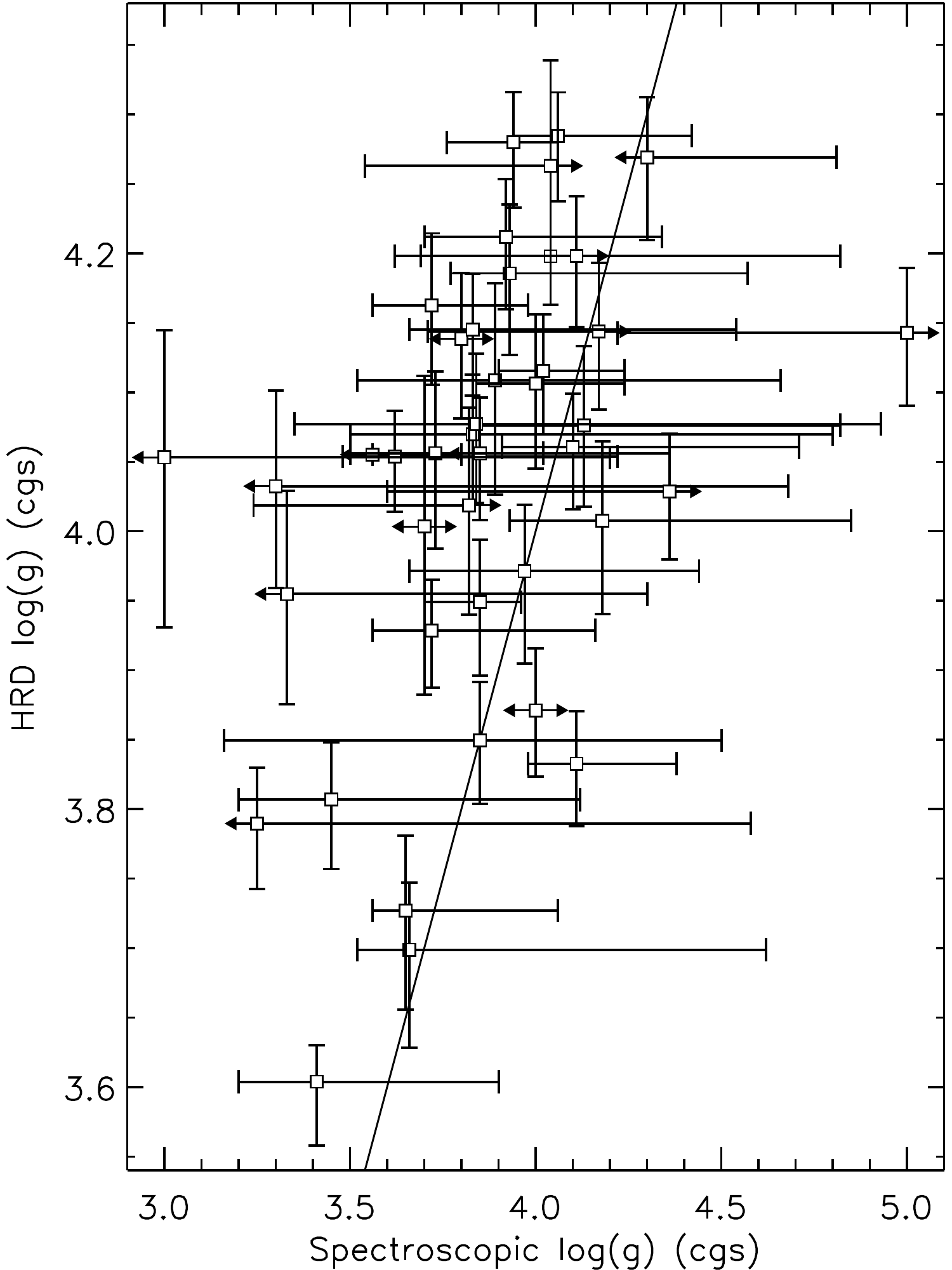}
	\caption{Surface gravity derived from H$\beta$ and H$\gamma$ (spectroscopic $\log{g}$) compared with 
	the values derived from each star's position on the theoretical HRD (HRD $\log{g}$). Each square 
	symbol corresponds to the best-fitting/most probable $\log{g}$ while the arrows indicate 
	uncertainties that extend beyond the $3.0\leq\log{g}\leq5.0$ (cgs) range permitted during the 
	fitting analysis. The black line corresponds to where the spectroscopic $\log{g}$ equals the HRD 
	$\log{g}$.}
	\label{fig:lgg_comp}
\end{figure}

\afterpage{
\onecolumn
\renewcommand{\arraystretch}{1.2}
\begin{center}
\begin{longtable}{@{\extracolsep{\fill}}l c c c c c c r@{\extracolsep{\fill}}}
\caption[asdf]{Derived fundamental parameters for the 52 confirmed mCP stars and 3 candidate mCP 
	stars (HD~15717, HD~32576, and HD~217831) in the sample. Columns (1) through (4) list parameters 
	derived from the spectroscopic and photometric observations. Columns (5) through (8) list parameters 
	derived using the evolutionary models generated by \citet{Mowlavi2012} and 
	\citet{Ekstrom2012}.} \label{tbl:fund_tbl} \\
	
\hline
\hline
\noalign{\vskip0.5mm}
	HD  & $T_{\rm eff}\,({\rm K})$ & $R\,(R_\odot)$ & $\log{L/L_\odot}$ & $M\,(M_\odot)$ & $\log{(g)}_{\rm HRD}$ & $\log{t/{\rm Gyrs}}$ & $\tau$ \\
	(1) & (2)                      & (3)            & (4)               & (5)            & (6)                   & (7)                  & (8)    \\
\noalign{\vskip0.5mm}
\hline
\noalign{\vskip0.5mm}
\endfirsthead

\multicolumn{8}{l}{continued from previous page}\\
\hline
\hline
\noalign{\vskip0.5mm}
	HD  & $T_{\rm eff}\,({\rm K})$ & $R\,(R_\odot)$ & $\log{L/L_\odot}$ & $M\,(M_\odot)$ & $\log{(g)}_{\rm HRD}$ & $\log{t/{\rm Gyrs}}$ & $\tau$ \\
	(1) & (2)                      & (3)            & (4)               & (5)            & (6)                   & (7)                  & (8)    \\
\noalign{\vskip0.5mm}
\hline	
\noalign{\vskip0.5mm}
\endhead

\noalign{\vskip0.5mm}\hline
\multicolumn{8}{l}{continued on next page}\\
\endfoot

\noalign{\vskip0.5mm}\hline
\endlastfoot

3980   &    $8620\pm420$ &   $1.85\pm0.19$ &   $1.23\pm0.04$ &    $1.96_{-0.21}^{+0.13}$ &             $4.20\pm0.10$ &    $8.67_{-0.70}^{+0.30}$ &    $0.41_{-0.33}^{+0.35}$ \\
11502  &   $10120\pm650$ &   $1.86\pm0.31$ &   $1.52\pm0.09$ &    $2.36_{-0.29}^{+0.17}$ &    $4.27_{-0.15}^{+0.10}$ &    $8.18_{-0.76}^{+0.52}$ &    $0.23_{-0.19}^{+0.45}$ \\
12446  &   $10000\pm710$ &   $2.66\pm0.45$ &   $1.80\pm0.08$ &    $2.64_{-0.28}^{+0.21}$ &             $4.01\pm0.14$ &    $8.52_{-0.27}^{+0.16}$ &    $0.68_{-0.29}^{+0.22}$ \\
15089  &    $8530\pm270$ &   $2.06\pm0.16$ &   $1.30\pm0.04$ &    $2.02_{-0.22}^{+0.14}$ &    $4.12_{-0.08}^{+0.07}$ &    $8.76_{-0.27}^{+0.20}$ &    $0.56_{-0.24}^{+0.27}$ \\
15144  &    $8510\pm240$ &   $2.07\pm0.17$ &   $1.31\pm0.06$ &    $2.02_{-0.22}^{+0.15}$ &    $4.11_{-0.08}^{+0.07}$ &    $8.77_{-0.24}^{+0.19}$ &    $0.57_{-0.23}^{+0.26}$ \\
15717  &   $6850\pm1020$ &   $1.70\pm0.22$ &   $0.76\pm0.20$ &    $1.49_{-0.24}^{+0.12}$ &    $4.15_{-0.13}^{+0.11}$ &    $9.17_{-0.50}^{+0.35}$ &    $0.63_{-0.42}^{+0.49}$ \\
18296  &   $10010\pm750$ &   $2.89\pm0.46$ &   $1.88\pm0.05$ &    $2.74_{-0.27}^{+0.19}$ &             $3.95\pm0.14$ &    $8.52_{-0.21}^{+0.13}$ &    $0.74_{-0.24}^{+0.20}$ \\
24712  &    $7150\pm170$ &   $1.80\pm0.11$ &   $0.88\pm0.03$ &    $1.60_{-0.16}^{+0.10}$ &    $4.13_{-0.07}^{+0.06}$ &    $9.06_{-0.22}^{+0.20}$ &    $0.60_{-0.25}^{+0.30}$ \\
27309  &   $11290\pm660$ &   $2.30\pm0.31$ &   $1.89\pm0.05$ &    $2.87_{-0.31}^{+0.20}$ &    $4.17_{-0.13}^{+0.11}$ &    $8.21_{-0.73}^{+0.30}$ &             $0.41\pm0.33$ \\
29305  &   $11810\pm450$ &   $3.19\pm0.28$ &   $2.25\pm0.04$ &    $3.41_{-0.32}^{+0.09}$ &    $3.96_{-0.08}^{+0.07}$ &    $8.26_{-0.08}^{+0.13}$ &    $0.72_{-0.12}^{+0.18}$ \\
32576  &    $8180\pm240$ &   $1.76\pm0.10$ &   $1.10\pm0.14$ &    $1.83_{-0.20}^{+0.13}$ &    $4.21_{-0.07}^{+0.05}$ &    $8.74_{-0.39}^{+0.28}$ &    $0.40_{-0.24}^{+0.34}$ \\
38823  &    $7750\pm360$ &   $1.54\pm0.23$ &   $0.88\pm0.10$ &    $1.65_{-0.20}^{+0.12}$ &    $4.28_{-0.13}^{+0.07}$ &    $8.58_{-0.76}^{+0.56}$ &    $0.21_{-0.17}^{+0.52}$ \\
40312  &   $10220\pm360$ &   $4.68\pm0.34$ &   $2.33\pm0.02$ &    $3.24_{-0.25}^{+0.23}$ &    $3.61_{-0.07}^{+0.08}$ &    $8.46_{-0.10}^{+0.02}$ &    $1.00_{-0.06}^{+0.01}$ \\
49976  &    $9460\pm380$ &   $1.95\pm0.20$ &   $1.44\pm0.06$ &    $2.22_{-0.25}^{+0.15}$ &    $4.20_{-0.10}^{+0.08}$ &    $8.48_{-0.62}^{+0.32}$ &    $0.38_{-0.29}^{+0.35}$ \\
54118  &   $10910\pm350$ &   $2.61\pm0.24$ &   $1.94\pm0.06$ &    $2.89_{-0.29}^{+0.20}$ &             $4.07\pm0.09$ &    $8.36_{-0.22}^{+0.17}$ &    $0.60_{-0.22}^{+0.23}$ \\
56022  &    $9660\pm240$ &   $2.00\pm0.10$ &   $1.50\pm0.02$ &    $2.29_{-0.24}^{+0.15}$ &    $4.20_{-0.07}^{+0.06}$ &    $8.45_{-0.43}^{+0.29}$ &    $0.39_{-0.24}^{+0.32}$ \\
62140  &    $7740\pm460$ &   $2.32\pm0.33$ &   $1.24\pm0.08$ &    $1.92_{-0.25}^{+0.14}$ &             $3.99\pm0.13$ &             $8.95\pm0.19$ &             $0.74\pm0.24$ \\
64486  &   $10490\pm460$ &   $2.61\pm0.28$ &   $1.88\pm0.05$ &    $2.77_{-0.28}^{+0.19}$ &    $4.05_{-0.10}^{+0.09}$ &    $8.43_{-0.23}^{+0.16}$ &    $0.63_{-0.24}^{+0.22}$ \\
65339  &    $8520\pm320$ &   $2.43\pm0.23$ &   $1.45\pm0.06$ &    $2.16_{-0.22}^{+0.15}$ &             $4.00\pm0.09$ &    $8.79_{-0.15}^{+0.13}$ &             $0.71\pm0.19$ \\
72968  &    $9310\pm490$ &   $2.32\pm0.26$ &   $1.56\pm0.05$ &    $2.32_{-0.25}^{+0.16}$ &             $4.07\pm0.11$ &    $8.63_{-0.28}^{+0.18}$ &    $0.60_{-0.27}^{+0.25}$ \\
74067  &   $10190\pm380$ &   $2.44\pm0.25$ &   $1.76\pm0.07$ &    $2.62_{-0.28}^{+0.18}$ &             $4.08\pm0.09$ &    $8.47_{-0.26}^{+0.18}$ &             $0.58\pm0.25$ \\
83368  &    $7660\pm220$ &   $2.00\pm0.15$ &   $1.09\pm0.04$ &    $1.78_{-0.17}^{+0.12}$ &    $4.09_{-0.08}^{+0.07}$ &    $8.97_{-0.20}^{+0.15}$ &    $0.62_{-0.21}^{+0.28}$ \\
96616  &    $9180\pm400$ &   $2.85\pm0.28$ &   $1.72\pm0.04$ &    $2.48_{-0.24}^{+0.16}$ &    $3.93_{-0.10}^{+0.09}$ &             $8.66\pm0.12$ &             $0.78\pm0.17$ \\
103192 &   $11750\pm380$ &   $3.86\pm0.38$ &   $2.41\pm0.07$ &    $3.65_{-0.53}^{+0.08}$ &    $3.83_{-0.11}^{+0.06}$ &    $8.25_{-0.03}^{+0.13}$ &    $0.89_{-0.05}^{+0.09}$ \\
108662 &    $9740\pm460$ &   $2.11\pm0.21$ &   $1.56\pm0.04$ &    $2.36_{-0.25}^{+0.16}$ &             $4.16\pm0.10$ &    $8.49_{-0.48}^{+0.27}$ &    $0.46_{-0.29}^{+0.32}$ \\
108945 &    $8670\pm230$ &   $2.64\pm0.16$ &   $1.55\pm0.03$ &    $2.27_{-0.22}^{+0.14}$ &    $3.95_{-0.07}^{+0.06}$ &             $8.76\pm0.10$ &    $0.76_{-0.14}^{+0.17}$ \\
109026 &   $11800\pm900$ &   $2.68\pm0.82$ &   $2.10\pm0.21$ &    $3.19_{-0.40}^{+0.29}$ &    $4.08_{-0.22}^{+0.20}$ &    $8.22_{-0.72}^{+0.21}$ &    $0.56_{-0.46}^{+0.31}$ \\
112185 &    $9350\pm330$ &   $3.99\pm0.28$ &   $2.04\pm0.02$ &    $2.94_{-0.41}^{+0.15}$ &    $3.70_{-0.10}^{+0.07}$ &    $8.54_{-0.05}^{+0.12}$ &    $0.94_{-0.08}^{+0.07}$ \\
112413 &   $11320\pm600$ &   $2.52\pm0.31$ &   $1.97\pm0.07$ &    $2.97_{-0.32}^{+0.22}$ &             $4.11\pm0.11$ &    $8.28_{-0.43}^{+0.22}$ &    $0.53_{-0.31}^{+0.28}$ \\
117025 &    $8590\pm470$ &   $2.12\pm0.25$ &   $1.34\pm0.04$ &    $2.06_{-0.21}^{+0.14}$ &             $4.10\pm0.11$ &    $8.76_{-0.32}^{+0.19}$ &    $0.58_{-0.29}^{+0.27}$ \\
118022 &    $9440\pm270$ &   $1.96\pm0.11$ &   $1.44\pm0.02$ &    $2.22_{-0.24}^{+0.14}$ &    $4.20_{-0.07}^{+0.06}$ &    $8.50_{-0.45}^{+0.30}$ &    $0.39_{-0.25}^{+0.32}$ \\
119213 &    $8620\pm470$ &   $1.99\pm0.23$ &   $1.29\pm0.04$ &    $2.02_{-0.22}^{+0.14}$ &             $4.14\pm0.11$ &    $8.73_{-0.43}^{+0.23}$ &             $0.51\pm0.31$ \\
120198 &    $9740\pm560$ &   $2.14\pm0.26$ &   $1.57\pm0.04$ &    $2.37_{-0.25}^{+0.17}$ &    $4.15_{-0.12}^{+0.11}$ &    $8.50_{-0.54}^{+0.25}$ &    $0.47_{-0.33}^{+0.31}$ \\
124224 &   $12260\pm480$ &   $2.05\pm0.19$ &   $1.93\pm0.04$ &    $3.05_{-0.33}^{+0.14}$ &    $4.30_{-0.10}^{+0.05}$ &    $7.63_{-0.51}^{+0.69}$ &    $0.13_{-0.09}^{+0.42}$ \\
128898 &    $7760\pm230$ &   $1.85\pm0.11$ &   $1.05\pm0.01$ &    $1.76_{-0.17}^{+0.12}$ &    $4.15_{-0.07}^{+0.06}$ &    $8.92_{-0.30}^{+0.20}$ &    $0.53_{-0.25}^{+0.33}$ \\
130559 &    $9120\pm420$ &   $2.30\pm0.25$ &   $1.52\pm0.05$ &    $2.26_{-0.23}^{+0.16}$ &             $4.07\pm0.10$ &    $8.67_{-0.26}^{+0.17}$ &    $0.62_{-0.26}^{+0.25}$ \\
137909 &    $7540\pm420$ &   $3.15\pm0.37$ &   $1.46\pm0.04$ &    $2.14_{-0.35}^{+0.13}$ &    $3.77_{-0.14}^{+0.11}$ &    $8.91_{-0.07}^{+0.16}$ &    $0.92_{-0.13}^{+0.10}$ \\
137949 &    $7370\pm270$ &   $2.16\pm0.24$ &   $1.09\pm0.07$ &    $1.77_{-0.28}^{+0.13}$ &    $4.02_{-0.12}^{+0.08}$ &    $9.04_{-0.15}^{+0.25}$ &    $0.72_{-0.19}^{+0.34}$ \\
140160 &    $9030\pm230$ &   $2.17\pm0.15$ &   $1.45\pm0.04$ &    $2.19_{-0.23}^{+0.15}$ &    $4.11_{-0.08}^{+0.06}$ &    $8.67_{-0.22}^{+0.19}$ &    $0.56_{-0.21}^{+0.26}$ \\
140728 &    $9730\pm550$ &   $2.44\pm0.30$ &   $1.69\pm0.03$ &    $2.49_{-0.25}^{+0.17}$ &    $4.06_{-0.12}^{+0.10}$ &    $8.56_{-0.27}^{+0.17}$ &    $0.62_{-0.26}^{+0.25}$ \\
148112 &    $9170\pm390$ &   $3.34\pm0.33$ &   $1.85\pm0.06$ &    $2.66_{-0.41}^{+0.16}$ &    $3.81_{-0.11}^{+0.09}$ &    $8.63_{-0.08}^{+0.17}$ &    $0.87_{-0.13}^{+0.14}$ \\
148898 &    $8150\pm250$ &   $2.97\pm0.19$ &   $1.54\pm0.02$ &    $2.25_{-0.35}^{+0.13}$ &    $3.85_{-0.10}^{+0.06}$ &    $8.82_{-0.07}^{+0.18}$ &    $0.86_{-0.11}^{+0.15}$ \\
151199 &    $8620\pm390$ &   $2.06\pm0.20$ &   $1.32\pm0.04$ &    $2.04_{-0.21}^{+0.14}$ &             $4.12\pm0.10$ &    $8.74_{-0.34}^{+0.21}$ &             $0.55\pm0.28$ \\
152107 &    $8840\pm190$ &   $2.30\pm0.11$ &   $1.47\pm0.03$ &    $2.19_{-0.22}^{+0.14}$ &    $4.06_{-0.06}^{+0.05}$ &             $8.72\pm0.16$ &    $0.64_{-0.18}^{+0.22}$ \\
170000 &   $11630\pm110$ &   $3.73\pm0.14$ &   $2.36\pm0.04$ &    $3.56_{-0.49}^{+0.04}$ &    $3.85_{-0.07}^{+0.03}$ &    $8.27_{-0.01}^{+0.12}$ &    $0.88_{-0.03}^{+0.09}$ \\
176232 &    $7440\pm370$ &   $2.69\pm0.30$ &   $1.30\pm0.04$ &    $1.96_{-0.40}^{+0.12}$ &    $3.87_{-0.12}^{+0.09}$ &    $8.99_{-0.09}^{+0.24}$ &    $0.86_{-0.15}^{+0.21}$ \\
187474 &    $9900\pm400$ &   $2.56\pm0.35$ &   $1.76\pm0.09$ &    $2.58_{-0.27}^{+0.20}$ &    $4.03_{-0.11}^{+0.10}$ &    $8.53_{-0.21}^{+0.15}$ &    $0.65_{-0.24}^{+0.23}$ \\
188041 &    $8520\pm340$ &   $2.32\pm0.19$ &   $1.41\pm0.04$ &    $2.11_{-0.22}^{+0.14}$ &             $4.03\pm0.08$ &    $8.79_{-0.18}^{+0.15}$ &             $0.67\pm0.21$ \\
201601 &    $7570\pm170$ &   $2.07\pm0.12$ &   $1.10\pm0.03$ &    $1.78_{-0.17}^{+0.12}$ &    $4.06_{-0.07}^{+0.06}$ &    $9.00_{-0.15}^{+0.13}$ &    $0.67_{-0.18}^{+0.25}$ \\
203006 &    $9240\pm480$ &   $2.35\pm0.34$ &   $1.56\pm0.09$ &    $2.32_{-0.26}^{+0.18}$ &             $4.06\pm0.12$ &    $8.64_{-0.28}^{+0.17}$ &    $0.62_{-0.28}^{+0.25}$ \\
217522 &    $6520\pm280$ &   $1.92\pm0.28$ &   $0.78\pm0.09$ &    $1.48_{-0.27}^{+0.13}$ &    $4.04_{-0.16}^{+0.10}$ &    $9.32_{-0.17}^{+0.30}$ &    $0.86_{-0.28}^{+0.49}$ \\
217831 &   $6900\pm1350$ &   $2.58\pm0.31$ &   $1.13\pm0.14$ &    $1.79_{-0.31}^{+0.12}$ &    $3.87_{-0.14}^{+0.12}$ &    $9.11_{-0.11}^{+0.22}$ &    $0.88_{-0.18}^{+0.23}$ \\
220825 &    $9180\pm370$ &   $1.73\pm0.14$ &   $1.28\pm0.03$ &    $2.06_{-0.22}^{+0.11}$ &    $4.27_{-0.09}^{+0.07}$ &    $8.36_{-0.80}^{+0.48}$ &    $0.23_{-0.20}^{+0.40}$ \\
221760 &    $8470\pm230$ &   $3.62\pm0.27$ &   $1.78\pm0.05$ &    $2.55_{-0.39}^{+0.15}$ &    $3.73_{-0.10}^{+0.07}$ &    $8.71_{-0.05}^{+0.14}$ &             $0.94\pm0.09$ \\
223640 &   $12240\pm440$ &   $2.30\pm0.22$ &   $2.03\pm0.05$ &    $3.17_{-0.34}^{+0.20}$ &    $4.21_{-0.10}^{+0.07}$ &    $7.98_{-0.74}^{+0.37}$ &    $0.32_{-0.26}^{+0.35}$ \\
\end{longtable}
\end{center}
\renewcommand{\arraystretch}{1.0}
\twocolumn
}

The modelling of H$\gamma$ and H$\beta$ carried out in Sect. \ref{sect:balmer_fit} is, to some extent, 
degenerate in $\log{g}$ and $T_{\rm eff}$: if the value of $\log{g}$ derived by fitting the 
Balmer lines is found to be inaccurate, $T_{\rm eff}$ may also be inaccurate despite having obtained a 
high-quality fit. In order to assess the accuracy of the spectroscopically-derived $\log{g}$ values 
(referred to as the spectroscopic $\log{g}$) and therefore, of the derived $T_{\rm eff}$ values, we 
compare with the $\log{g}$ values derived from the masses and radii associated with each star's 
position on the theoretical HRD (referred to as the HRD $\log{g}$). We derived spectroscopic surface 
gravities for 42 of the confirmed mCP stars in the sample and compare these with the HRD $\log{g}$ 
values; the comparison is shown in Fig. \ref{fig:lgg_comp}.

Thirty-seven out of the fourty-two (88~per~cent) spectroscopic $\log{g}$ values are in agreement with 
the HRD surface gravities within their uncertainties. The discrepancies between the five stars whose 
spectroscopic and HRD $\log{g}$ values are not in agreement range from 7 to 21~per~cent. The overall 
agreement that was yielded by the analysis is to be expected considering the large uncertainties in the 
spectroscopic $\log{g}$ values. This is particularly true of the cooler stars having 
$T_{\rm eff}\lesssim10\,000\,{\rm K}$, which exhibit median $\log{g}$ uncertainties $\gtrsim1$ -- more 
than twice that of the hotter stars ($T_{\rm eff}>10\,000{\rm K}$) for which the median uncertainty is 
$0.5$. The increase in spectroscopic $\log{g}$ uncertainties with decreasing $T_{\rm eff}$ reflects a 
decrease in the sensitivity of the Balmer line profiles to changes in $\log{g}$.

It is apparent from Fig. \ref{fig:lgg_comp} that the spectroscopic $\log{g}$ values are systematically 
smaller than the HRD $\log{g}$ values. There are a number of possible explanations for this 
discrepancy. It could be related to the choice of evolutionary model grids: if the stars exhibit a 
metallicity that is systematically higher or lower than the adopted solar value, the HRD $\log{g}$ will 
be affected. Furthermore, the HRD $\log{g}$ values are associated with non-rotating models -- 
it is possible that including the rotational velocities of each of the sample mCP stars may serve to 
reduce this discrepancy. The Lorentz force that exists within the strongly magnetic atmospheres of mCP 
stars is predicted to modify the pressure stratification \citep{Valyavin2004}. This effect is manifest 
in the profiles of H$\alpha$, H$\beta$, and H$\gamma$ lines, which may vary over the star's rotational 
period for those stars exhibiting field structures that are not symmetric about the star's rotational 
axis (e.g. stars hosting dipole fields with non-zero obliquity angles).

Given that the sample consists only of relatively bright stars, effective temperatures of the majority 
of the confirmed mCP stars have been derived in previously published studies. In particular, both 
\citet{Kochukhov2006} (henceforth referred to as KB2006) and \citet{Hubrig2007} (henceforth referred to 
as HNS2007) derived $T_{\rm eff}$ values for 87~per~cent and 58~per~cent of the stars in this sample, 
respectively. We expect the effective temperatures yielded by these studies to be consistent with those 
derived here considering that all three studies attempt to take into account the presence of chemical 
peculiarities.

\begin{figure}
	\centering
	\includegraphics[width=1.0\columnwidth]{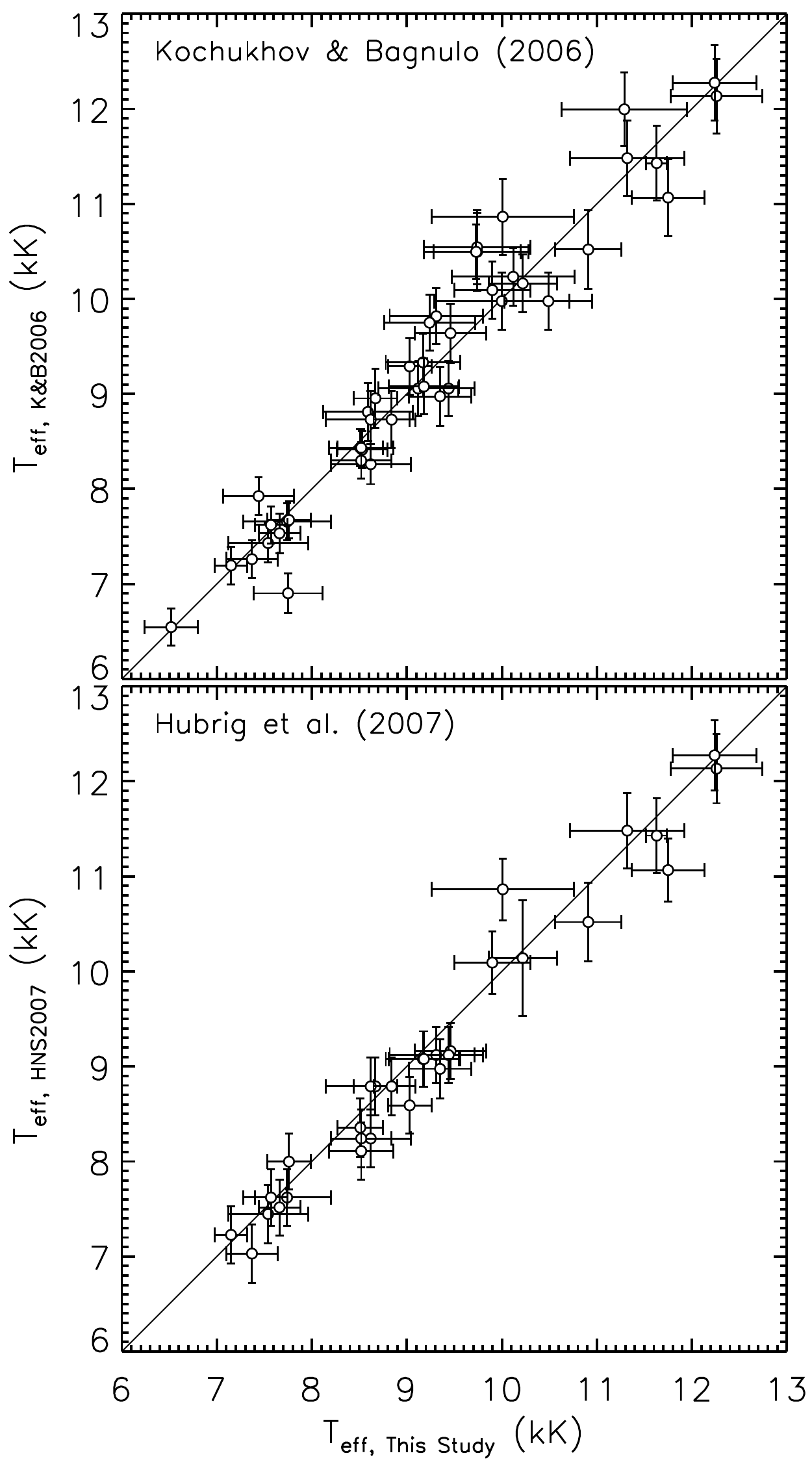}
	\caption{Comparison of $T_{\rm eff}$ values derived in this study with those derived by 
	\citet{Kochukhov2006} (top) and \citet{Hubrig2007} (bottom).}
	\label{fig:Teff_comp}
\end{figure}

In Fig. \ref{fig:Teff_comp}, we compare the values of $T_{\rm eff}$ derived in this study with those 
derived by KB2006 and HNS2007. We find that 87~per~cent and 97~per~cent of the effective temperatures 
derived for the mCP stars included in both this study and those published by KB2006 and HNS2007, 
respectively, are consistent within their associated uncertainties. HD~38823 exhibits the largest 
discrepancy ($12\pm5$~per~cent) with derived $T_{\rm eff}$ values of $7\,750\pm360\,{\rm K}$ (this 
study) and $6\,900\pm210\,{\rm K}$ (KB2006); we find that a lower $T_{\rm eff}$ of 
$7\,350\pm800\,{\rm K}$ is derived when a solar metallicity is adopted and no chemical peculiarities or 
magnetic fields are considered (see Section \ref{sect:mag_sed}).

\subsection{Mass distribution}\label{sect:mass_dist}

\begin{figure}
	\centering
	\includegraphics[width=1.0\columnwidth]{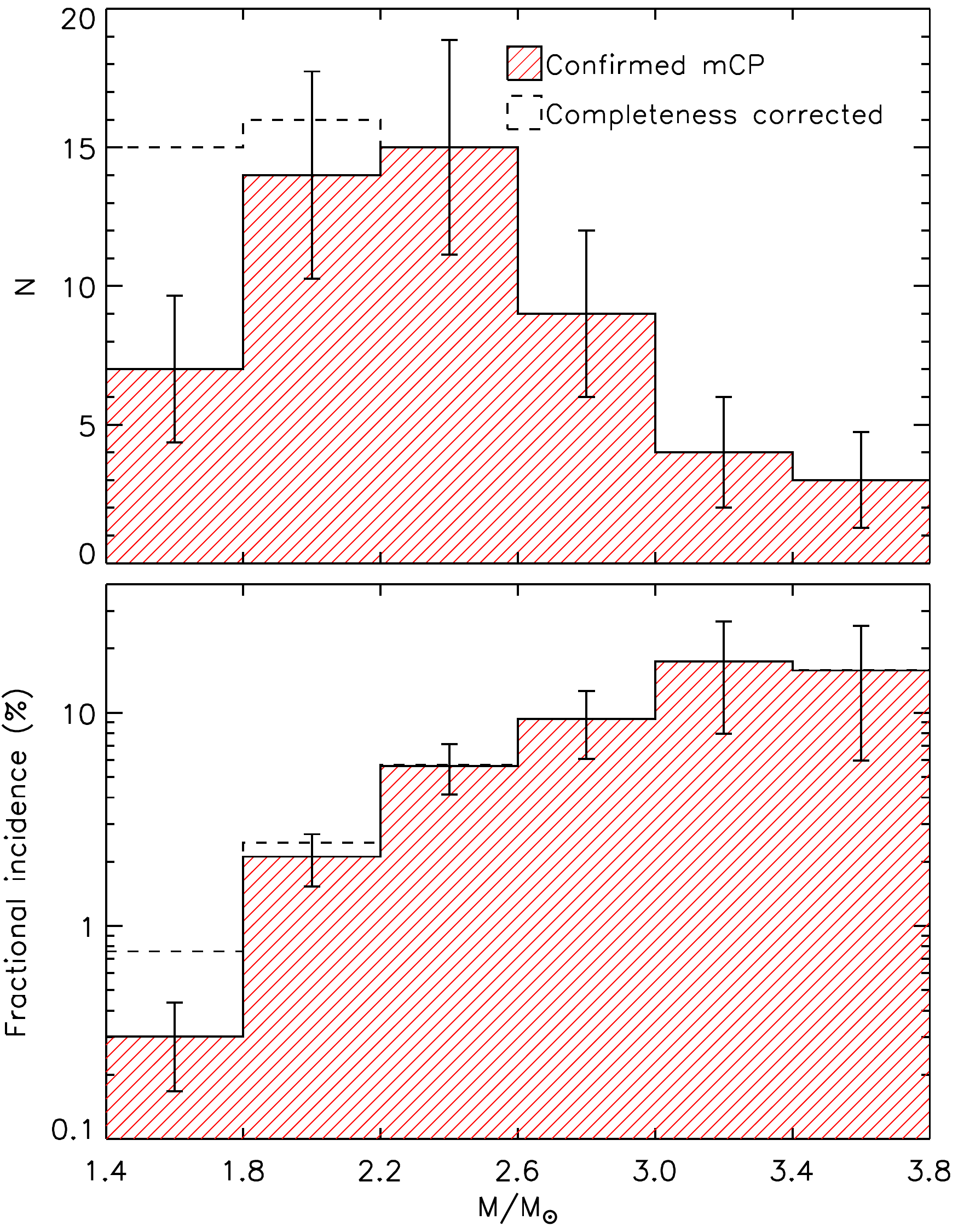}
	\caption{\emph{Top:} Distribution of masses of the mCP stars in the sample (red hatched). 
	Correcting for completeness yields the black dashed distribution.
	\emph{Bottom:} Incidence rate of mCP stars with respect to stellar mass. The large uncertainty 
	associated with the highest mass bin ($5<M/M_\odot<5.4$) is due to the fact that it contains only 
	three stars. Correcting for completeness increases the incidence rates of the first two bins, 
	as indicated by the black dashed distribution.}
	\label{fig:mag_mass}
\end{figure}

Based on the comparisons with the evolutionary tracks carried out in Sect. \ref{sect:HRD}, we find 
that the sample's 52 confirmed mCP stars span a range of masses from $1.48$ to $3.65\,M_\odot$. The 
mass distribution is shown in Fig. \ref{fig:mag_mass} (top). Our volume-limited sample is affected by 
two sources of incompleteness: (1) the completeness of the identified mCP stars is 
$\lesssim80$~per~cent for those stars having $M\lesssim1.5\,M_\odot$ and is $\gtrsim97$~per~cent for 
stars with $M\gtrsim1.9\,M_\odot$ (Fig. \ref{fig:comp_dm}); (2) the completeness of the volume-limited 
sample of stars extracted from the Hipparcos sample increases with mass from $\sim87$~per~cent 
to $\sim100$~per~cent. The mass distribution can be approximately corrected for completeness if the 
actual incidence rate of mCP stars within individual mass bins is known. We estimated the actual 
incidence rate by taking the ratio of the number of confirmed mCP stars within each mass bin to the 
bin's total number of stars (both mCP and non-mCP) which have been observationally confirmed or 
rejected as mCP members (see Sect. \ref{sect:complete}). For instance, in the smallest mass bin 
($1.4<M/M_\odot<1.8\,M_\odot$) there are $1\,063$ stars with observational constraints on their mCP 
status, $968$ stars without such constraints, and $\approx294$ stars that may be missing 
from the sample due to the incompleteness of the Hipparcos catalogue. Seven confirmed mCP stars are 
found in this bin implying an estimated mCP incidence rate of $7/1\,063=0.66\times10^{-2}$; therefore, 
we can expect approximately 8 ($[968+294]\times0.66\times10^{-2}$) additional mCP stars to be found 
within the subsample of $968+294=1\,262$ stars that have not been confirmed or rejected as mCP members 
within this mass bin. Applying this estimation to the mass distribution shown in Fig. 
\ref{fig:mag_mass} yields an increase of 8 and 2 stars to the two smallest mass bins; the higher mass 
bins are unaffected.

The incidence rate of mCP stars as a function of mass can be derived using the masses of both the mCP 
and non-mCP stars within the sample; this is shown in Fig. \ref{fig:mag_mass} (bottom). We find that the 
fractional incidence increases monotonically from $0.3\pm0.1$~per~cent at $M<1.8\,M_\odot$ to 
approximately 11~per~cent at $3.4<M/M_\odot<3.8$. As shown in Fig. \ref{fig:mag_mass}, correcting for 
completeness increases the estimated incidence rate in the two smallest mass bins to $0.8$~per~cent 
and $0.3$~per~cent, respectively, while the changes to the higher mass bins are negligible.

As discussed in Fig. \ref{sect:HRD}, the metallicities of the mCP stars cannot be confidently 
determined and, as a result, a relatively large uncertainty is adopted based on the solar 
neighbourhood ($Z=0.014\pm0.009$; see Sect. \ref{sect:HRD}). Accounting for this $\sigma_Z$, we find 
that the uncertainties in mass exhibit median and maximum values of $0.15$ and $0.37\,M_\odot$, 
respectively (4 and 10~per~cent of the sample's mass range of $1.4\leq M/M_\odot\leq5$). Therefore, 
we conclude that the properties of both the mCP mass distribution and the fractional incidence rate 
of mCP stars as a function of mass presented here are likely robust.

\subsection{Evolutionary state}\label{sect:evo_state}

Unlike the distribution of masses presented in Sect. \ref{sect:mass_dist}, the absolute ages 
($t_{\rm age}$) and fractional MS ages ($\tau$) derived using theoretical evolutionary tracks 
vary significantly depending on the adopted metallicity. Furthermore, the uncertainties in 
$t_{\rm age}$ and $\tau$ are significantly larger for those stars positioned closer to the 
TAMS compared to the ZAMS as a result of their more rapid evolution 
\citep[e.g.][]{Kochukhov2006,Landstreet2007}. These properties are reflected in the large 
uncertainties derived in Sect. \ref{sect:HRD}.

\begin{figure}
	\centering
	\includegraphics[width=1.0\columnwidth]{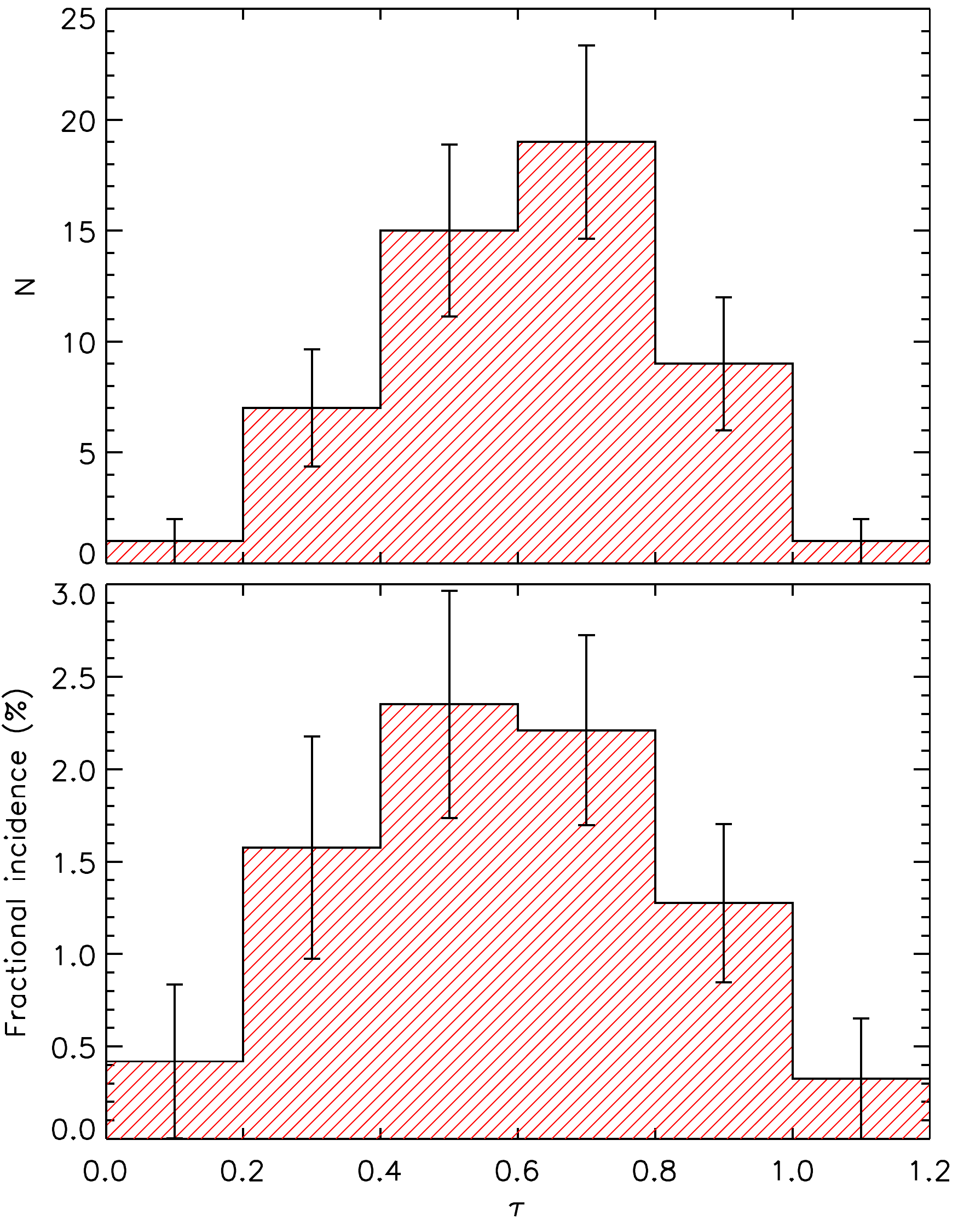}
	\caption{\emph{Top:} Distribution of fractional MS ages of the mCP stars in the sample.
	\emph{Bottom:} Incidence rate of mCP stars with respect to fractional MS age.}
	\label{fig:MS_age_dist}
\end{figure}

In Fig. \ref{fig:MS_age_dist} (top), we show the mCP subsample's distribution of $\tau$. The 
distribution is found to peak near the middle of the MS band at approximately $\tau=0.7$. 
This is consistent with the $\tau$ distributions of mCP stars with $M<3\,M_\odot$ reported by 
KB2006 and HNS2007. These two studies also identified differences in the low-mass ($M<3\,M_\odot$) and 
high-mass ($M\geq3\,M_\odot$) $\tau$ distributions; only 7 mCP stars in our sample have 
$M\geq3\,M_\odot$ and so their $\tau$ distribution cannot be meaningfully compared with these 
results.

For the non-mCP stars, the $\tau$ distribution is characteristically similar to that of the mCP stars.
In Fig. \ref{fig:MS_age_dist} (bottom) we show the fractional incidence of mCP stars within the total 
sample as a function of $\tau$. It is evident that the incidence rate is approximately symmetric about 
$\tau=0.6$ increasing from a rate $\lesssim0.5$~per~cent where $\tau<0.2$ and $\tau>1$ to a peak rate 
of $\approx~2.3$~per~cent. We compared the mCP and non-mCP $\tau$ distributions by computing a 
Kolmogorov-Smirnov (KS) test statistic, which was found to be 0.16. The significance of this result can 
be evaluated by estimating the uncertainty in the KS test statistic. The two most important 
contributions to the uncertainty are likely (1) the relatively small sample size of the mCP sample and 
(2) the large uncertainties in $\tau$ that may be attributed to the fact that each stars' metallicity 
is effectively unknown.

The first of the factors contributing to the uncertainty in the KS test statistic can be evaluated 
using the case resampling bootstrapping method, which yielded a 1$\sigma$ uncertainty of 0.05. We 
evaluated the second factor using a Monte Carlo (MC) simulation. This involved generating $1\,000$ 
mCP and non-mCP $\tau$ distributions in which, for each star, we add on a random value, $\Delta\tau$, 
that is sampled from the associated $\tau$ uncertainty (i.e. $\tau\to\tau+\Delta\tau$ where 
$\Delta\tau$ is randomly drawn from a Gaussian distribution defined by the most probable value and 
the upper and lower error limits). We then computed the KS test statistic comparing each of the pairs 
of simulated mCP and non-mCP $\tau$ distributions. The width of the resulting distribution of test 
statistics was then interpreted as the approximate uncertainty introduced by the large $\tau$ 
uncertainties associated with each star; the MC analysis yielded a 1$\sigma$ uncertainty of 0.07. 
Combining the two uncertainties estimated from the bootstrapping and MC analyses yields a 3$\sigma$ 
uncertainty of 0.24. Comparing with the derived KS statistic of 0.16 suggests that the apparent 
differences between the mCP and non-mCP $\tau$ distributions is likely not statistically significant.

\subsection{Chemical abundances}\label{sect:chem_abund}

\begin{figure}
	\centering
	\includegraphics[width=1.0\columnwidth]{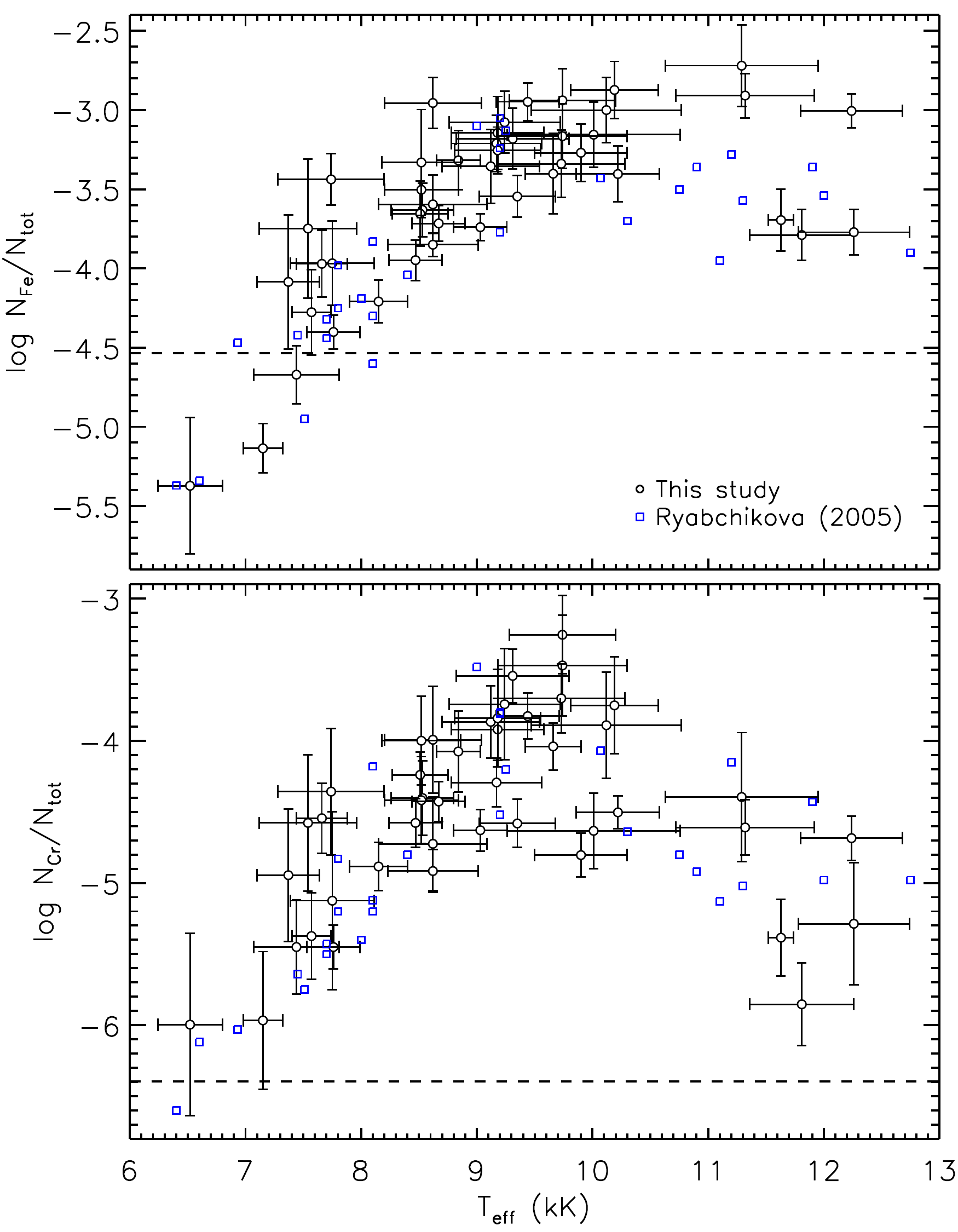}
	\caption{Mean Fe (top) and Cr (bottom) abundances as a function $T_{\rm eff}$ derived in this 
	study (open black circles) and derived by \citet{Ryabchikova2005a} (open blue squares). The 
	horizontal dashed lines correspond to the Fe and Cr solar abundances of $-4.54$ and 
	$-6.40\,{\rm dex}$, respectively \citep{Asplund2009}.}
	\label{fig:abund_Teff}
\end{figure}

In Fig. \ref{fig:abund_Teff}, we show mean Fe (top) and Cr (bottom) abundances as a function of 
$T_{\rm eff}$ derived for 45 of the mCP stars. These are compared with the mean abundances and 
effective temperatures of those 104 stars included in the study published by \citet{Ryabchikova2005a} 
but not included in our volume-limited sample. In general, both data sets appear to be in 
agreement in terms of the overall trends: $\log{N_{\rm Fe}/N_{\rm tot}}$ and 
$\log{N_{\rm Cr}/N_{\rm tot}}$ tend to increase with $T_{\rm eff}$ and reach a peak value at 
approximately $9\,500\,{\rm K}$. We also compared how the mean Si and Ca abundances derived in 
both studies vary with $T_{\rm eff}$ and find similar agreement: a slight increase in the Si 
abundances with increasing $T_{\rm eff}$ is found while the Ca abundances do not appear strongly 
correlated with $T_{\rm eff}$. The overall agreement between the Fe, Cr, Si, and Ca abundances 
derived in this study and those appearing in \citet{Ryabchikova2005a} suggests that the final 
atmospheric parameters (e.g. $T_{\rm eff}$, $\log{g}$, and abundances) adopted in our sample are 
reasonably accurate. It is noted that the abundances compiled in their study typically include the 
effects of magnetic fields \citep[e.g.][]{Ryabchikova2004}.

\begin{figure*}
	\centering
	\includegraphics[width=1.9\columnwidth]{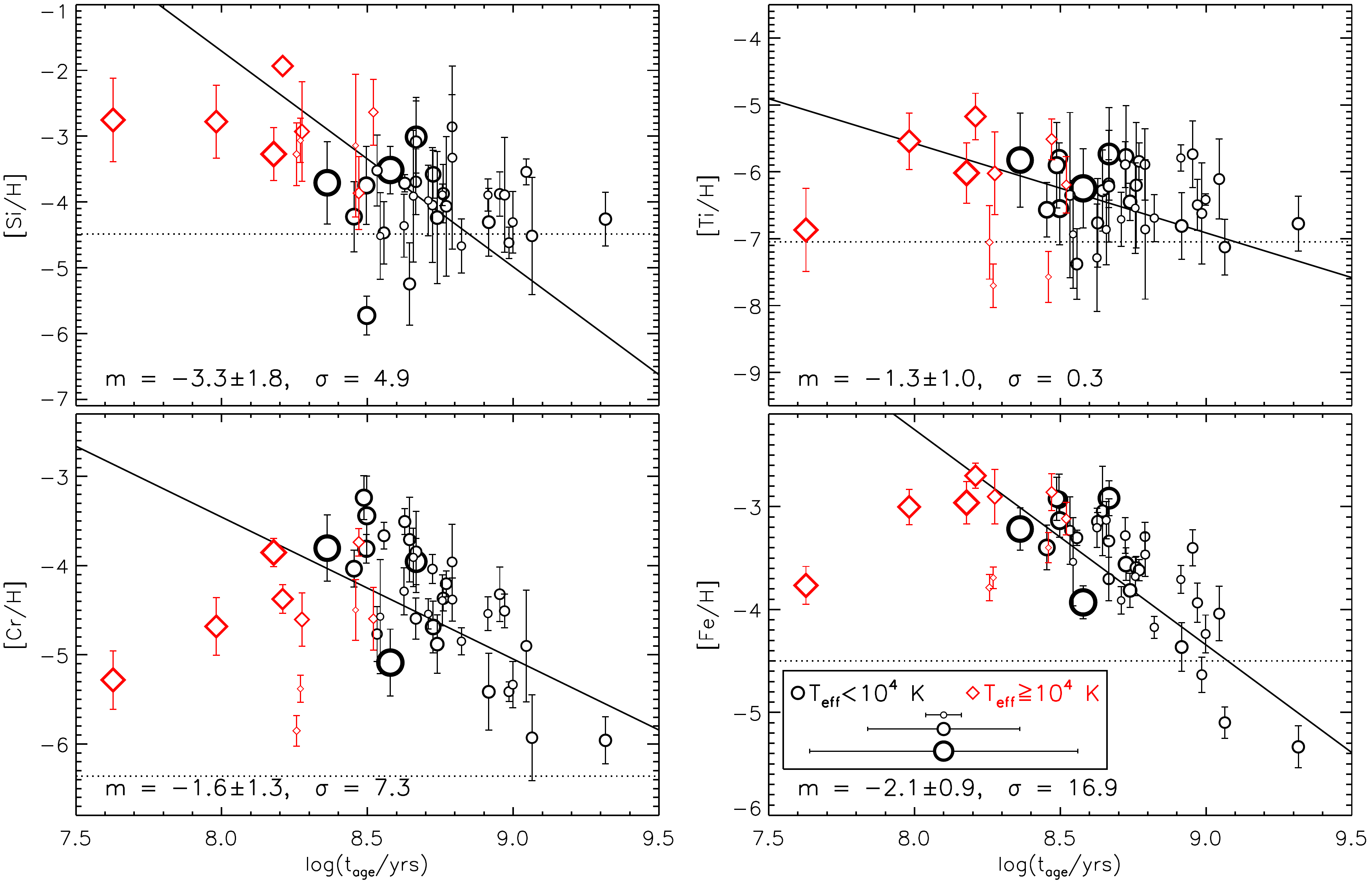}
	\caption{Mean Si (top left), Ti (top right), Cr (bottom left), and Fe (bottom right) abundances 
	as a function of absolute stellar age. The sizes of each data point correspond to the widths of the 
	mean uncertainty in $t_{\rm age}$ (i.e. half the full width of the error interval); the relation 
	between point size and error bar width is shown in the bottom right panel. The black circles 
	correspond to the cooler stars having $T_{\rm eff}<10^4\,{\rm K}$ and the red diamonds correspond to 
	the hotter stars having $T_{\rm eff}\geq10^4\,{\rm K}$. The solid lines correspond to the 
	derived best-fitting linear functions (yielding a slope and fit significance of $m$ and $\sigma$). 
	The horizontal dotted lines indicate the solar abundances \citep{Asplund2009} of each element.}
	\label{fig:abund_age}
\end{figure*}

We were also able to identify correlations between certain chemical abundances and absolute stellar 
age. In Fig. \ref{fig:abund_age}, we show the abundances of Si, Ti, Cr, and Fe for 45 of the mCP stars 
as a function of $\log{t_{\rm age}}$. No $t_{\rm age}$ error bars are plotted because of the relatively 
high density of data points and their large error bars which, in some cases, span nearly 1/3 of the 
$x$-axis. A best-fitting linear function was calculated for each element using the fitting routine 
described by \citet{Williams2010}, which considers errors in both $x$ and $y$ coordinates. Symmetric 
$\log t_{\rm age}$ errors were input into the fitting routine by taking the average of the positive and 
negative error intervals. The statistical significance of each fit was calculated by comparing the 
$\chi^2$ values obtained from the first order linear fit with that of a zeroth order fit. The 
slope ($m$) and significance ($\sigma$) of each fit is shown in Fig. \ref{fig:abund_age}.

The linear fits shown in Fig. \ref{fig:abund_age} were found to have slopes ranging from $m=-3.3$ 
to $-1.3$ indicating overall decreases in the mean abundances with age. Although the significance of 
the apparent [Ti/H] trend is relatively low at $0.3\,\sigma$, [Si/H], [Cr/H], and [Fe/H] all exhibit 
statistically significant trends ($\sigma\geq4.9$). All of the other abundances that were studied -- 
including O, Mg, and S -- yielded significance levels $<3\,\sigma$.

Similar decreases in the mean abundances with age were previously discovered by \citet{Bailey2014} 
within a sample of cluster Bp stars. The significance of the [Si/H], [Cr/H], and [Fe/H] trends they 
reported are lower than those found in this study while they report higher significance levels 
associated with O, Mg, and Ti. Comparing the slopes of the [Si/H], [Cr/H], and [Fe/H] trends, we 
find that those derived in this study are significantly steeper; only the linear fits to [Cr/H] 
derived in both studies exhibit slopes which are in agreement within error. By identifying the stars 
in our sample having $T_{\rm eff}\geq10^4\,{\rm K}$ \citep[i.e. the minimum $T_{\rm eff}$ of the stars 
studied by][]{Bailey2014} it is clear that the trends detected in our study are dominated by comparably 
cooler stars. Furthermore, the hotter stars appear to exhibit mean Cr, Fe, and Ti abundances which 
deviate from the best-fitting linear trends towards lower abundances. This is confirmed by comparing 
reduced $\chi^2$ values associated with the hotter ($\chi^2_{\rm hot}$) and cooler 
($\chi^2_{\rm cool}$) stars: for [Ti/H], [Cr/H], and [Fe/H], $\chi^2_{\rm hot}/\chi^2_{\rm cool}$ is 
found to be 2.1, 6.7, and 5.1, respectively. Considering that the minimum age of the cooler stars in 
our sample corresponds roughly to the maximum age of the hotter stars, it is unclear whether or not 
the observed deviation in these particular abundances is unique to the hotter stars. A more probable 
explanation is that the deviations arise from vertical and/or horizontal stratification, which is 
known to significantly affect both Fe and Cr spectral line profiles 
\citep[e.g.][]{Bagnulo2001,Wade2001a,Kochukhov2006a}.

\section{Discussion and conclusions}\label{sect:discussion}

The results presented here constitute the first detailed, analytically homogeneous study of the 
fundamental parameters and chemical abundances of a volume-limited sample of mCP stars. Such a study 
provides two important and unique contributions to our understanding of mCP stars: (1) it allows for 
basic conclusions to be drawn while minimizing the observer biases inherent in most previously 
published surveys; and (2) it comprises of a population of non-mCP stars which can be compared against 
the sample's mCP stars. By adopting a distance limit of $100\,{\rm pc}$, we are restricted to a stellar 
population with relatively precise Hipparcos parallax measurements and generally extensive photometric 
measurements.

The most significant findings presented here are summarized as follows.
\begin{enumerate}
	\item The incidence rate of mCP stars amongst the population of non-mCP stars is found to 
	sharply increase with mass from $\approx0.3$~per~cent at $M\approx1.6\,M_\odot$ to 
	$\approx11$ per cent at $3\lesssim M\lesssim5\,M_\odot$. It is noted that this trend is not 
	strongly influenced by sample incompleteness or the uncertainty in metallicity.
	\item The mCP stars are more frequently found within the middle of the MS band 
	($0.4<\tau<0.8$) as opposed to near the ZAMS or the TAMS. This result confirms the 
	findings of \citet{Kochukhov2006} and \citet{Hubrig2007}, whose studies both used stellar ages 
	derived from the older, less-sophisticated evolutionary models calculated by \citet{Schaller1992} 
	and \citet{Schaerer1993} compared to the newer models calculated by \citet{Ekstrom2012} and 
	\citet{Mowlavi2012} that are used in this study. Our conclusion is supported by the fact that both 
	the mCP stars' fractional MS ages and their incidence rates with respect to the non-mCP subsample 
	exhibit similar peaks near $\tau\approx0.7$. However, we note that the relatively small mCP sample 
	size and the large errors in $\tau$ suggests that the apparent differences to the distributions of 
	mCP and non-mCP $\tau$ values may not be statistically significant.
	\item The mean Cr and Fe abundances appear to decrease with increasing absolute stellar age. 
	Similar trends are found from mean Si and Ti abundances albeit with lower statistical significance 
	levels. The overall trends are consistent with those reported by \citet{Bailey2014} based on their 
	sample of Bp stars ($T_{\rm eff}\geq10^4\,{\rm K}$); however, the estimated abundance decay rates 
	of Ti, Cr, and Fe are greater. This discrepancy between the derived abundance decay rates may be 
	related to the fact that our sample is predominantly populated by cooler stars: lower abundance 
	decay rates of Cr and Fe associated with the cooler ($T_{\rm eff}<10^4\,{\rm K}$) stars were 
	found compared to those of the hotter ($T_{\rm eff}\geq10^4\,{\rm K}$) stars in our sample.
\end{enumerate}

This paper's submission date approximately coincides with the second Gaia data release (GDR2) 
\citep{Gaia2016}. GDR2 contains parallax measurements for 49/52 of the mCP stars in our sample. 
Comparing the reported parallax uncertainties associated with the measurements obtained using 
Hipparocs and Gaia, we find that the formal uncertainties of the Gaia measurements are typically 
$\approx50$~per~cent of the Hipparcos values; however, we find that for 9 out of the 49 stars, the 
uncertainties associated with the Gaia measurements are larger than those associated with the Hipparcos 
measurements. Moreover, the two sets of parallax measurements are only in agreement within the reported 
uncertainties for 22/49 stars \citep[this may not necessarily imply that the distances are also not in 
agreeement,][]{BailerJones2018}. We note that the majority of the mCP stars in our sample (42/52) are 
brighter than Gaia's optimal magnitude limit of $G>6\,{\rm mag}$. These factors have motivated our 
decision to not replace the Hipparcos parallax measurements used in the preceding analysis with the 
GDR2 parallax measurements.

When deriving the masses and ages of both the mCP and non-mCP stars, we adopted a solar metallicity 
\citep{Asplund2009} with an uncertainty estimated from the star-to-star variations of F-, G-, and 
K-type dwarfs within $100\,{\rm pc}$ \citep[${\rm [M/H]}=0.0\pm0.3$, as inferred from the catalogue 
published by][]{Casagrande2011}. However, considering that these low-mass stars are generally expected 
to be older than the early-F, A-, and late-B-type dwarfs included in this study, there is no reason to 
expect both populations to exhibit similar metallicity distributions. That is, the older stars may 
have formed within dramatically different environments compared to those of the younger stars 
because of (1) the chemical evolution of the solar neighbourhood and (2) the fact that these stars may 
have originated from further away. In Fig. 16 of \citet{Casagrande2011}, the metallicity distributions 
(in terms of [Fe/H]) of nearby low-mass stars are divided into old ($t_{\rm age}\geq5\,{\rm Gyrs}$), 
middle-aged ($1\leq t_{\rm age}\leq5\,{\rm Gyrs}$), and young ($t_{\rm age}<1\,{\rm Gyrs}$) 
populations. It is evident that the dispersions of [Fe/H] associated with these sub-populations 
increase with age suggesting that our adopted 2$\sigma$ [M/H] uncertainty -- estimated using the full 
sample -- may by somewhat conservative. While this does not discount the possibility of systematic 
[M/H] biases in our sample of hotter stars, it does provide further evidence that the populations of 
low- and intermediate-mass stars exhibit characteristically similar metallicities.

Assuming that the Cr and Fe abundance trends are real and not the result of systematic bias, it is 
somewhat surprising that such a correlation was detected considering the typically large 
uncertainties derived for $t_{\rm age}$. This may be an indication that the intermediate-mass MS 
stars within the solar neighbourhood exhibit a lower star-to-star variation in metallicity than was 
adopted in our analysis. Considering that the derived effective temperatures and luminosities are, 
in general, consistent with those reported by \citet{Kochukhov2006} and \citet{Hubrig2007}, such a 
detection may have been facilitated by the use of the modern, high-density evolutionary model 
grids calculated by \citet{Ekstrom2012} and \citet{Mowlavi2012}.

We estimate that the completeness of the subsample of mCP stars within $100\,{\rm pc}$ is largely 
complete ($\gtrsim91$~per~cent). However, our study is not sensitive to those stars which may host 
ultra-weak fields \citep[e.g.][]{Lignieres2009,Petit2011,Blazere2016a} (i.e. no magnetic constraints 
are obtained for the vast majority of the non-mCP stars). \citet{Petit2011a} report that Vega, which is 
the first A-type star discovered to host an ultra-weak field, exhibits a complex magnetic field 
structure \citep[characteristically similar brightness spots were subsequently 
discovered,][]{Petit2017}. The authors note that the observed field complexity is generally 
incompatible with the fossil field theory and that Vega's field -- and perhaps all MS A-type stars 
hosting ultra-weak fields -- may have an origin that is distinct from that of the field's hosted by mCP 
stars.

The ultra-weak fields detected on the Am stars Sirius A \citep{Petit2011}, $\beta$~UMa, and 
$\theta$~Leo \citep{Blazere2016a} all exhibit atypical asymmetric circularly polarized Zeeman 
signatures: the signals have a strong positive lobe and no obvious negative lobe, which differs from 
the generally symmetric signatures associated with mCP stars. This may suggest that 
the fields hosted by these stars are similar to Vega -- both in terms of their structure and their 
origin. On the other hand, the ultra-weak Zeeman signature detected on the surface of the Am star 
Alhena has a symmetric profile that is indistinct from many signatures detected on strongly magnetic 
mCP stars. This suggests that there may exist a larger population of Am stars that host fields produced 
by similar processes to those of the mCP stars. At this point, the connection between the fields 
hosted by Am stars and those hosted by mCP stars is unclear. We note that no Am stars hosting 
fields $\gtrsim10\,{\rm G}$ have been discovered despite repeated searches 
\citep[e.g.][]{Auriere2010}. If the fields hosted by Am and mCP stars share the same origin (e.g. 
fossil fields), it is perplexing why they appear to differ so significantly in terms of their 
strengths.

\section*{Acknowledgments}

GAW acknowledges support in the form of a Discovery Grant from the Natural Science and 
Engineering Research Council (NSERC) of Canada. We are grateful to Dr. Andrew Tkachenko, 
Dr. Vadim Tsymbal, and Dr. Colin Folsom for their assistance using the {\sc llmodels}, {\sc gssp}, 
and {\sc zeeman} LMA codes.

\bibliography{ApBp_survey,ApBp_survey-other}
\bibliographystyle{mn2e}

\clearpage

\renewcommand{\arraystretch}{1.2}
\begin{table*}
	\caption{Derived chemical abundances for the 45/52 mCP stars in the sample with available 
	spectroscopic measurements.}
	\label{tbl:abund_tbl}
	\begin{center}
	\begin{tabular*}{2.0\columnwidth}{@{\extracolsep{\fill}}l c c c r@{\extracolsep{\fill}}}
		\noalign{\vskip-0.2cm}
		\hline
		\hline
		\noalign{\vskip0.5mm}
		HD  & [Si/H] & [Ti/H] & [Cr/H] & [Fe/H] \\
		\noalign{\vskip0.5mm}
		\hline	
		\noalign{\vskip0.5mm}
3980   &            $-3.01\pm0.36$ &            $-5.73\pm0.59$ &            $-3.96\pm0.37$ &            $-2.92\pm0.16$ \\
11502  &            $-3.27\pm0.62$ &            $-6.02\pm0.70$ &            $-3.85\pm0.37$ &            $-2.96\pm0.20$ \\
15089  &            $-3.87\pm0.55$ &            $-6.20\pm0.69$ &            $-4.37\pm0.26$ &            $-3.59\pm0.17$ \\
15144  &            $-4.07\pm0.40$ &            $-5.84\pm0.45$ &            $-4.20\pm0.16$ &            $-3.62\pm0.20$ \\
18296  &            $-2.64\pm0.41$ &            $-6.19\pm0.41$ &            $-4.60\pm0.26$ &            $-3.12\pm0.20$ \\
24712  &            $-4.52\pm0.89$ &            $-7.13\pm0.42$ &            $-5.93\pm0.48$ &            $-5.10\pm0.15$ \\
27309  &            $-1.93\pm0.51$ &            $-5.17\pm0.50$ &            $-4.38\pm0.43$ &            $-2.70\pm0.24$ \\
29305  &            $-3.27\pm0.36$ &            $-7.06\pm0.76$ &            $-5.85\pm0.29$ &            $-3.79\pm0.16$ \\
38823  &            $-3.51\pm0.20$ &            $-6.25\pm0.60$ &            $-5.09\pm0.63$ &            $-3.93\pm0.26$ \\
40312  &            $-3.14\pm0.24$ &            $-7.57\pm0.75$ &            $-4.50\pm0.11$ &            $-3.40\pm0.17$ \\
56022  &            $-4.22\pm0.59$ &            $-6.57\pm0.23$ &            $-4.03\pm0.16$ &            $-3.40\pm0.25$ \\
62140  &            $-3.88\pm0.30$ &            $-5.74\pm0.54$ &            $-4.32\pm0.44$ &            $-3.40\pm0.16$ \\
65339  &            $-3.33\pm0.34$ &            $-6.86\pm0.50$ &            $-4.38\pm0.30$ &            $-3.47\pm0.18$ \\
72968  &            $-3.72\pm0.87$ &            $-6.76\pm0.38$ &            $-3.51\pm0.19$ &            $-3.14\pm0.19$ \\
74067  &            $-3.87\pm1.00$ &            $-5.51\pm0.28$ &            $-3.74\pm0.33$ &            $-2.86\pm0.17$ \\
83368  &            $-3.89\pm0.53$ &            $-6.50\pm0.64$ &            $-4.51\pm0.25$ &            $-3.93\pm0.21$ \\
96616  &            $-3.91\pm0.47$ &            $-6.86\pm0.41$ &            $-3.91\pm0.21$ &            $-3.13\pm0.22$ \\
108662 &                         - &            $-5.90\pm0.09$ &            $-3.24\pm0.26$ &            $-2.92\pm0.18$ \\
108945 &            $-3.89\pm0.44$ &            $-6.54\pm0.94$ &            $-4.39\pm0.14$ &            $-3.68\pm0.11$ \\
112185 &            $-4.52\pm0.12$ &            $-6.93\pm0.35$ &            $-4.57\pm0.16$ &            $-3.54\pm0.12$ \\
112413 &            $-2.93\pm0.24$ &            $-6.02\pm0.20$ &            $-4.60\pm0.19$ &            $-2.90\pm0.14$ \\
118022 &            $-3.75\pm1.06$ &            $-5.79\pm0.28$ &            $-3.81\pm0.15$ &            $-2.93\pm0.10$ \\
119213 &            $-3.58\pm0.64$ &            $-5.77\pm0.62$ &            $-4.69\pm0.33$ &            $-3.56\pm0.18$ \\
120198 &            $-5.73\pm0.55$ &            $-6.55\pm0.42$ &            $-3.44\pm0.32$ &            $-3.14\pm0.17$ \\
124224 &            $-2.75\pm0.49$ &            $-6.87\pm0.53$ &            $-5.28\pm0.42$ &            $-3.77\pm0.14$ \\
128898 &            $-4.31\pm0.41$ &            $-6.81\pm0.35$ &            $-5.42\pm0.15$ &            $-4.36\pm0.11$ \\
130559 &            $-3.08\pm0.50$ &            $-6.22\pm0.36$ &            $-3.85\pm0.23$ &            $-3.33\pm0.21$ \\
137909 &            $-3.90\pm0.63$ &            $-5.79\pm0.61$ &            $-4.54\pm0.47$ &            $-3.71\pm0.43$ \\
137949 &            $-3.54\pm0.68$ &            $-6.11\pm0.83$ &            $-4.90\pm0.45$ &            $-4.04\pm0.41$ \\
140160 &            $-3.69\pm0.13$ &            $-6.18\pm0.66$ &            $-4.59\pm0.15$ &            $-3.70\pm0.09$ \\
140728 &            $-4.47\pm1.39$ &            $-7.38\pm1.04$ &            $-3.66\pm0.24$ &            $-3.30\pm0.21$ \\
148112 &            $-4.36\pm0.87$ &            $-7.29\pm0.34$ &            $-4.29\pm0.16$ &            $-3.21\pm0.17$ \\
148898 &            $-4.67\pm0.54$ &            $-6.69\pm0.41$ &            $-4.85\pm0.17$ &            $-4.17\pm0.13$ \\
151199 &            $-4.24\pm0.47$ &            $-6.45\pm0.53$ &            $-4.88\pm0.15$ &            $-3.81\pm0.08$ \\
152107 &            $-4.05\pm0.16$ &            $-5.89\pm0.68$ &            $-4.04\pm0.29$ &            $-3.28\pm0.19$ \\
170000 &            $-3.07\pm0.48$ &            $-7.70\pm0.80$ &            $-5.38\pm0.26$ &            $-3.69\pm0.19$ \\
176232 &            $-4.62\pm1.00$ &            $-6.62\pm0.53$ &            $-5.41\pm0.33$ &            $-4.63\pm0.18$ \\
187474 &            $-3.52\pm0.55$ &            $-6.35\pm0.30$ &            $-4.77\pm0.15$ &            $-3.23\pm0.18$ \\
188041 &            $-2.86\pm0.54$ &            $-5.89\pm1.24$ &            $-3.96\pm0.31$ &            $-3.29\pm0.33$ \\
201601 &            $-4.31\pm0.76$ &            $-6.42\pm0.62$ &            $-5.33\pm0.30$ &            $-4.24\pm0.27$ \\
203006 &            $-5.25\pm0.50$ &            $-6.28\pm0.42$ &            $-3.71\pm0.35$ &            $-3.04\pm0.16$ \\
217522 &            $-4.26\pm0.66$ &            $-6.78\pm0.81$ &            $-5.96\pm0.64$ &            $-5.33\pm0.43$ \\
220825 &            $-3.71\pm1.08$ &            $-5.82\pm0.38$ &            $-3.80\pm0.34$ &            $-3.22\pm0.15$ \\
221760 &            $-3.98\pm0.47$ &            $-6.71\pm0.55$ &            $-4.54\pm0.17$ &            $-3.91\pm0.13$ \\
223640 &            $-2.78\pm0.34$ &            $-5.54\pm0.32$ &            $-4.68\pm0.15$ &            $-3.00\pm0.10$ \\
		\noalign{\vskip0.5mm}
		\hline \\
	\end{tabular*}
	\end{center}
\end{table*}
\renewcommand{\arraystretch}{1.0}

\end{document}